\documentclass[apj]{emulateapj}
\usepackage{subfigure}
\usepackage{graphics}

\def\spose#1{\hbox to 0pt{#1\hss}}
\def\simlt{\mathrel{\spose{\lower 3pt\hbox{$\mathchar"218$}}
    \raise 2.0pt\hbox{$\mathchar"13C$}}}
\def\simgt{\mathrel{\spose{\lower 3pt\hbox{$\mathchar"218$}}
    \raise 2.0pt\hbox{$\mathchar"13E$}}}

\newcommand{\oiiiw}{\mbox{[\ion{O}{3}] $\lambda$5007} $\,$}
\newcommand{\oiiiwn}{\mbox{[\ion{O}{3}] $\lambda$5007}}

\newcommand{\hbn}{\mbox{H$\beta$}}

\newcommand{\han}{\mbox{H$\alpha$}}

\newcommand{\feii}{\mbox{[\ion{Fe}{2}]} $\,$} 
\newcommand{\hi}{\mbox{\ion{H}{1}} $\,$}
\newcommand{\oiiis}{\mbox{[\ion{O}{3}] $\lambda$4363} $\,$}
\newcommand{\oiiisn}{\mbox{[\ion{O}{3}] $\lambda$4363}}
\newcommand{\oiidouble}{\mbox{[\ion{O}{2}] $\lambda\lambda$3725, 3727} $\,$} 
\newcommand{\oiiidouble}{\mbox{[\ion{O}{3}] $\lambda\lambda$4959, 5007}}
\newcommand{\oidouble}{\mbox{[\ion{O}{1}] $\lambda\lambda$6300, 6363} $\,$}
\newcommand{\oidoublen}{\mbox{[\ion{O}{1}] $\lambda\lambda$6300, 6363}}
\newcommand{\niidouble}{\mbox{[\ion{N}{2}] $\lambda\lambda$6548, 6583}}
\newcommand{\siidouble}{\mbox{[\ion{S}{2}] $\lambda\lambda$6716, 6730} $\,$}
\newcommand{\oi}{\mbox{[\ion{O}{1}] $\lambda$6300} $\,$} 
\newcommand{\oin}{\mbox{[\ion{O}{1}] $\lambda$6300}} 
\newcommand{\hii}{\mbox{\ion{H}{2}} $\,$}

\slugcomment{\it Published in ApJ}
\shortauthors{Comerford et al.}
\shorttitle{Shocks and Spatially Offset Active Galactic Nuclei Produce Velocity Offsets in Emission Lines}

\begin{document}

\title{Shocks and Spatially Offset Active Galactic Nuclei \\ Produce Velocity Offsets in Emission Lines}

\author{Julia M. Comerford\altaffilmark{1}, R. Scott Barrows\altaffilmark{1}, Jenny E. Greene\altaffilmark{2}, and David Pooley\altaffilmark{3}}

\affil{$^1$Department of Astrophysical and Planetary Sciences, University of Colorado, Boulder, CO 80309, USA}
\affil{$^2$Department of Astrophysical Sciences, Princeton University, Princeton, NJ 08544, USA}
\affil{$^3$Department of Physics and Astronomy, Trinity University, San Antonio, TX 78212, USA}

\begin{abstract}
While 2\% of active galactic nuclei (AGNs) exhibit narrow emission lines with line-of-sight velocities that are significantly offset from the velocity of the host galaxy's stars, the nature of these velocity offsets is unknown.  We investigate this question with {\it Chandra}/ACIS and {\it Hubble Space Telescope}/Wide Field Camera 3 observations of seven velocity-offset AGNs at $z<0.12$, and all seven galaxies have a central AGN but a peak in emission that is spatially offset by $<$ kpc from the host galaxy's stellar centroid.  These spatial offsets are responsible for the observed velocity offsets and are due to shocks, either from AGN outflows (in four galaxies) or gas inflowing along a bar (in three galaxies).  We compare our results to a velocity-offset AGN whose velocity offset originates from a spatially offset AGN in a galaxy merger.  The optical line flux ratios of the offset AGN are consistent with pure photoionization, while the optical line flux ratios of our sample are consistent with contributions from photoionization and shocks.  We conclude that these optical line flux ratios could be efficient for separating velocity-offset AGNs into subgroups of offset AGNs -- which are important for studies of AGN fueling in galaxy mergers -- and central AGNs with shocks -- where the outflows are biased towards the most energetic outflows that are the strongest drivers of feedback.
\end{abstract}


\section{Introduction}
\label{intro}

Galaxies and their supermassive black holes are linked in their evolution, resulting in surprisingly tight observational correlations between parameters such as supermassive black hole mass, stellar velocity dispersion, and host galaxy mass (\citealt{HE14.1} for a review).  Active galactic nuclei (AGNs) have emerged as key players in this coevolution, by the primary mechanisms of AGN fueling and AGN feedback.  Supermassive black holes build up mass by accreting gas during AGN fueling, while AGN outflows are a crucial regulator of star formation that controls the mass growth of the galaxies (e.g., \citealt{DI05.1,CR06.1,FA12.2}).  

In recent years, double-peaked narrow emission lines in AGN host galaxies have been studied as a population (e.g., \citealt{LI10.1,CO11.2,BA12.1,CO12.1,FU12.1,BA13.1,MC15.1}), and have been shown to be signatures of both AGN fueling and AGN outflows.  Some of these double-peaked emission lines are produced by dual AGNs, which are a pair of AGNs being fueled during a galaxy merger \citep{FU11.3,LI13.1,CO15.1,MU15.1}, and the majority of double-peaked emission lines are produced by AGN outflows (e.g., \citealt{RO10.1,GR12.2,NE16.1}).

Analogous to the AGNs with double-peaked narrow emission lines, there is also a population of galaxy spectra with single-peaked narrow AGN emission lines that exhibit a statistically significant line-of-sight velocity offset relative to the velocity of the host galaxy's stars; 2\% of AGNs exhibit these velocity offsets \citep{CO09.1,CO13.1}.  These objects have been much less well studied than the AGNs with double-peaked narrow emission lines, and numerical simulations of galaxy mergers show that velocity-offset emission lines can be produced by offset AGNs, which are off-nuclear AGNs in ongoing galaxy mergers (e.g., \citealt{BL13.1,ST16.1}).  Inflows or outflows of gas could also produce velocity-offset AGN emission lines (e.g., \citealt{AL15.1}).  

Here, we investigate the origins of the velocity-offset narrow emission lines observed in the Sloan Digital Sky Survey (SDSS) spectra of seven AGNs at $z<0.12$.  We observe each galaxy with the {\it Chandra X-ray Observatory} ACIS ({\it Chandra}/ACIS), to pinpoint the location of the AGN, and the {\it Hubble Space Telescope} Wide Field Camera 3 ({\it HST}/WFC3), to obtain high spatial resolution maps of the stellar continuum and the ionized gas.  Our goal is to determine the nature of each galaxy and whether its velocity-offset emission lines are tracers of AGN fueling (via inflows or offset AGNs) or AGN feedback (via outflows). 

This paper is organized as follows: In Section 2 we describe the sample selection and characteristics.  In Section 3 we describe the observations of the sample (SDSS spectra, Keck/OSIRIS integral-field spectroscopy for three of the seven galaxies, {\it Chandra} observations, and {\it HST}/WFC3 multiband imaging), the astrometry, and our analyses of the data.  Section 4 presents our results, including the nature of each velocity-offset AGN.  Finally, our conclusions are summarized in Section 5.

We assume a Hubble constant $H_0 =70$ km s$^{-1}$ Mpc$^{-1}$, $\Omega_m=0.3$, and $\Omega_\Lambda=0.7$ throughout, and all distances are given in physical (not comoving) units.

\section{The Sample}

We begin with a parent sample of 18,314 Type 2 AGNs at $z<0.21$ in SDSS, which were identified as AGNs via their optical emission line ratios \citep{BR04.1} and the requirement that the fits to the absorption and emission line systems in the SDSS spectra are robust (by examining the signal, residual noise, and statistical noise; \citealt{OH11.1}).  The line-of-sight velocity offsets of the emission lines relative to the stellar absorption lines were then measured.  From the parent sample of 18,314 Type 2 AGNs, the velocity-offset AGNs were the systems that fulfilled the following four criteria: 1) the velocity offsets of the forbidden emission lines and the Balmer emission lines are the same to within $1\sigma$; 2) the velocity offsets of the emission lines are greater than $3\sigma$ in significance; 3) the emission line profiles are symmetric; 4) the systems do not have double-peaked emission lines.  

The 351 AGNs that meet these criteria are the velocity-offset AGNs \citep{CO14.1}.  From these 351 velocity-offset AGNs, we select seven systems with low redshifts ($z<0.12$) and high estimated 2-10 keV fluxes ($>5 \times 10^{-14}$ erg cm$^{-2}$ s$^{-1}$).  We estimate the 2-10 keV fluxes from the \oiiiw fluxes of the AGNs (which are $>1.3 \times 10^{-14}$ erg cm$^{-2}$ s$^{-1}$ for this sample of seven systems) and the established Type 2 AGN \oiiiw to X-ray scaling relation \citep{HE05.1}.  The low redshifts maximize the physical spatial resolution that we can achieve with {\it Chandra} and {\it HST}, while the high 2-10 keV fluxes minimize the observing time necessary for X-ray detections.  The seven systems are listed in Table~\ref{tbl-1}.

\section{Observations and Analysis}

\subsection{Optical SDSS Observations}
\label{sdss}

For each of the seven velocity-offset AGNs, the host galaxy redshift (based on the stellar absorption features), the line-of-sight velocity offset of the emission lines, and the \oiiiw luminosity were determined from the SDSS spectrum \citep{CO14.1}.  Three of the AGNs have emission lines with redshifted velocity offsets, and four have emission lines with blueshifted velocity offsets.  The absolute values of the velocity offsets range from 50 to 113 km s$^{-1}$ (Table~\ref{tbl-1}).

\begin{deluxetable*}{llll}
\tabletypesize{\scriptsize}
\tablewidth{0pt}
\tablecolumns{4}
\tablecaption{Measurements from SDSS Observations} 
\tablehead{
\colhead{SDSS Designation} &
\colhead{$z$} &
\colhead{$\Delta v$} &
\colhead{$L_{\oiiiwn}$} \\ 
& & (km s$^{-1}$) & ($10^{40}$ erg s$^{-1}$) 
}
\startdata
SDSS J013258.92$-$102707.0 & $0.03222 \pm 0.00002$ & \phn \phn \phd $56 \pm 10$ & \phn $5.5 \pm 0.6$  \\
SDSS J083902.97+470756.3 & $0.05236 \pm 0.00006$ & \phn $-50 \pm 10$ & $17.7 \pm 1.2$  \\
SDSS J105553.64+152027.4 & $0.09201 \pm 0.00002$ & $-113 \pm 10$ & $48.1 \pm 5.7$ \\ 
SDSS J111729.22+614015.2 & $0.11193 \pm 0.00001$ & \phn \phn \phd $85 \pm 12$ & $44.5 \pm 7.2$  \\ 
SDSS J134640.79+522836.6 & $0.02918 \pm 0.00001$ & \phn $-52 \pm 10$ & \phn $7.5 \pm 0.8$  \\ 
SDSS J165430.72+194615.5 & $0.05367 \pm 0.00001$ & \phn \phn \phd $66 \pm 11$ & $20.7 \pm 2.8$  \\
SDSS J232328.01+140530.2 & $0.04142 \pm 0.00007$ & \phn $-53 \pm 10$ & $10.2 \pm 1.3$ 
\enddata
\tablecomments{Column 2: host galaxy redshift, based on stellar absorption features.  Column 3: line-of-sight velocity offset of emission lines relative to host galaxy systemic.  Column 4: observed \oiiiw luminosity.}
\label{tbl-1}
\end{deluxetable*}

\subsection{Keck/OSIRIS Near-infrared IFU Observations}
\label{osiris}

Three of the velocity-offset AGNs were observed with Keck Laser Guide Star Adaptive Optics with OH-Suppressing Infra-Red Imaging Spectrograph (OSIRIS) integral-field spectroscopy \citep{MU16.1}.  In each galaxy (SDSS J1055+1520, SDSS J1117+6140, and SDSS J1346+5228), the peak of the line emission (Pa$\alpha$, Pa$\alpha$, and \feii in each galaxy, respectively) was spatially offset from the galaxy center by $0\farcs1$ (0.2 kpc), $0\farcs2$ (0.5 kpc), and $0\farcs3$ (0.2 kpc), respectively.  Based on the kinematics of the gas in the OSIRIS observations, \cite{MU16.1} found that SDSS J1055+1520 and SDSS J1346+5228 host AGN outflows while SDSS J1117+6140 has gas inflow along a bar.  They concluded that the spatially-offset peaks in line emission are the result of the outflows or inflows driving shocks into off-nuclear gas.

\subsection{Chandra/ACIS X-ray Observations}
\label{chandra}

The seven velocity-offset AGNs were observed with {\it Chandra}/ACIS for the program GO4-15113X (PI: Comerford).  Our exposure times were derived from the observed \oiiiw flux for each system (Table~\ref{tbl-1}) and the scaling relation between \oiiiw flux and hard X-ray (2-10 keV) flux for Type 2 AGNs, which has a scatter of 1.06 dex \citep{HE05.1}.  We selected exposure times that would ensure a firm detection of at least 10 counts for each AGN, even in the case of the actual X-ray flux falling in the low end of the 1.06 dex scatter.  The galaxies were observed with exposure times of 10 ks to 20 ks (Table~\ref{tbl-2}).

\begin{deluxetable*}{lllllll}
\tabletypesize{\scriptsize}
\tablewidth{0pt}
\tablecolumns{7}
\tablecaption{Summary of Chandra and HST Observations} 
\tablehead{
\colhead{SDSS Name} &
\colhead{{\it Chandra}/ACIS} & 
\colhead{{\it Chandra}/ACIS} & 
\colhead{{\it HST}/WFC3} & 
\colhead{{\it HST}/WFC3} &
\colhead{{\it HST/WFC3}} & 
\colhead{{\it HST/WFC3}} \\ 
 & exp. time (s) & obs. date (UT) & F160W & F606W & F438W & obs. date (UT)  \\
 & & & exp. time (s) & exp. time (s) & exp. time (s) & 
}
\startdata
J0132$-$1027 & 14871 & 2014-08-23 & 147 & 900 & 1047 & 2014-06-24 \\
J0839+4707 & 9927 & 2014-09-03 & 147 & 945 & 1050 & 2014-09-06 \\
J1055+1520 & 14869 & 2015-02-04 & 147 & 900 & 957 & 2014-10-25 \\
J1117+6140 & 19773 & 2015-02-03 & 147 & 1062 & 1065 & 2014-07-03 \\ 
J1346+5228 & 9937 & 2014-08-29 & 147 & 996 & 1050 & 2015-02-05 \\
J1654+1946 & 9937 & 2014-07-23 & 147 & 900 & 957 & 2014-07-27 \\
J2323+1405 & 14868 & 2014-08-31 & 147 & 900 & 954 & 2014-06-08 
\enddata
\tablecomments{Column 2: exposure time for the {\it Chandra}/ACIS observation.  Column 3: UT date of the {\it Chandra}/ACIS observation.  Columns 4 -- 6: exposure times for the {\it HST}/WFC3 F160W, F606W, and F438W observations.  Column 7: UT date of the {\it HST}/WFC3 observations.}
\label{tbl-2}
\end{deluxetable*}

The galaxies were observed with the telescope aimpoint on the ACIS S3 chip in ``timed exposure'' mode and telemetered to the ground in ``faint'' mode.  We reduced the data with the latest {\it Chandra} software (CIAO\,4.6.1) in combination with the most recent set of calibration files (CALDB\,4.6.2).

\begin{figure*}
\begin{center}
\includegraphics[height=1.5in]{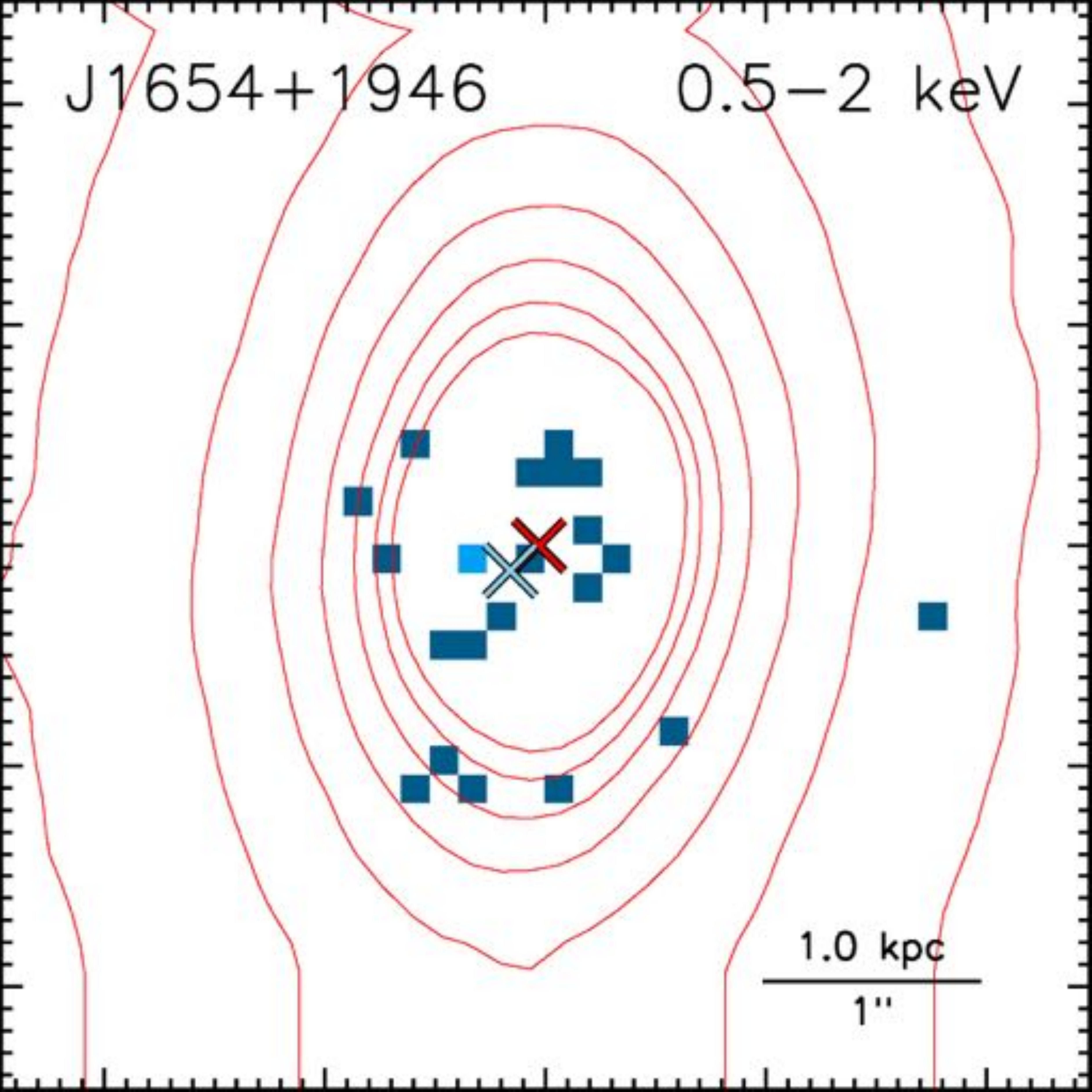}
\includegraphics[height=1.5in]{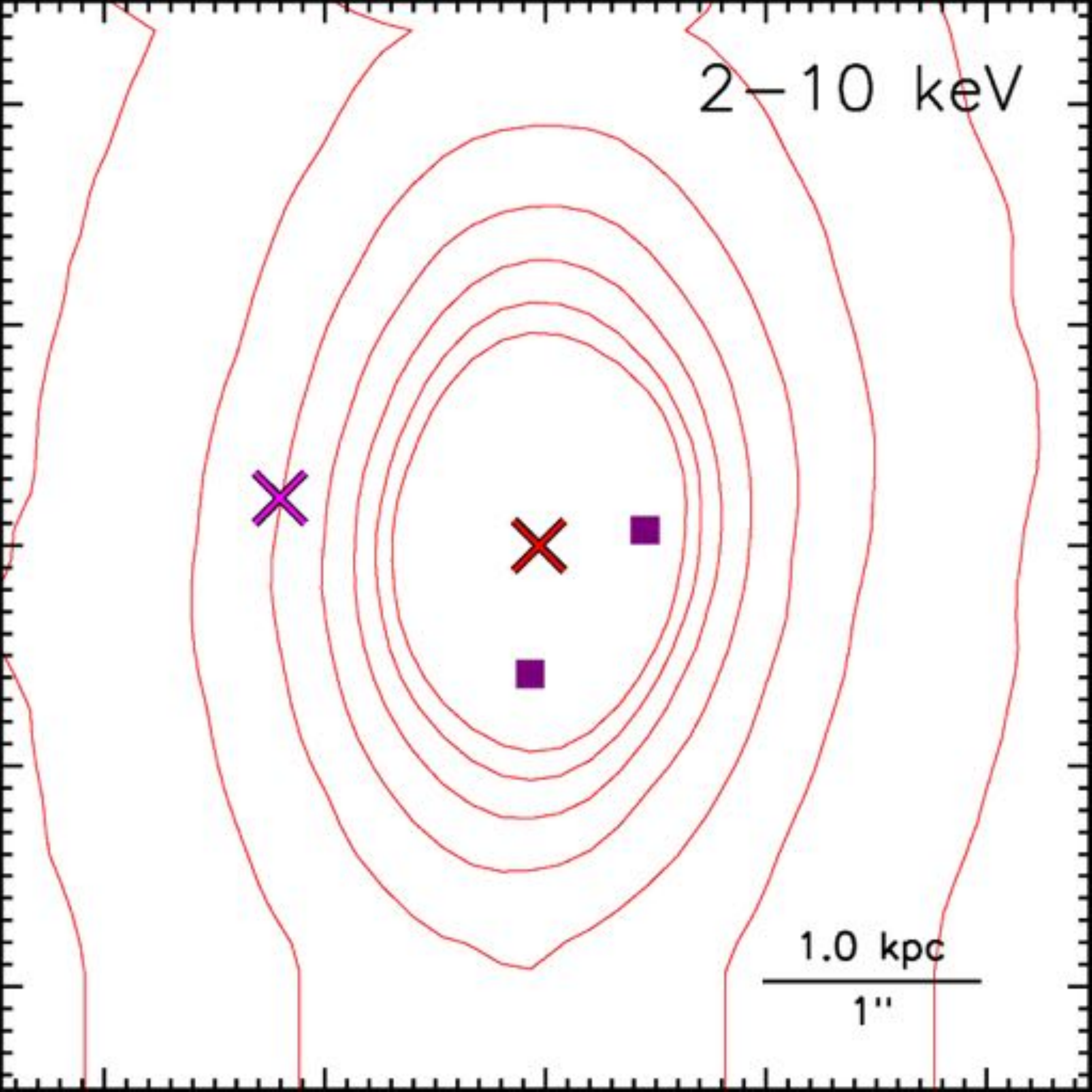}
\includegraphics[height=1.5in]{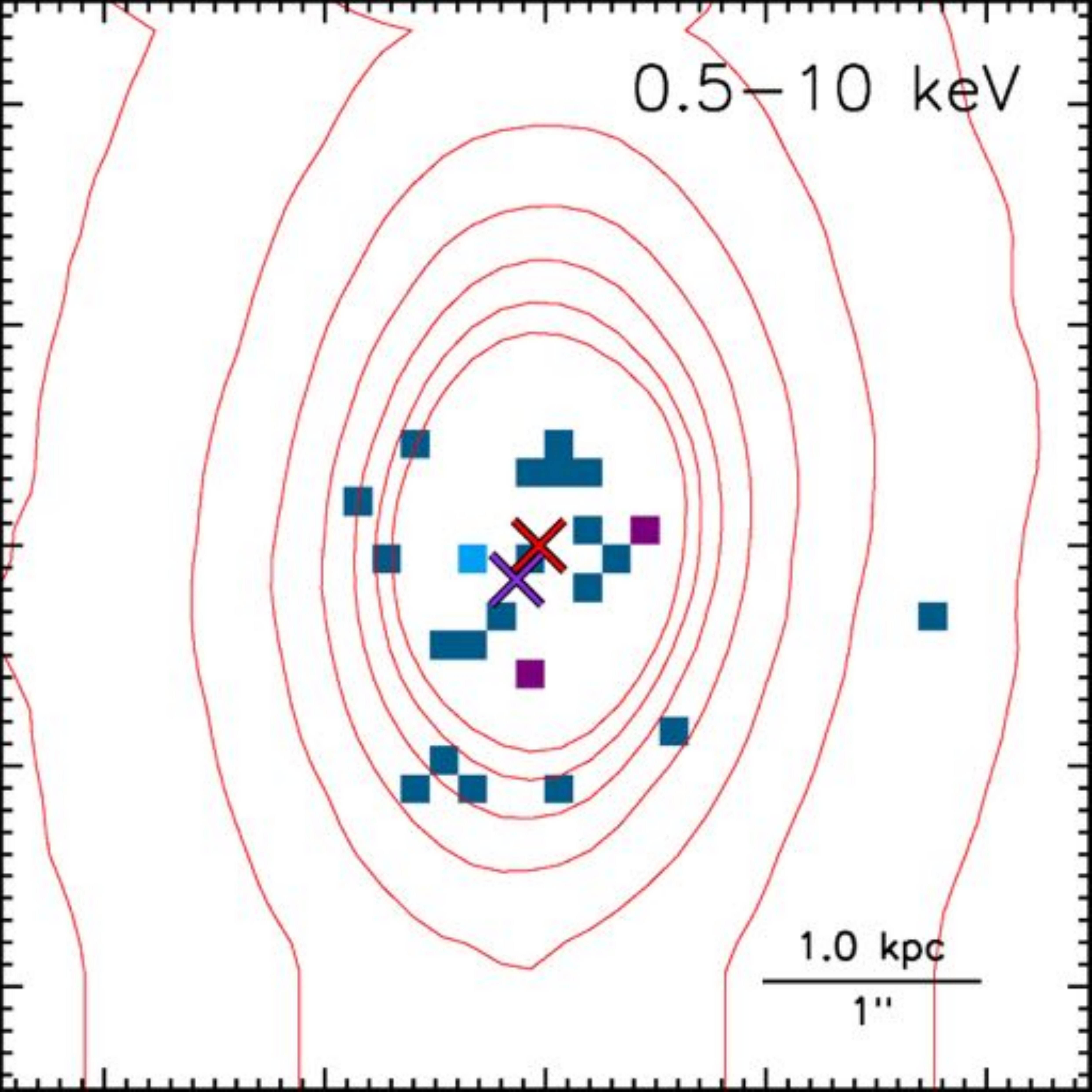}
\includegraphics[height=1.5in]{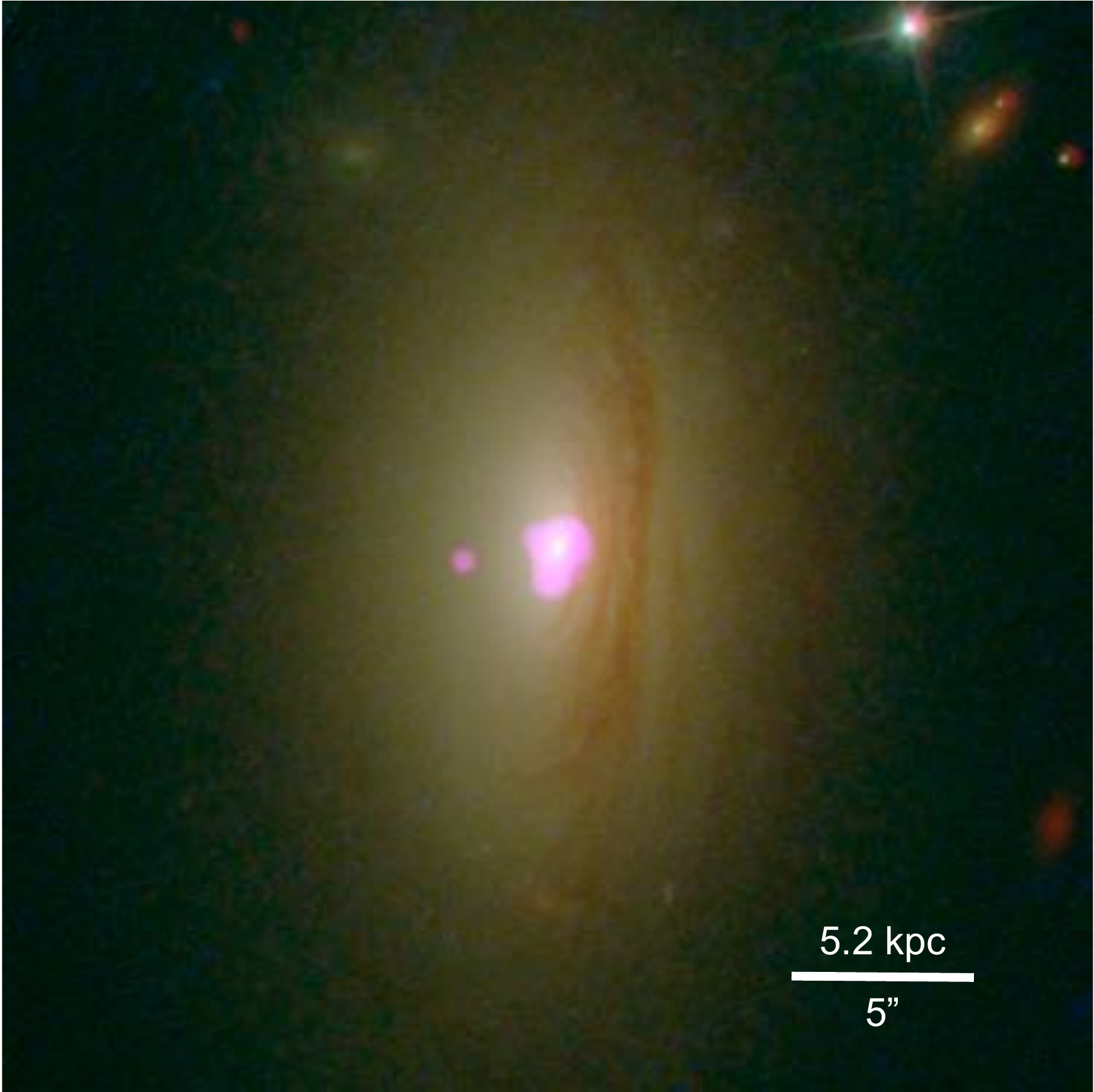}
\includegraphics[height=1.5in]{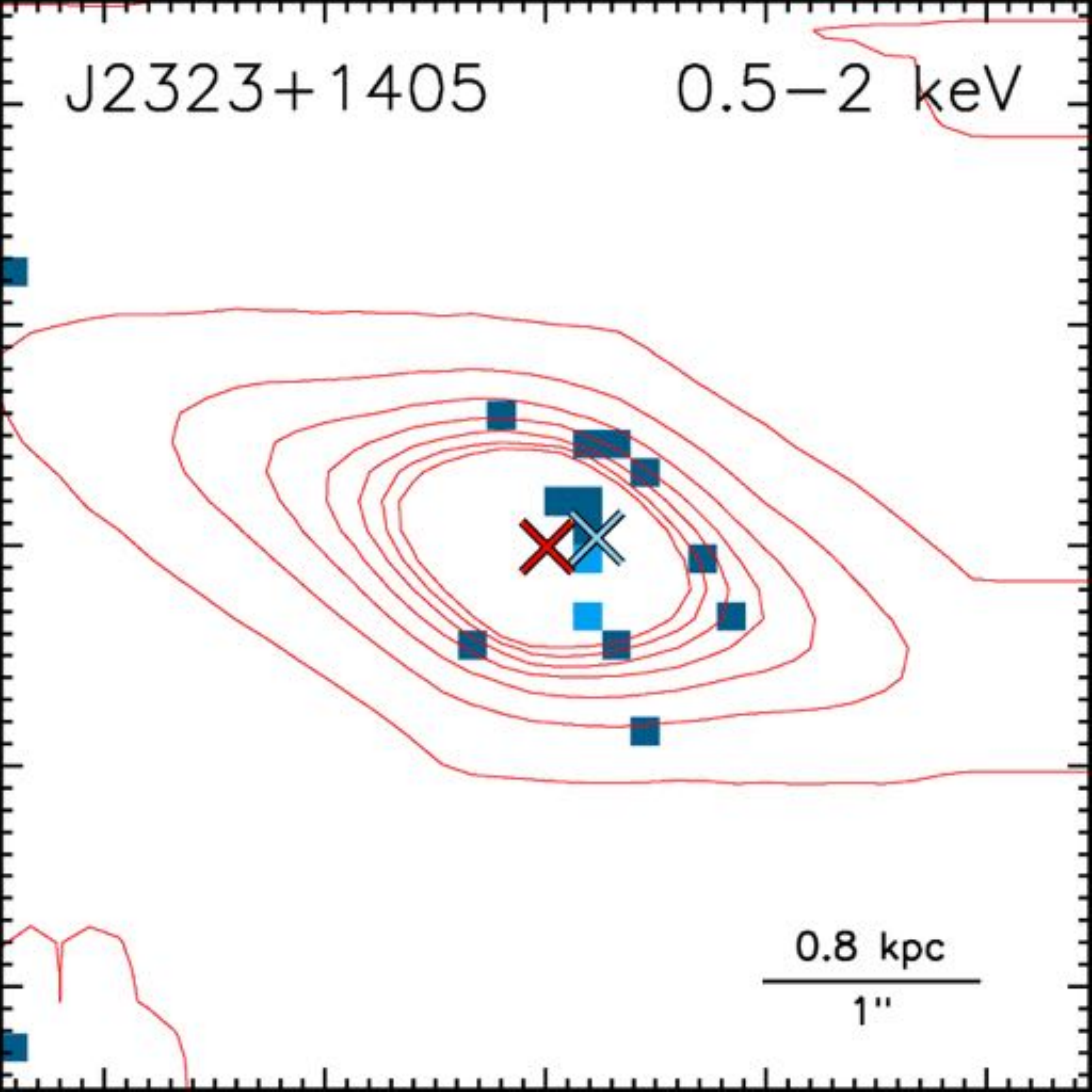}
\includegraphics[height=1.5in]{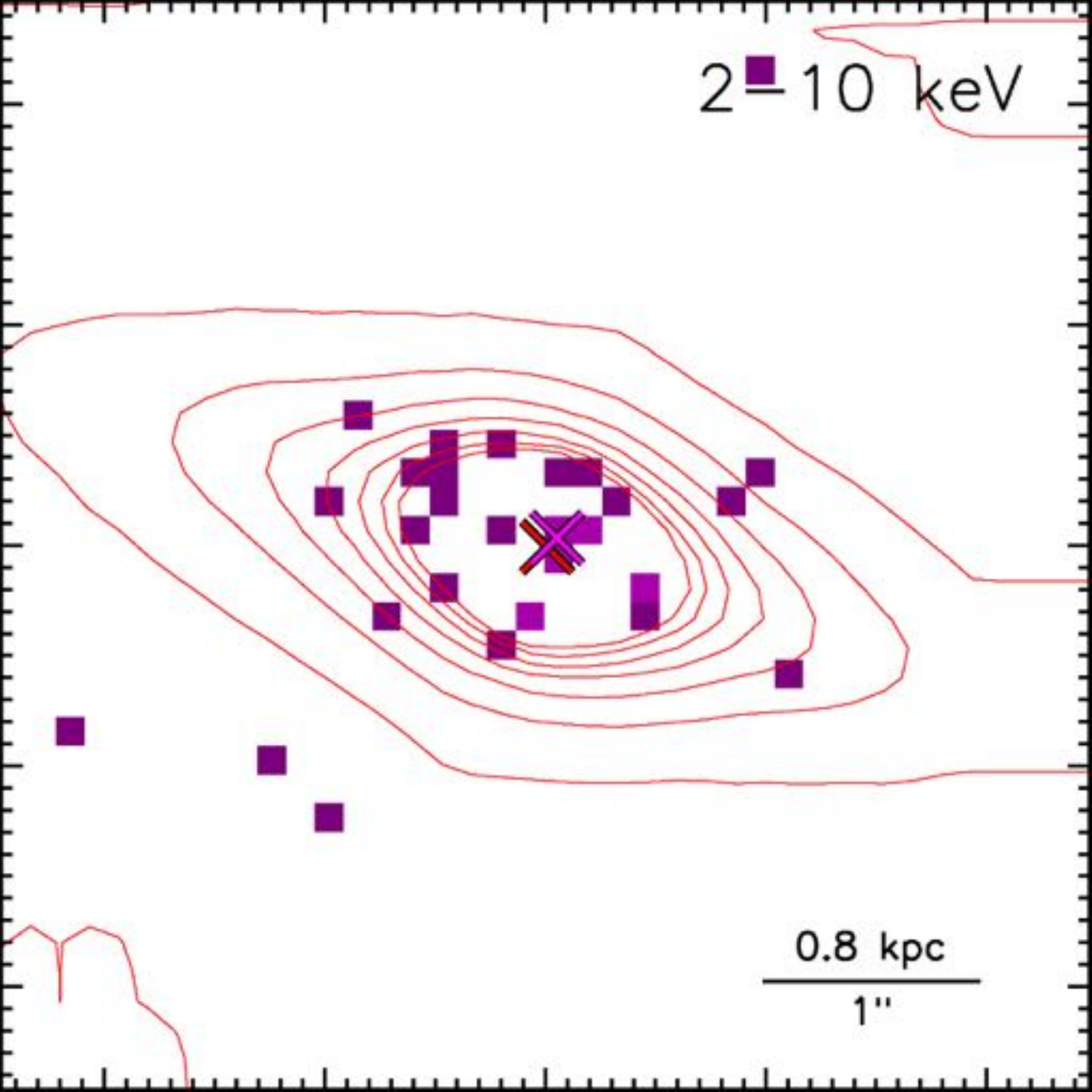}
\includegraphics[height=1.5in]{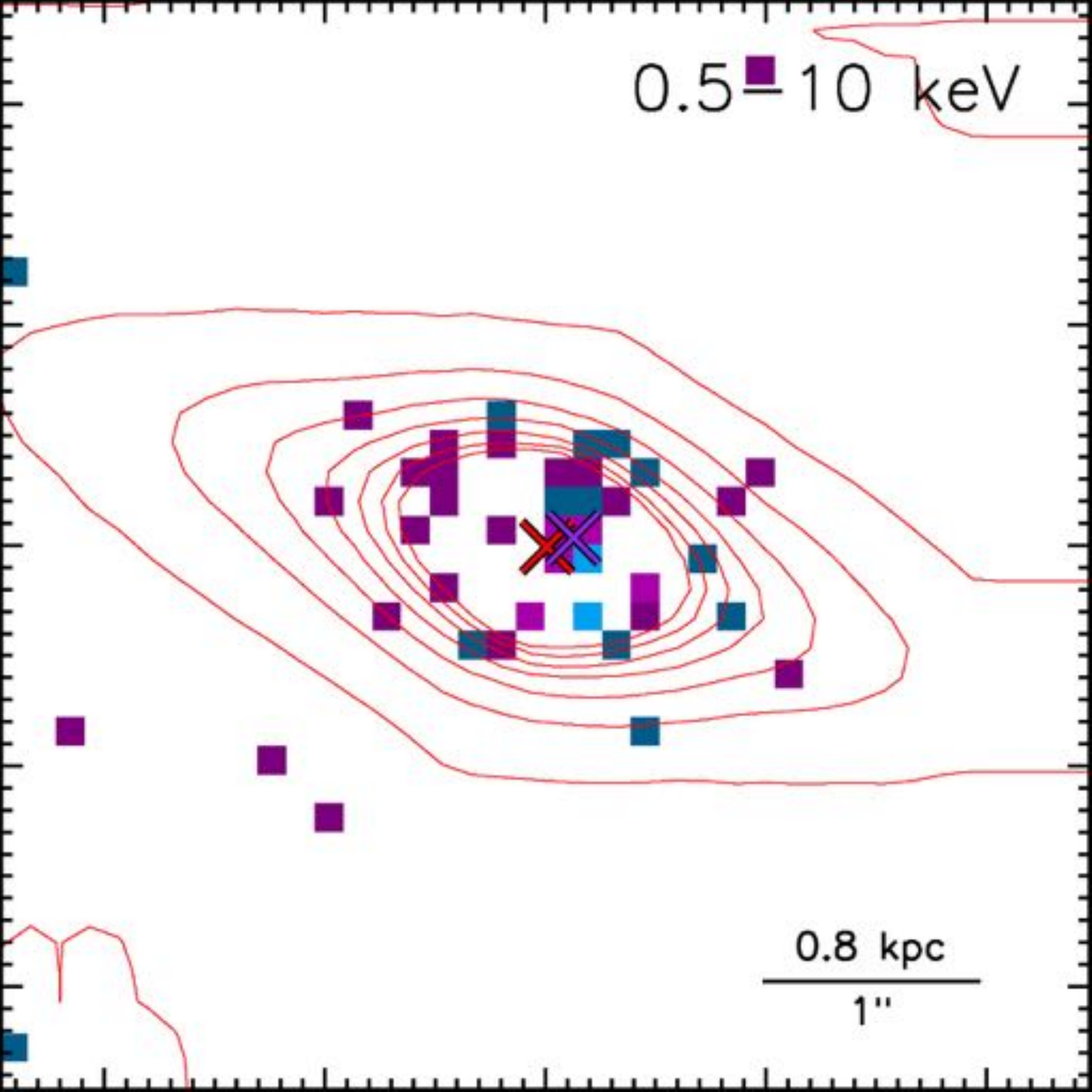}
\includegraphics[height=1.5in]{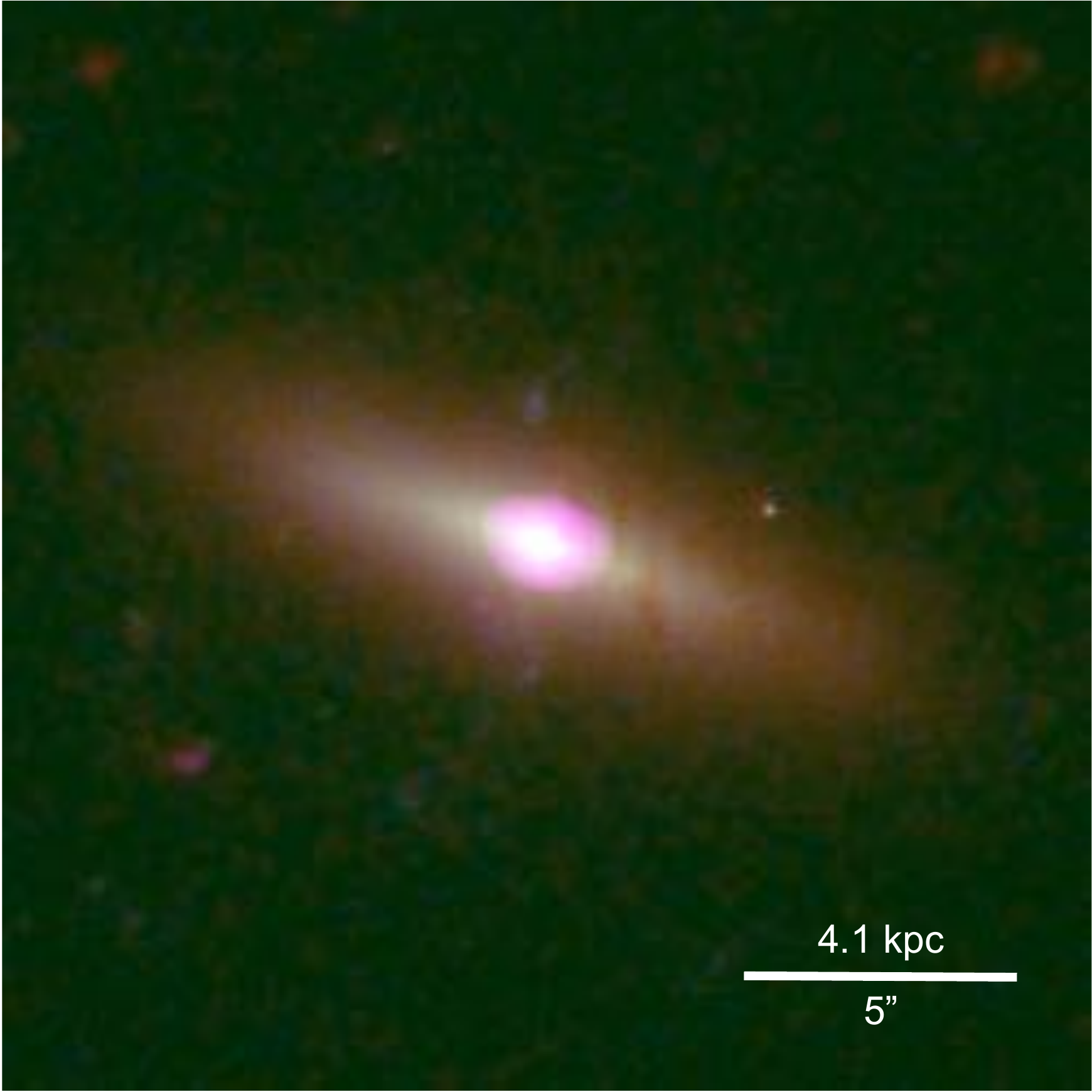}
\end{center}
\caption{From left to right: {\it Chandra} restframe $0.5-2$ keV (soft) observations, {\it Chandra} restframe $2-10$ keV (hard) observations, {\it Chandra} restframe $0.5-10$ keV (total) observations, and four-color combined {\it Chandra} and {\it HST} observations for each velocity-offset AGN.   The leftmost three panels show $5^{\prime\prime} \times 5^{\prime\prime}$ images centered on the host galaxy (red crosses indicate the centers of the host galaxies, based on {\it HST}/F160W observations).  The red curves show linear contours for the {\it HST}/F160W observations.  The pixels are one-fourth size {\it Chandra} pixels (soft X-rays shown in blue and hard X-rays shown in magenta), and the blue, magenta, and purple crosses indicate the best-fit locations of the soft, hard, and total X-ray sources, respectively.  The rightmost panels show four-color images of {\it HST} F160W (red), F606W (green), F438W (blue), and {\it Chandra} restframe $0.5-10$ keV (purple, one-twelfth size pixels smoothed with a 16 pixel radius Gaussian kernel) observations, where the {\it HST} and {\it Chandra} images have been aligned using the astrometric shifts described in Section~\ref{astrometry}.  In all panels, north is up and east is to the left.}
\label{fig:chandraresults}
\end{figure*}

For each galaxy, we used \texttt{dmcopy} to make a sky image of the field in the rest-frame soft ($0.5-2$ keV), hard ($2-10$ keV) and total ($0.5-10$ keV) energy ranges.  Using the modeling facilities in \texttt{Sherpa}, we simultaneously modeled the source as a two-dimensional Lorenztian function (\texttt{beta2d}: $f(r)=A(1+[r/r_{0}]^{2})-\alpha$) and the background as a fixed count rate estimated using a source-free adjacent circular region of $30^{\prime\prime}$ radius.  We used the SDSS galaxy coordinates as the initial position of the \texttt{beta2d} component, and then we allowed the model to fit a region of 3 times the PSF size (estimated with \texttt{psfSize}) at that location.  We determined the best-fit model parameters with \texttt{Sherpa}'s implementation of the `Simplex' minimization algorithm \citep{LA98.1}, by minimizing the Cash statistic.  We also attempted a two-component \texttt{beta2d} model to test for additional sources, but all secondary components were detected with $<1\sigma$ significance.  Therefore, none of the systems require a secondary component, and Table~\ref{tbl-3} and Figure~\ref{fig:chandraresults} show the best-fit positions of the X-ray source in each galaxy.  Table~\ref{tbl-3} also gives the spatial separations between each X-ray source and the host galaxy's stellar nucleus.  The errors on these separations are dominated by the astrometric uncertainties in aligning the {\it Chandra} and {\it HST} images.  These astrometric erros are calculated in Section~\ref{astrometry}, and the median astrometric error is $0\farcs5$.

\begin{deluxetable*}{lllllllll}
\tabletypesize{\scriptsize}
\tablewidth{0pt}
\tablecolumns{9}
\tablecaption{Chandra and HST/F160W Positions of Each Source} 
\tablehead{
\colhead{SDSS Name} &
\colhead{RA$_{HST/F160W}$} &
\colhead{DEC$_{HST/F160W}$} &
\colhead{{\it Chandra} Energy} &
\colhead{RA$_{Chandra}^a$} & 
\colhead{DEC$_{Chandra}^a$} &
\colhead{$\Delta \theta (^{\prime\prime})^b$} &
\colhead{$\Delta x$ (kpc)$^b$} &
\colhead{Sig.} \\
& & & Range (keV) & & & &
}
\startdata
J0132$-$1027 & 01:32:58.927 & $-$10:27:07.05 & $0.5-2$ & 01:32:58.924 & $-$10:27:06.87 & $0.18 \pm 0.33$ & $0.12 \pm 0.21$ & $0.6\sigma$ \\
 & & & \phn \phd $2-10$ & 01:32:58.917 & $-$10:27:07.05 & $0.14 \pm 0.46$ & $0.09 \pm 0.29$ & $0.3\sigma$ \\
 & & & $0.5-10$ & 01:32:58.922 & $-$10:27:07.02 & $0.08 \pm 0.43$ & $0.05 \pm 0.27$ & $0.2\sigma$ \\
\hline
J0839+4707 & 08:39:02.949 & +47:07:55.88 & $0.5-2$ & 08:39:02.944 & +47:07:55.95 & $0.09 \pm 0.29$ & $0.09 \pm 0.30$ & $0.3\sigma$ \\
 & & & \phn \phd $2-10$ & 08:39:02.961 & +47:07:55.84 & $0.13 \pm 0.19$ & $0.13 \pm 0.19$ & $0.7\sigma$ \\
 & & & $0.5-10$ & 08:39:02.961 & +47:07:55.88 & $0.12 \pm 0.18$ & $0.12 \pm 0.18$ & $0.7\sigma$ \\
\hline
J1055+1520 & 10:55:53.644 & +15:20:27.87 & $0.5-2$ & 10:55:53.653 & +15:20:27.40 & $0.49 \pm 0.84$ & $0.83 \pm 1.44$ & $0.6\sigma$ \\
 & & & \phn \phd $2-10$ &  10:55:53.682 & +15:20:27.16 & $0.90 \pm 0.84$ & $1.54 \pm 1.44$ & $1.1\sigma$ \\
 & & & $0.5-10$ &  10:55:53.662 & +15:20:27.30 & $0.62 \pm 0.84$ & $1.07 \pm 1.44$ & $0.7\sigma$ \\
\hline
J1117+6140 & 11:17:29.208 & +61:40:15.38 & $0.5-2$ & 11:17:29.193 & +61:40:16.06 & $0.69 \pm 0.73$ & $1.41 \pm 1.49$ & $0.9\sigma$ \\
 & & & \phn \phd $2-10$ & 11:17:29.287 & +61:40:15.63 & $0.62 \pm 0.37$ & $1.26 \pm 0.76$ & $1.7\sigma$ \\
 & & & $0.5-10$ & 11:17:29.268 & +61:40:15.56 & $0.46 \pm 0.36$ & $0.94 \pm 0.74$ & $1.3\sigma$ \\
\hline
J1346+5228 & 13:46:40.812 & +52:28:36.22 & $0.5-2$ & 13:46:40.816 & +52:28:36.15 & $0.08 \pm 0.36$ & $0.05 \pm 0.21$ & $0.2\sigma$ \\
 & & & \phn \phd $2-10$ & 13:46:40.821 & +52:28:35.76 & $0.48 \pm 0.35$ & $0.28 \pm 0.20$ & $1.4\sigma$ \\
 & & & $0.5-10$ & 13:46:40.816 & +52:28:35.76 & $0.47 \pm 0.35$ & $0.27 \pm 0.21$ & $1.3\sigma$ \\
\hline
J1654+1946 & 16:54:30.724 & +19:46:15.56 & $0.5-2$ & 16:54:30.734 & +19:46:15.45 & $0.17 \pm 0.48$ & $0.18 \pm 0.50$ & $0.4\sigma$ \\
 & & & \phn \phd $2-10$ & 16:54:30.806 & +19:46:15.78 & $1.17 \pm 0.44$ & $1.22 \pm 0.46$ & $2.6\sigma$ \\
 & & & $0.5-10$ & 16:54:30.732 & +19:46:15.42 & $0.18 \pm 0.44$ & $0.18 \pm 0.46$ & $0.4\sigma$ \\
\hline
J2323+1405 & 23:23:28.010 & +14:05:30.08 & $0.5-2$ & 23:23:27.996 & +14:05:30.12 & $0.21 \pm 0.26$ & $0.17 \pm 0.22$ & $0.8\sigma$ \\
 & & & \phn \phd $2-10$ & 23:23:28.008 & +14:05:30.12 & $0.05 \pm 0.22$ & $0.04 \pm 0.18$ & $0.2\sigma$ \\
 & & & $0.5-10$ & 23:23:28.003 & +14:05:30.12 & $0.11 \pm 0.21$ & $0.09 \pm 0.17$ & $0.5\sigma$ 
\enddata
\tablecomments{Columns 2 and 3: coordinates of the host galaxy's stellar nucleus, measured from {\it HST}/WFC3/F160W observations. Column 4: rest-frame energy range of {\it Chandra} observations.  Columns 5 and 6: coordinates of the X-ray AGN source, measured from {\it Chandra}/ACIS observations in the energy range given in Column 4. Columns 7 and 8: angular and physical separations between the positions of the host galaxy's stellar nucleus and the X-ray AGN source, where the error includes uncertainties in the positions of the {\it HST} and {\it Chandra} sources as well as the astrometric uncertainty.  Column 9: significance of the separation between the host galaxy's stellar nucleus and the X-ray AGN source.}
\tablenotetext{a}{The astrometric shifts described in Section~\ref{astrometry} have been applied to the {\it Chandra} source positions.}
\tablenotetext{b}{The errors are dominated by the astrometric uncertainties, which range from $0\farcs2$ to $0\farcs8$.}
\label{tbl-3}
\end{deluxetable*}

\begin{figure*}
\begin{center}
\includegraphics[height=1.5in]{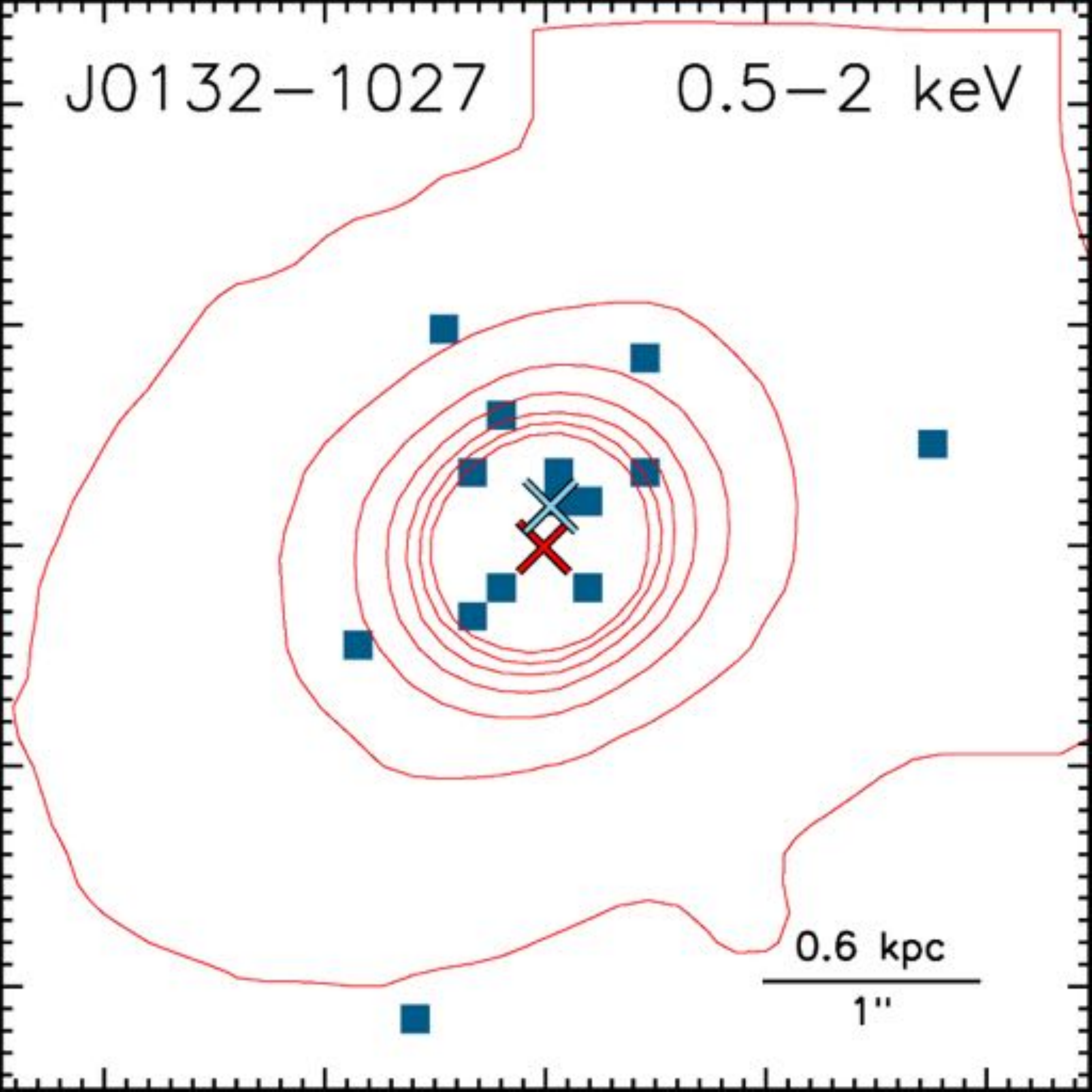}
\includegraphics[height=1.5in]{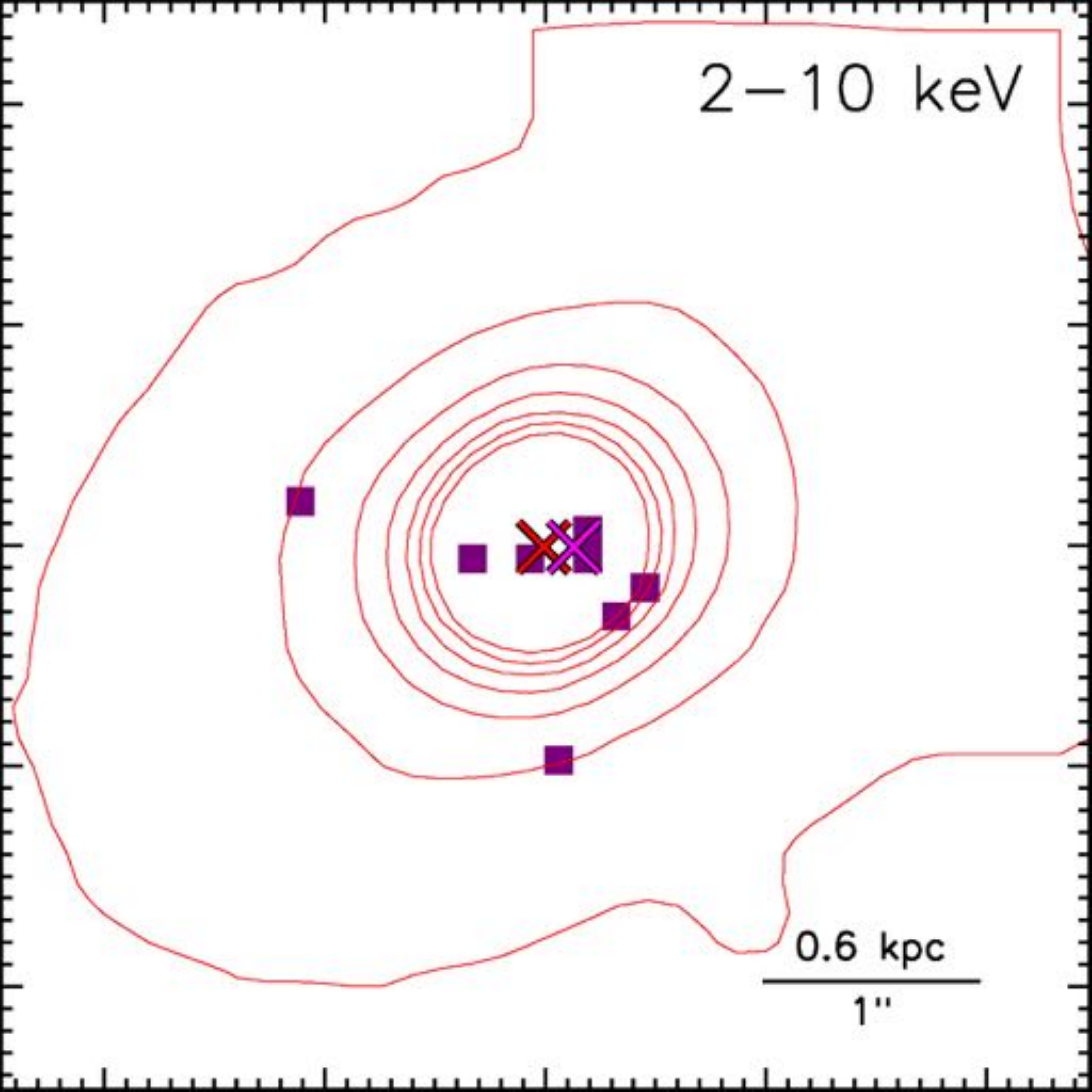}
\includegraphics[height=1.5in]{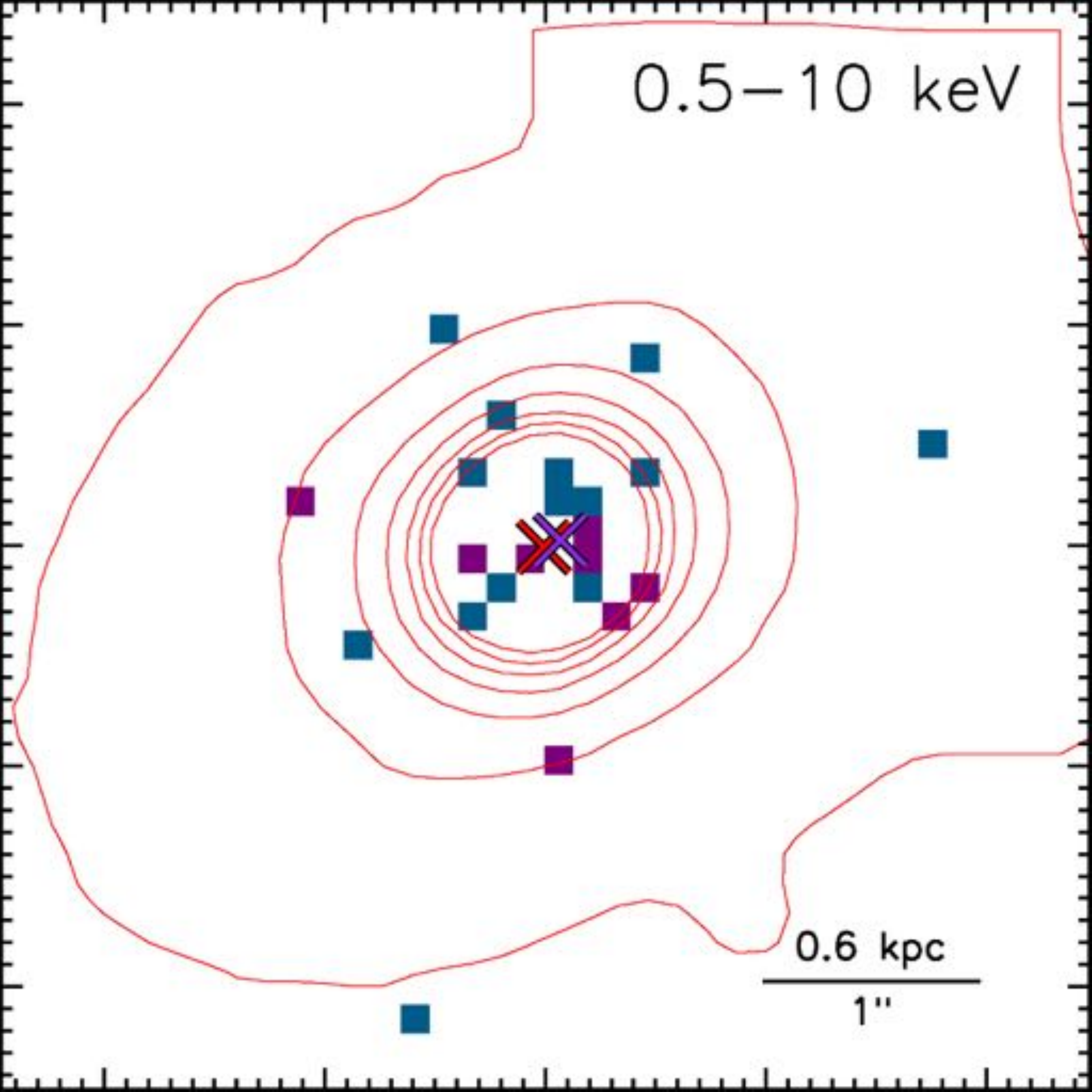}
\includegraphics[height=1.5in]{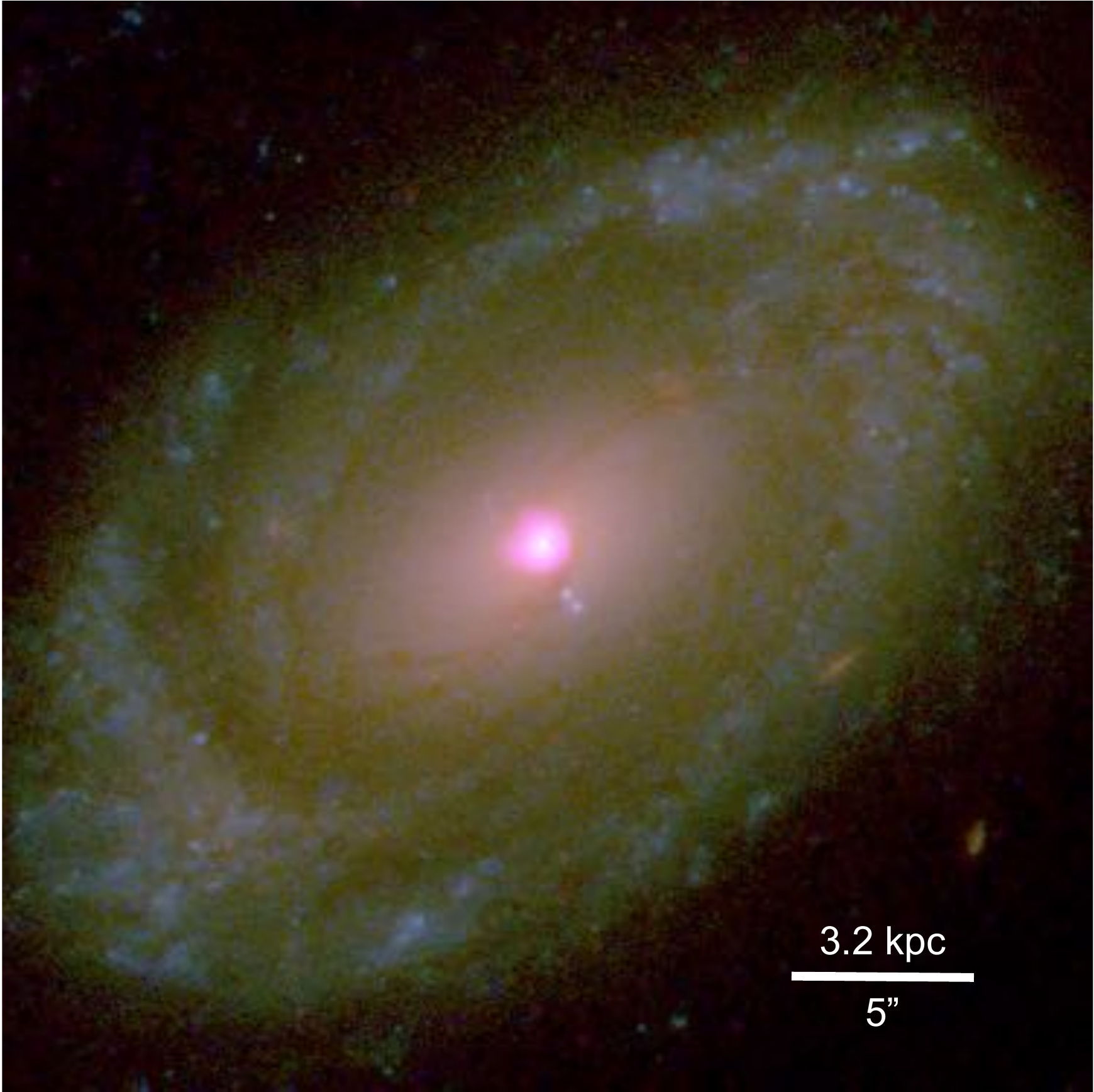}
\includegraphics[height=1.5in]{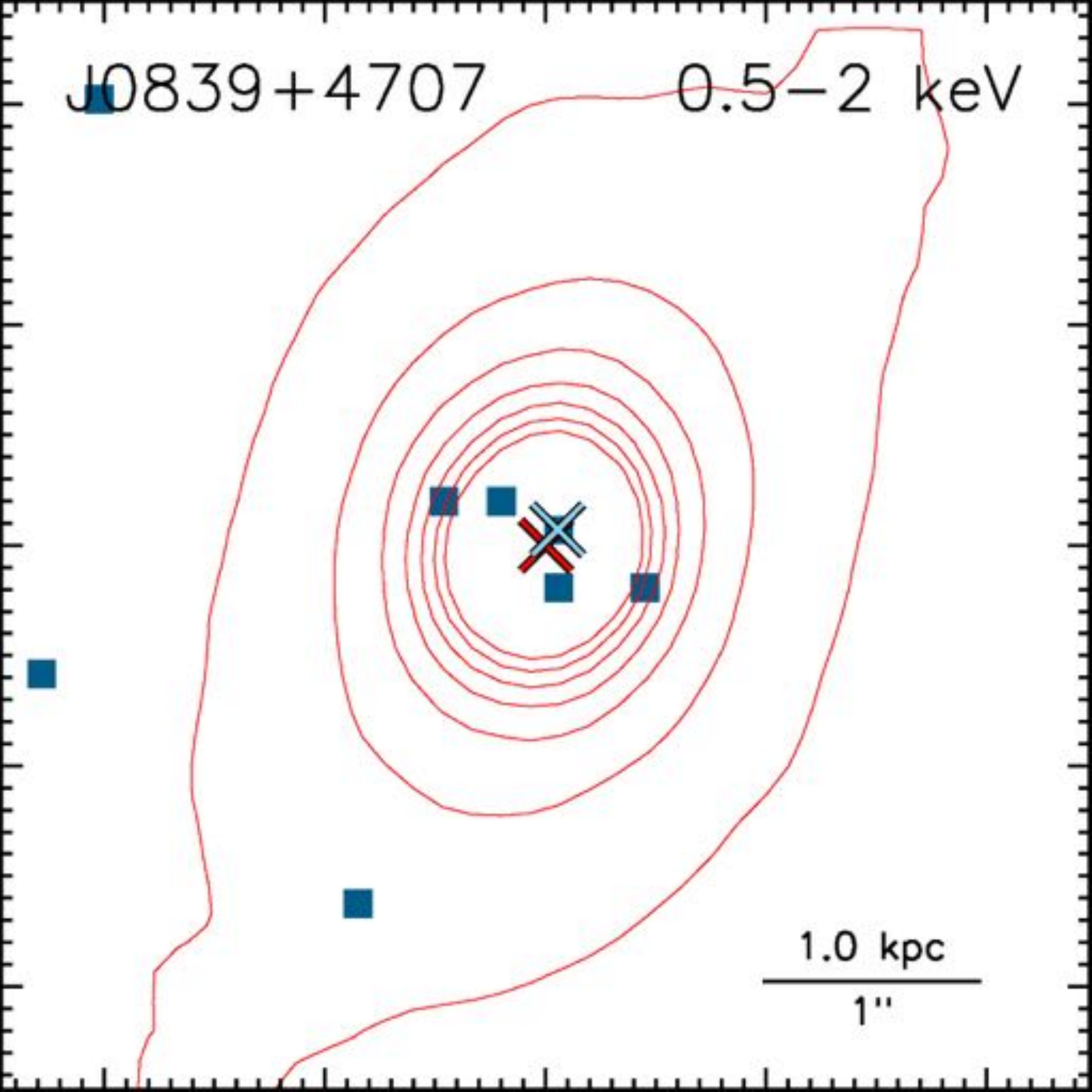}
\includegraphics[height=1.5in]{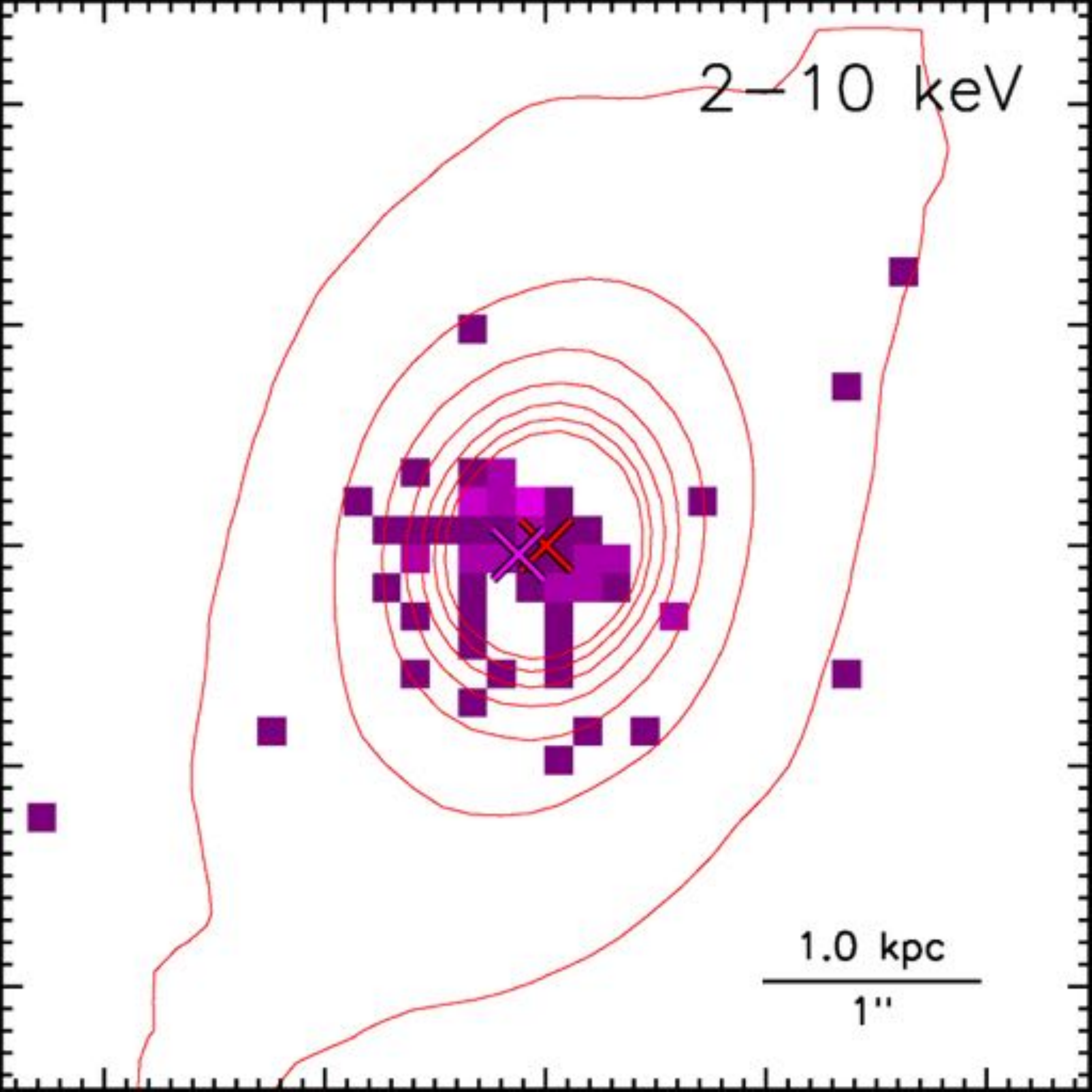}
\includegraphics[height=1.5in]{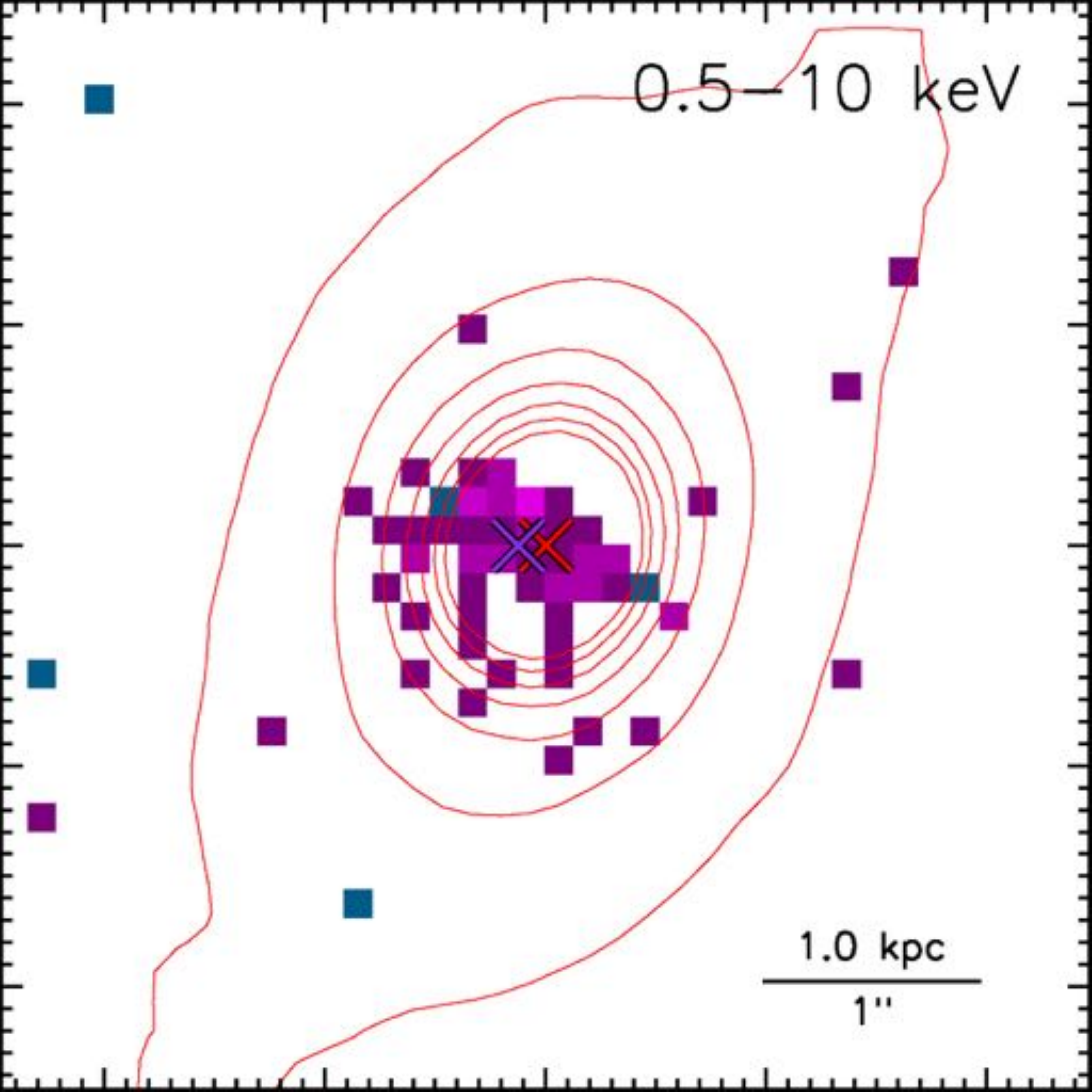}
\includegraphics[height=1.5in]{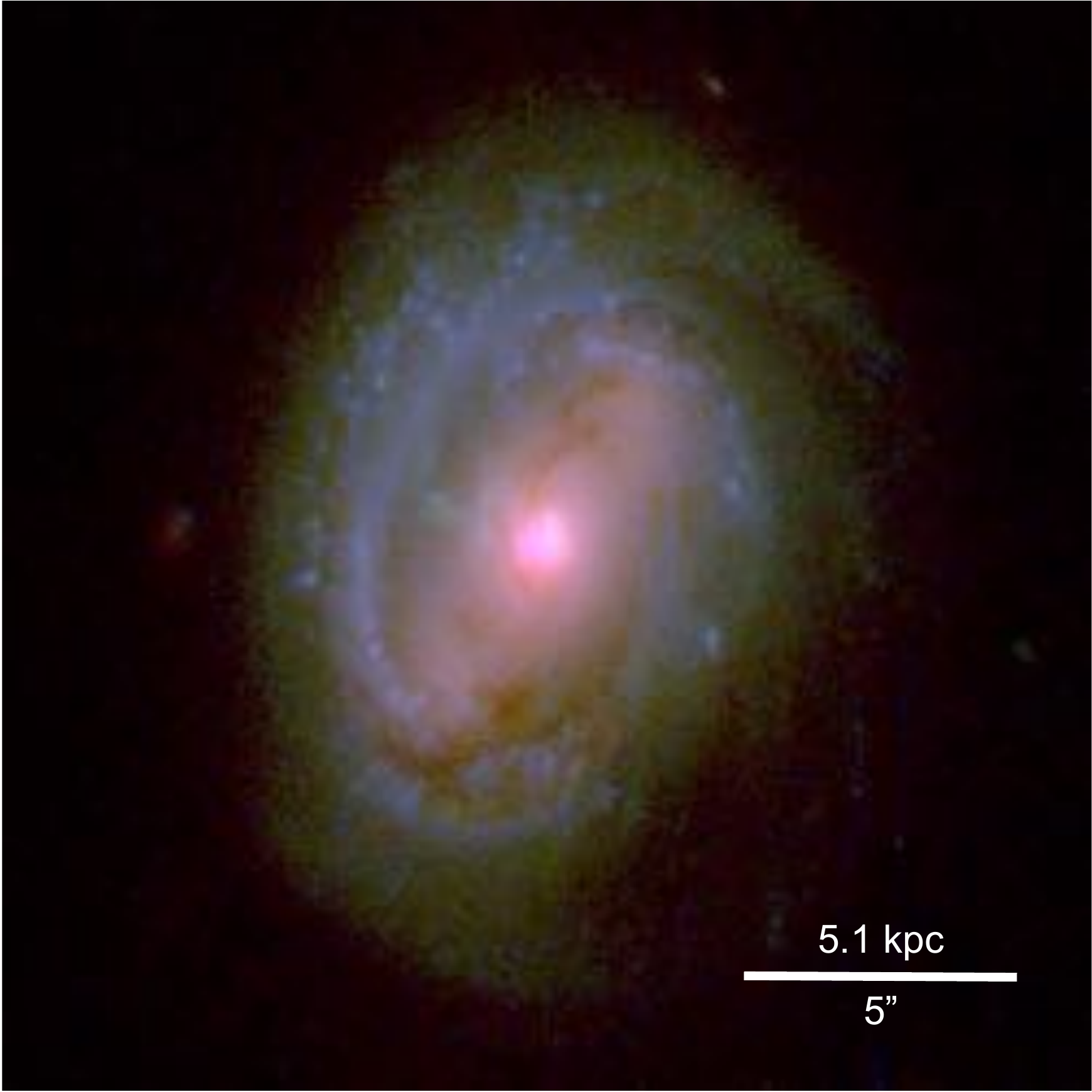}
\includegraphics[height=1.5in]{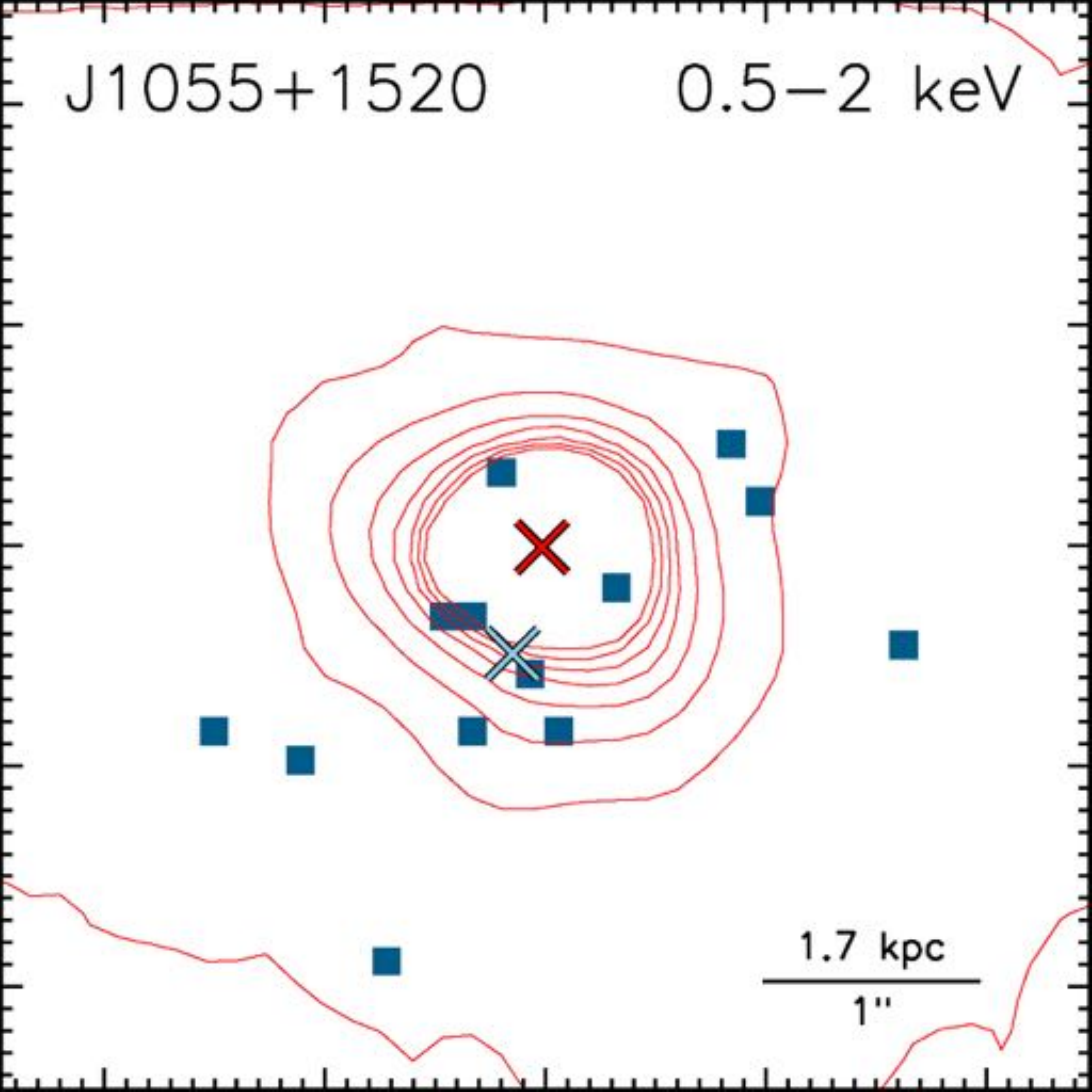}
\includegraphics[height=1.5in]{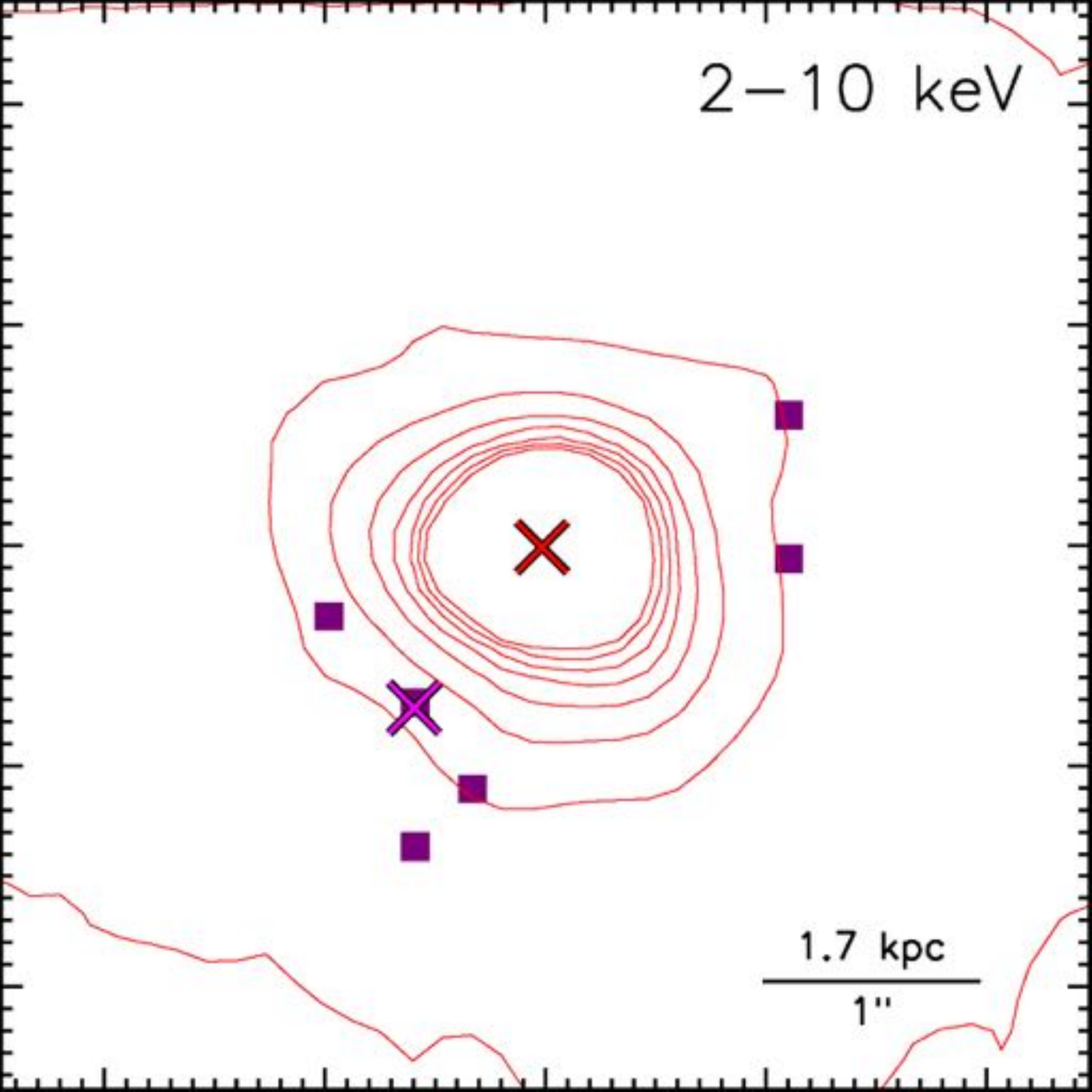}
\includegraphics[height=1.5in]{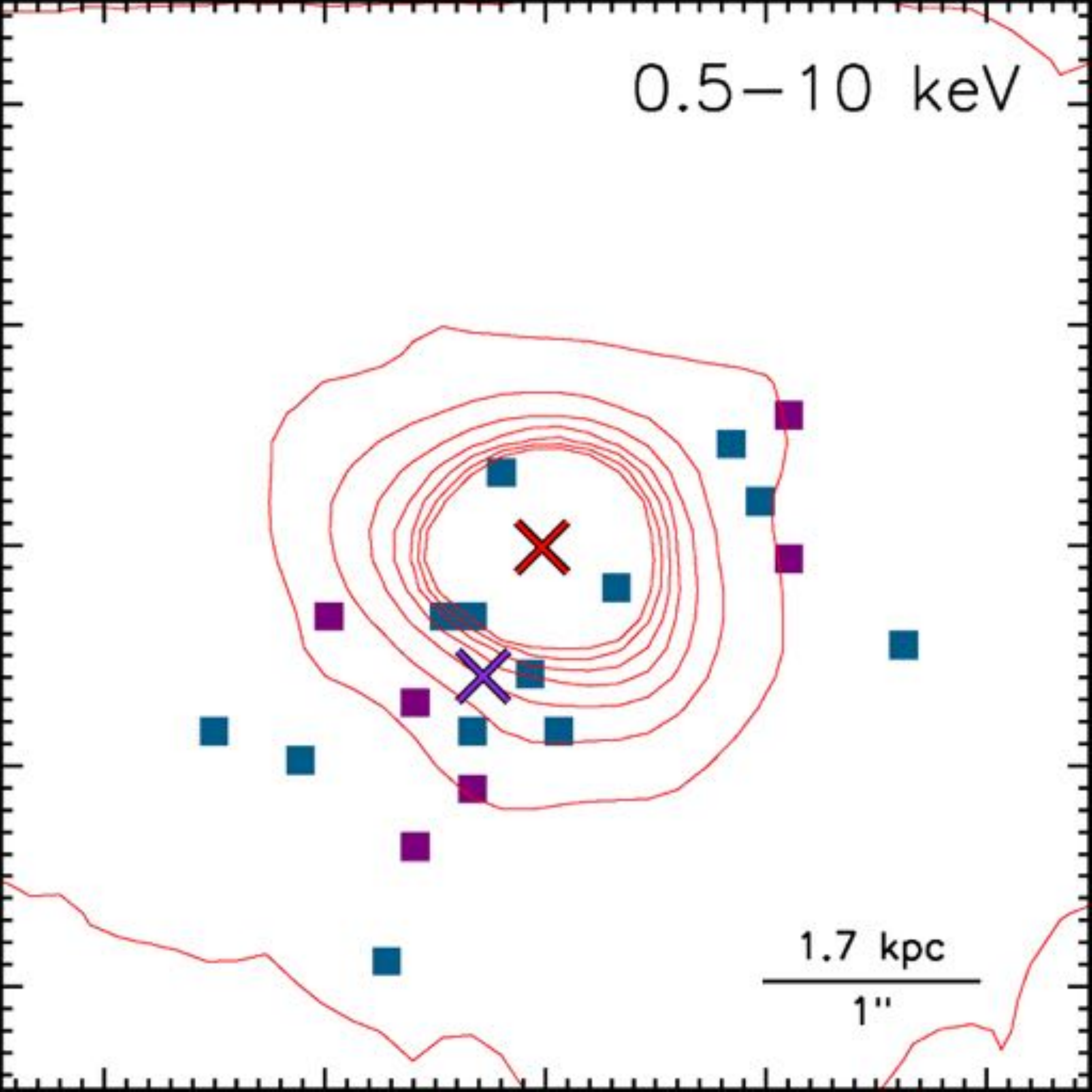}
\includegraphics[height=1.5in]{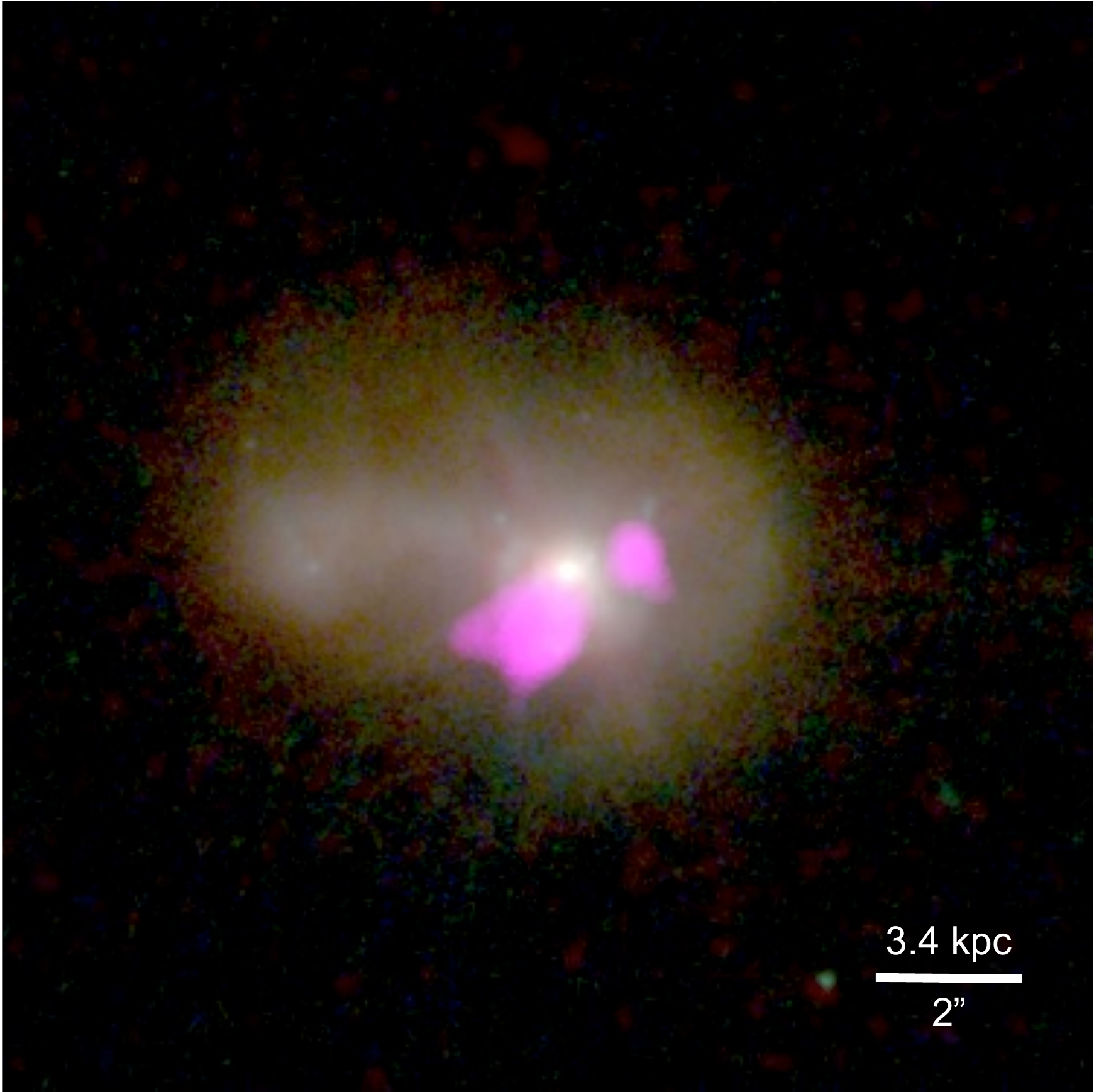}
\includegraphics[height=1.5in]{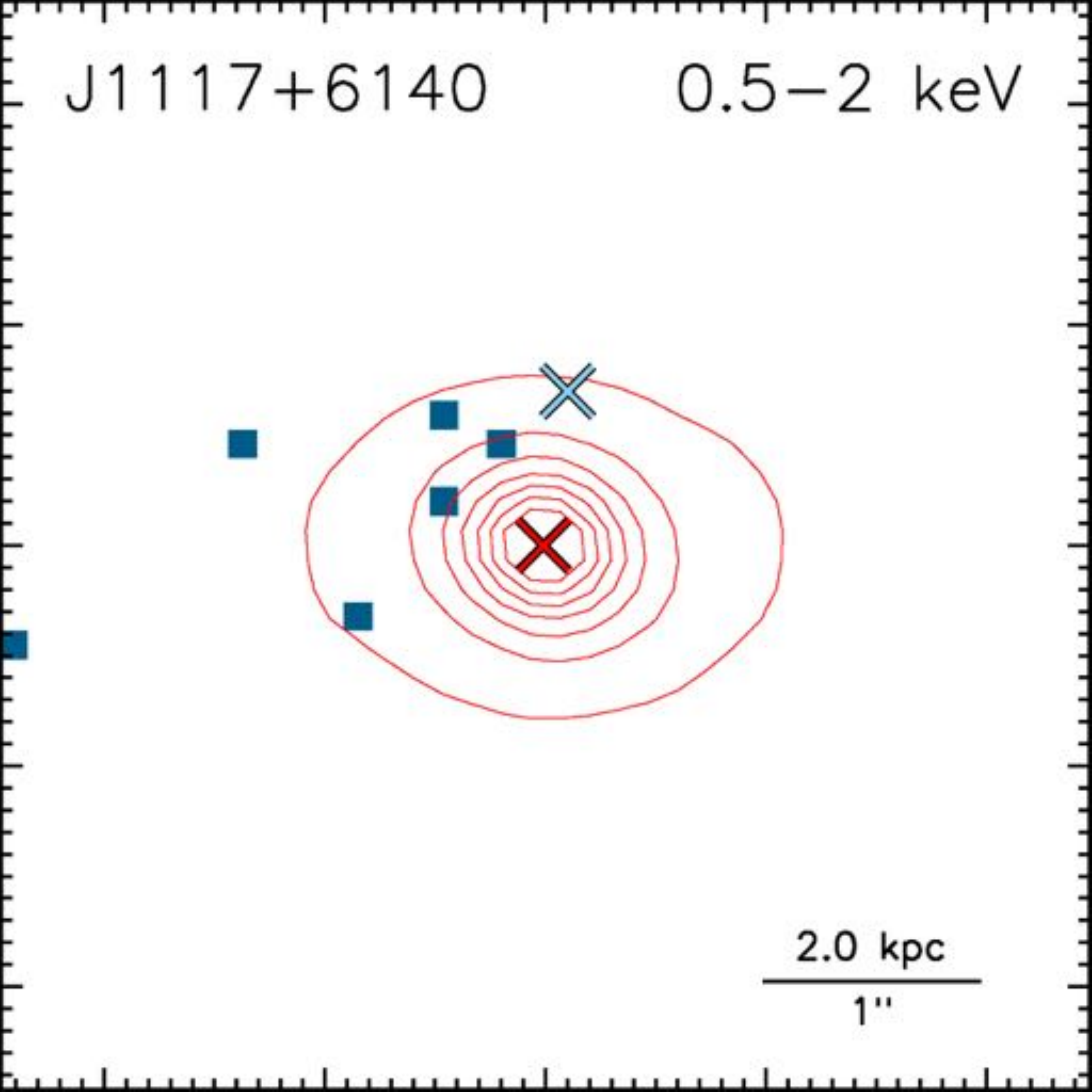}
\includegraphics[height=1.5in]{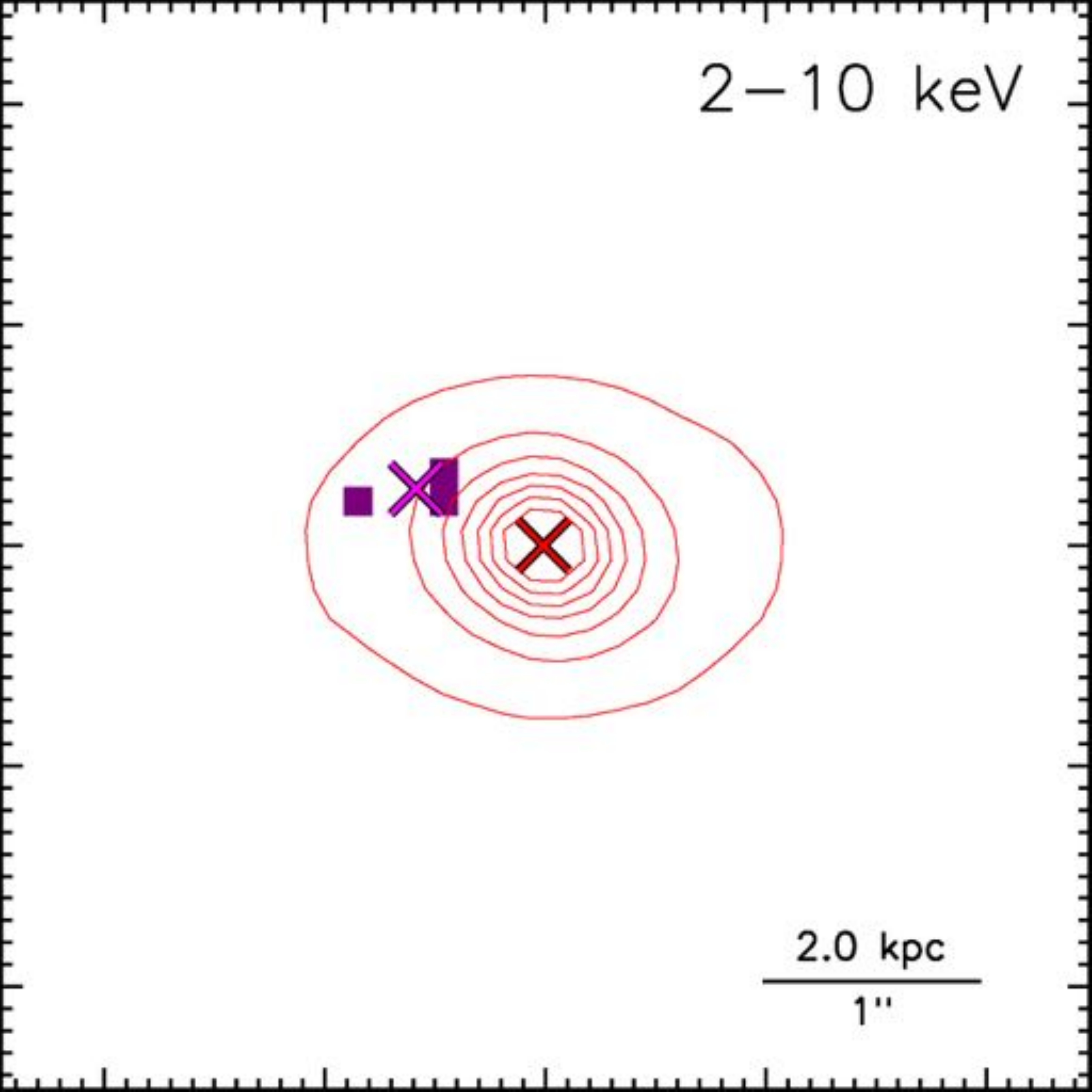}
\includegraphics[height=1.5in]{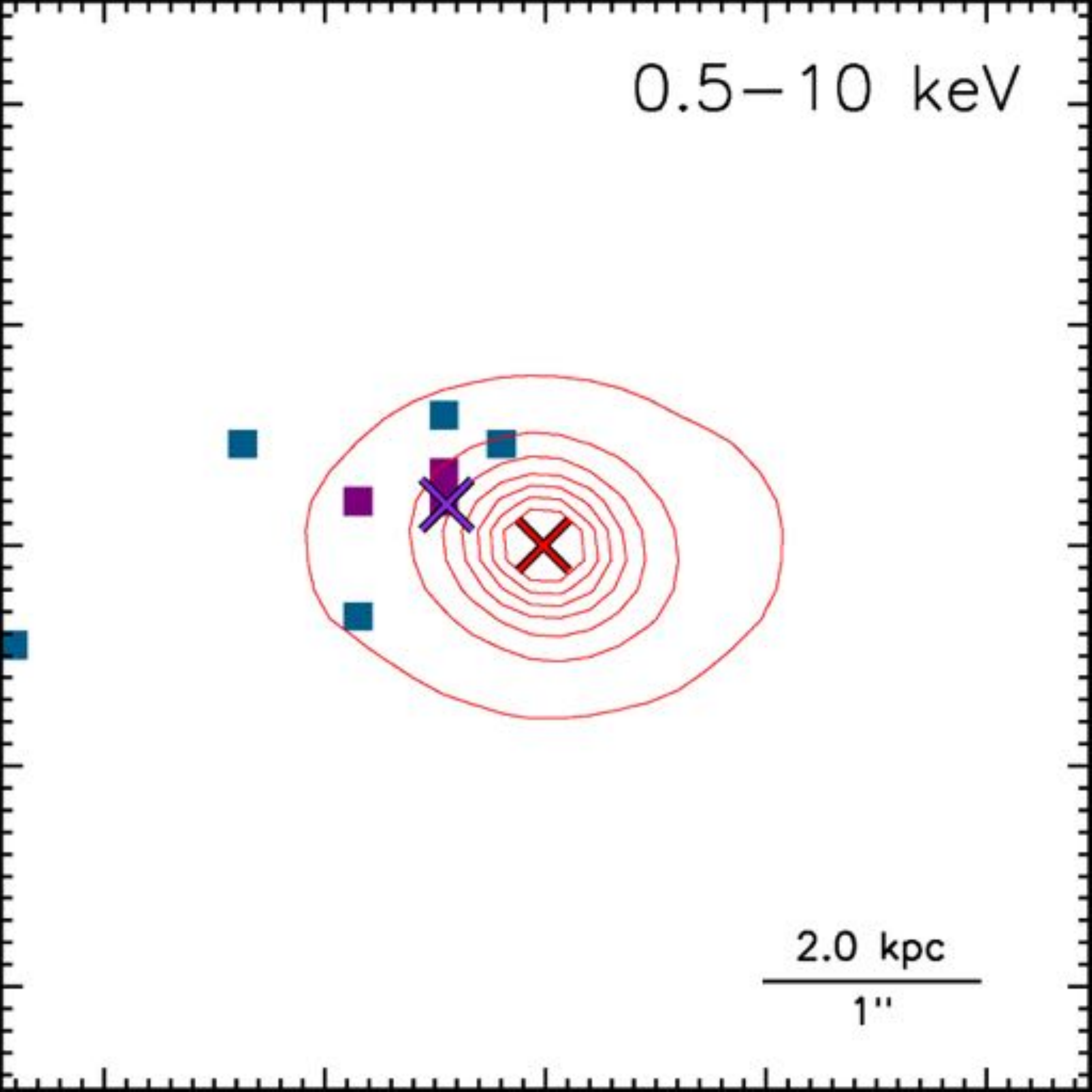}
\includegraphics[height=1.5in]{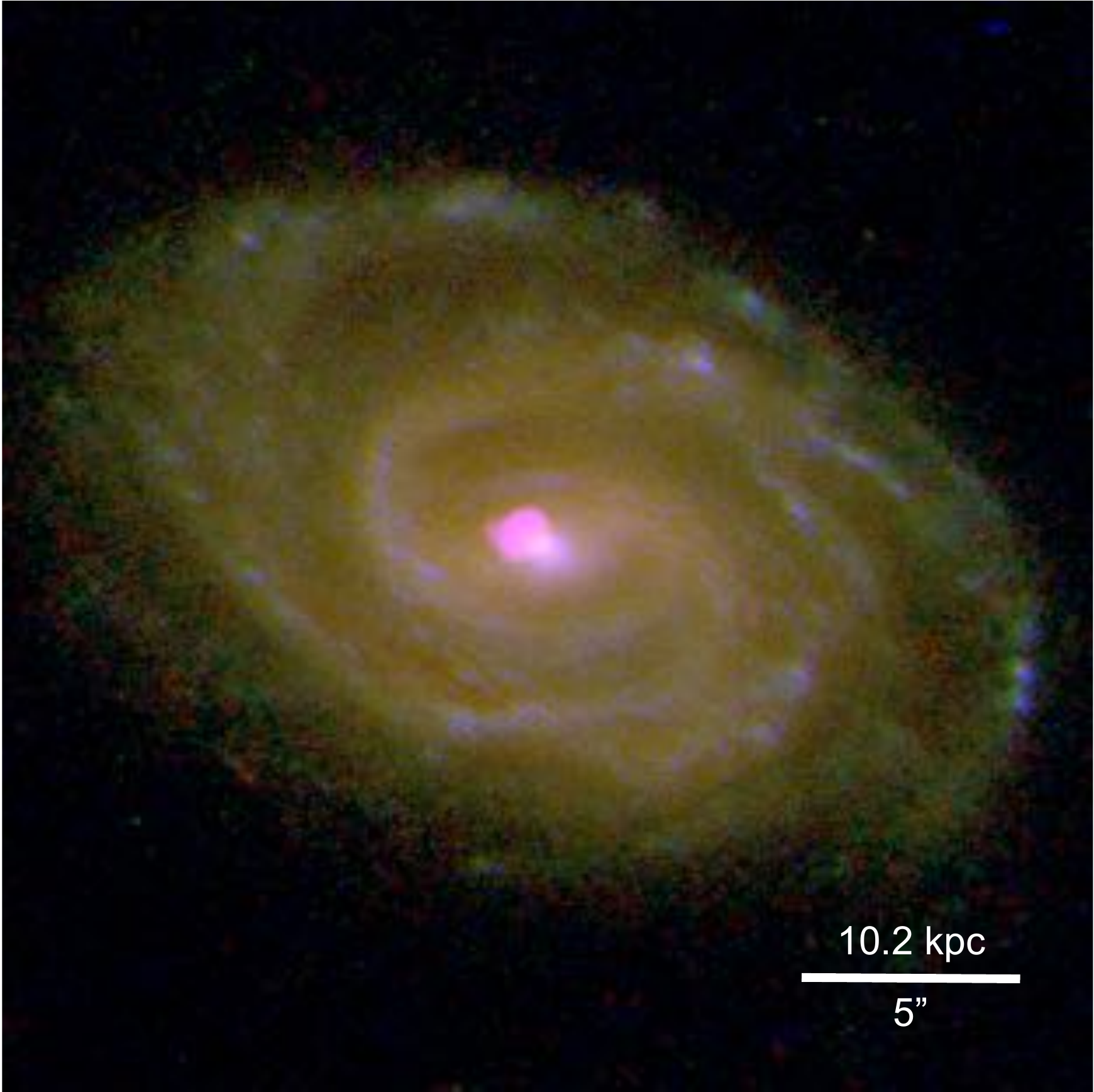}
\includegraphics[height=1.5in]{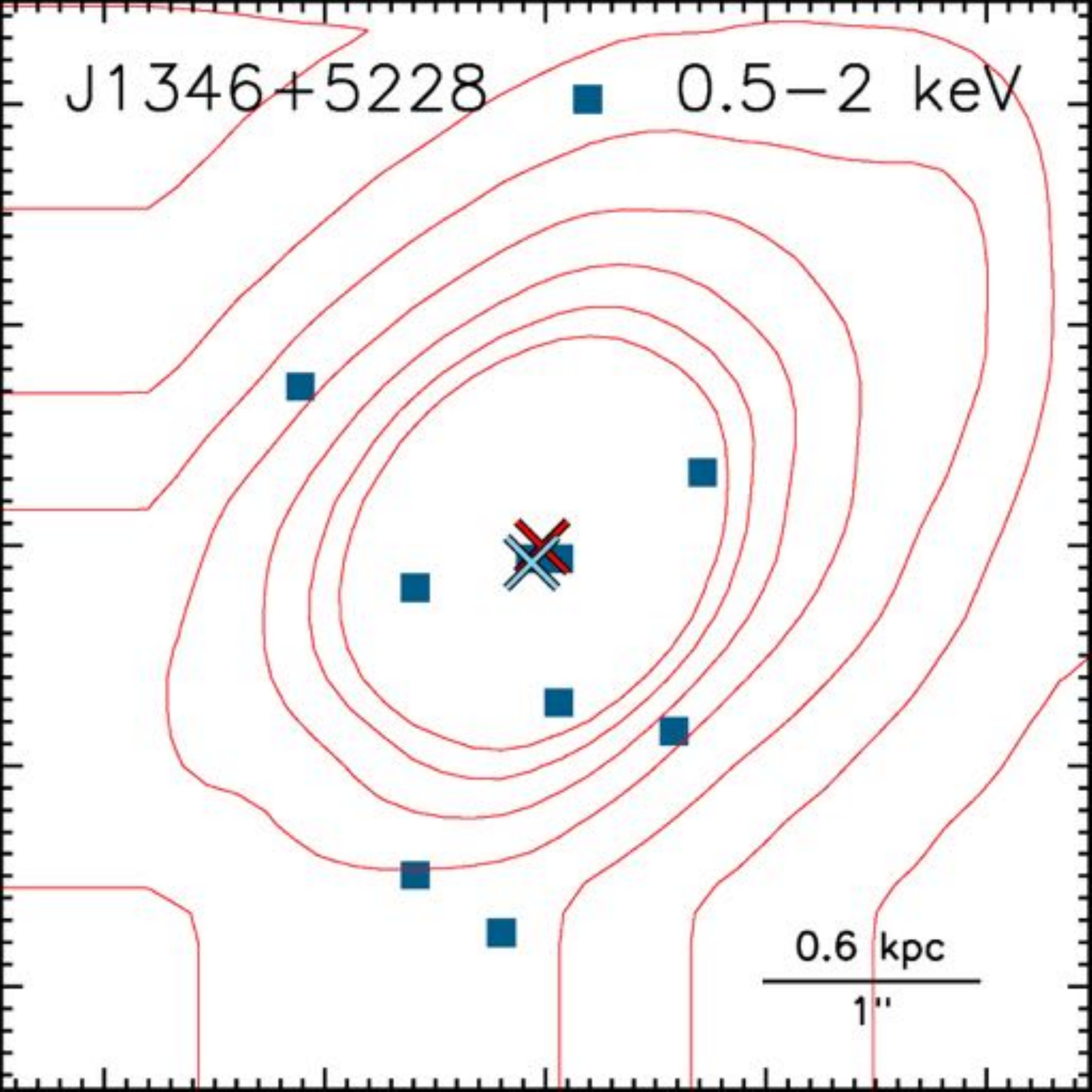}
\includegraphics[height=1.5in]{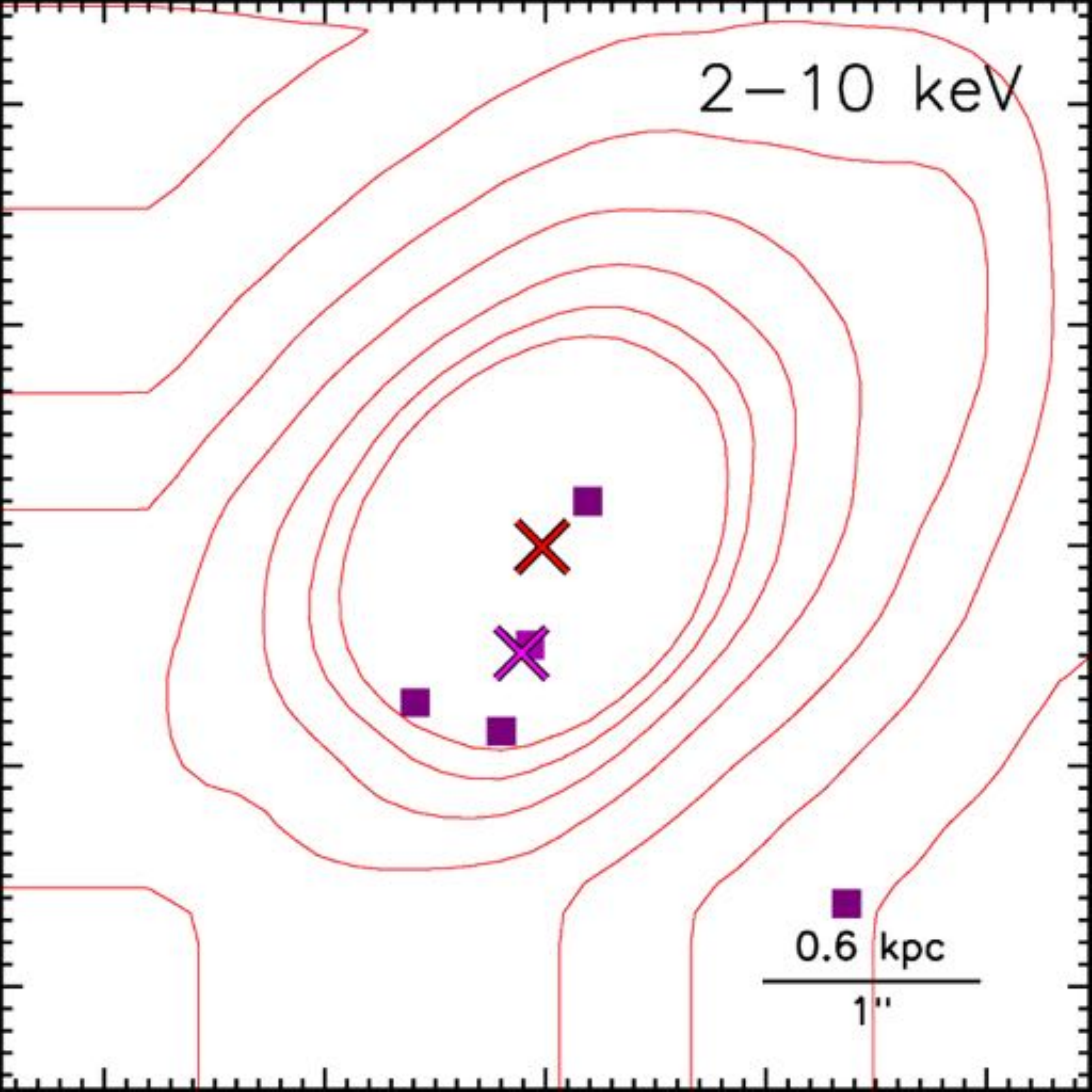}
\includegraphics[height=1.5in]{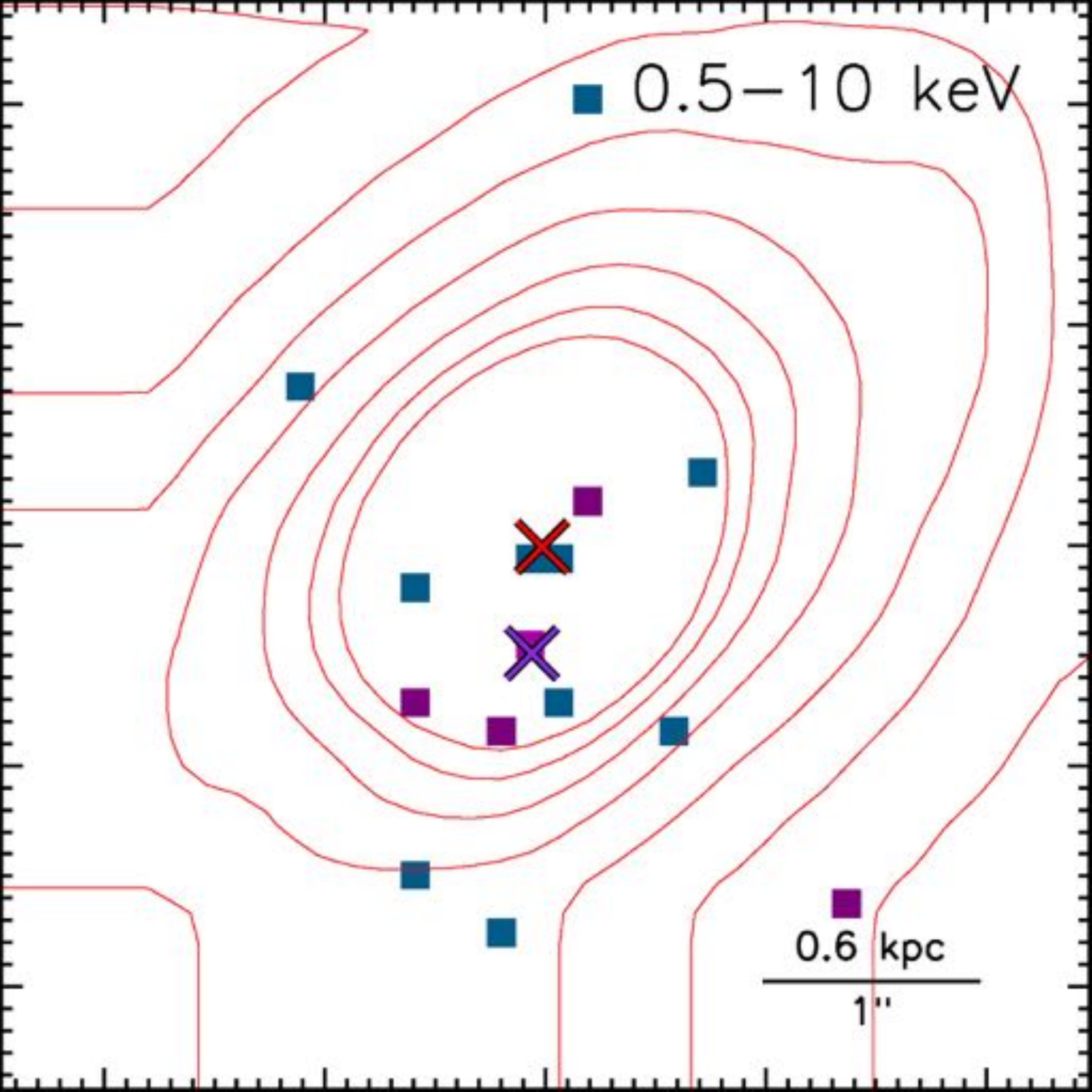}
\includegraphics[height=1.5in]{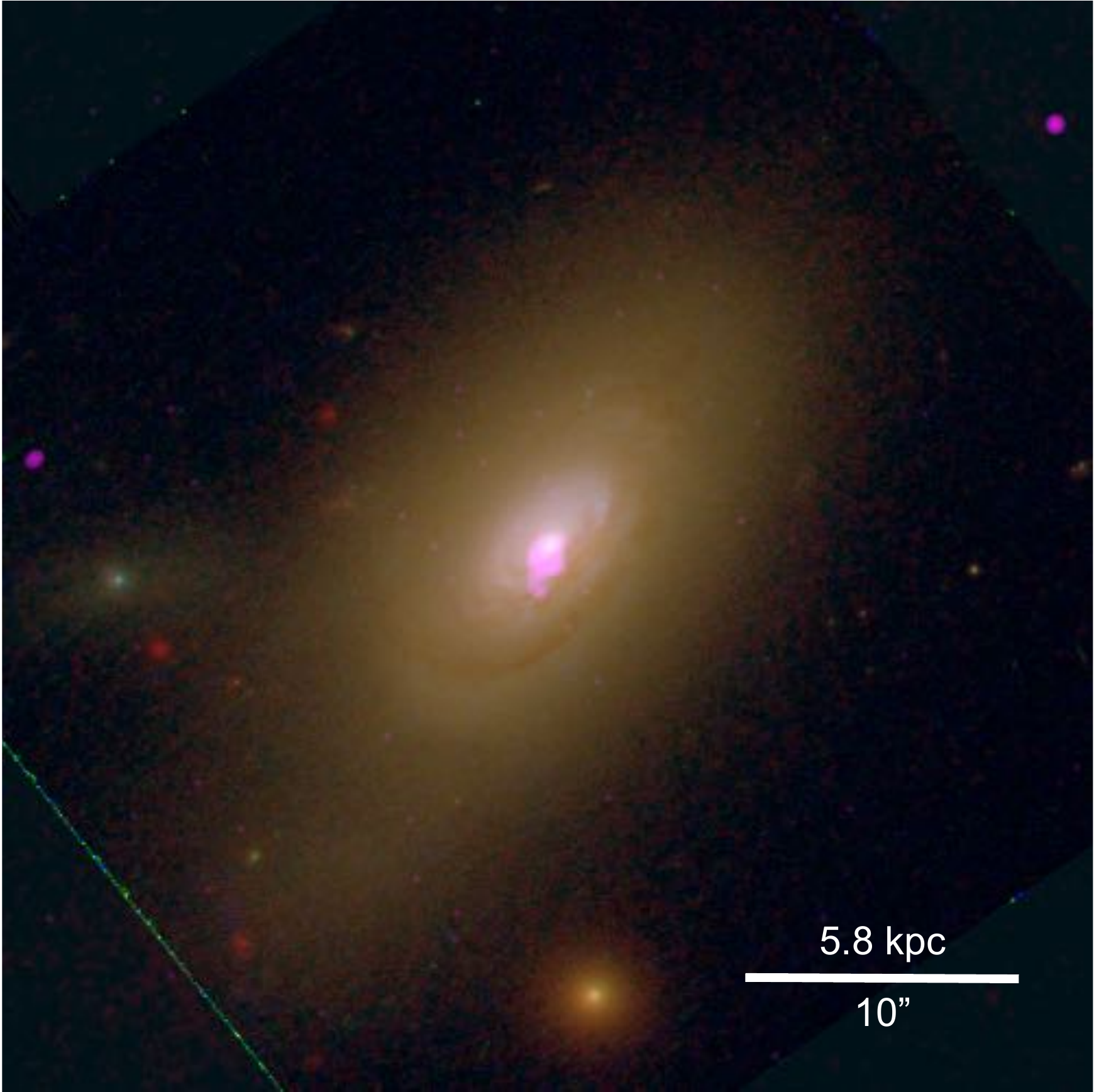}
\end{center}
\end{figure*}

Then, we used the Bayesian Estimation of Hardness Ratios (\texttt{BEHR}) code \citep{PA06.1} to measure the rest-frame soft, hard, and total counts in each X-ray source.  We used \texttt{calc$\_$data$\_$sum} to determine the number of observed soft and hard counts from both the source region and a background region, and then \texttt{BEHR} used a Bayesian approach to estimate the expected values and uncertainties of the rest-frame soft counts, hard counts, total counts, and hardness ratio.  Table~\ref{tbl-4} shows these values, and we estimated errors on the counts assuming Poisson noise. 
 
 \begin{deluxetable*}{llllllll} 
\tabletypesize{\scriptsize}
\tablewidth{0pt}
\tablecolumns{8}
\tablecaption{X-ray Counts and Spectral Fits} 
\tablehead{
\colhead{SDSS Name} &
\colhead{Soft Counts} & 
\colhead{Hard Counts} &
\colhead{Total Counts} & 
\colhead{Hardness} &  
\colhead{$n_{H,exgal}$} & 
\colhead{$\Gamma$} &
\colhead{Reduced} \\
  & ($0.5-2$ keV) & ($2-10$ keV) & ($0.5-10$ keV) & Ratio & ($10^{22}$ cm$^{-2}$) & & C-stat 
 }
 \startdata
J0132$-$1027 & $14.9^{+3.1}_{-4.2}$ & $8.2^{+2.3}_{-3.4}$ & $23.1^{+4.2}_{-5.2}$ & $-0.30^{+0.19}_{-0.22}$ & $<0.02$ & $1.73^{+0.41}_{-0.38}$ & 0.24 \\ 
J0839+4707 & $8.7^{+2.1}_{-3.4}$ & $66.1^{+7.4}_{-8.6}$ & $74.8^{+7.9}_{-9.1}$ & $0.77^{+0.09}_{-0.06}$ & $8.18^{+2.21}_{-0.29}$ & $1.70$ (fixed)$^a$ &0.59 \\ 
J1055+1520 & $15.8^{+3.2}_{-4.5}$ & $6.8^{+2.0}_{-3.3}$ & $22.7^{+4.1}_{-5.3}$ & $-0.40^{+0.19}_{-0.22}$ & $<0.03$ & $1.71^{+0.39}_{-0.47}$ & 0.22 \\ 
J1117+6140 & $7.6^{+2.0}_{-3.3}$ & $3.1^{+1.1}_{-2.5}$ & $10.7^{+2.6}_{-4.0}$ & $-0.43^{+0.25}_{-0.35}$ & $<0.10$ & $1.87^{+0.78}_{-0.76}$ & 0.12 \\ 
J1346+5228 & $10.8^{+2.4}_{-3.9}$ & $7.2^{+1.9}_{-3.3}$ & $18.0^{+3.8}_{-4.7}$ & $-0.21^{+0.23}_{-0.25}$ & $<0.12$ & $1.35^{+0.54}_{-0.85}$ & 0.19 \\ 
J1654+1946  & $23.9^{+4.2}_{-5.3}$ & $4.2^{+1.4}_{-2.7}$ & $28.1^{+4.4}_{-6.0}$ & $-0.71^{+0.11}_{-0.17}$ & $<0.02$ & $1.70$ (fixed)$^a$ & 0.20 \\ 
J2323+1405 & $17.4^{+3.5}_{-4.7}$ & $32.5^{+5.2}_{-6.2}$ & $49.9^{+6.3}_{-7.6}$ & $0.30^{+0.15}_{-0.13}$ & $0.41^{+0.18}_{-0.16}$ & $1.70$ (fixed)$^a$ & 0.52 
\enddata
\tablecomments{Column 2: soft X-ray (restframe 0.5-2 keV) counts (S).  Column 3: hard X-ray (restframe 2-10 keV) counts (H).  Column 4: total X-ray (restframe 0.5-10 keV) counts.  Column 5: hardness ratio HR = (H-S)/(H+S).  Column 6: extragalactic column density. Column 7: best-fit spectral index. Column 8: reduced Cash statistic of the fit.}
\tablenotetext{a}{The best-fit spectrum had a spectral index of $\Gamma <1$ or $\Gamma>3$, so we redid the fit by freezing the spectral index to $\Gamma=1.70$.}
\label{tbl-4}
\end{deluxetable*}

To model the energy spectra of the extracted regions over the observed energy range 2-8 keV, we used \texttt{Sherpa}. We fit each unbinned spectrum with a redshifted power law, $F\sim E^{-\Gamma}$ (which represents the intrinsic AGN X-ray emission at the SDSS spectroscopic redshift $z$).  This spectrum is attenuated by passing through two absorbing column densities of neutral Hydrogen.  One of these is fixed to the Galactic value, $n_{H,Gal}$, and the other is assumed to be intrinsic to the source, $n_{H,exgal}$, at the redshift $z$.  We determined $n_{H,Gal}$ using an all-sky interpolation of the \hi in the Galaxy \citep{DI90.1}.

For our first fit to each spectrum, we allowed $\Gamma$ and $n_{H,exgal}$ to vary freely.  If the best-fit value of $\Gamma$ was not within the typical range of observed power-law indices, i.e. $1\le \Gamma \le 3$ \citep{IS10.1,PI05.1,NA94.2,RE00.1}, then we fixed $\Gamma$ at a value of 1.7, which is a typical value for the continuum of Seyfert galaxies, and ran the fit again.  

To determine the best-fit model parameters for each spectrum, we used \texttt{Sherpa}'s implementation of the Levenberg-Marquardt optimization method \citep{BE69.1} to minimize the Cash statistic.  Table~\ref{tbl-4} shows the results of these spectral fits.  All fluxes are $k$-corrected, and we calculated the observed flux values from the model sum (including the absorbing components) and the intrinsic flux values from the unabsorbed power law component.  Finally, we used the redshift to determine the distance to each system and convert the X-ray fluxes to X-ray luminosities (Table~\ref{tbl-5}). 

\begin{deluxetable*}{llll} 
\tabletypesize{\scriptsize}
\tablewidth{0pt}
\tablecolumns{4}
\tablecaption{X-ray Luminosities} 
\tablehead{
\colhead{SDSS Name} &
\colhead{$L_{X, 0.5-2 \mathrm{keV}}$ abs (unabs)} &
\colhead{$L_{X, 2-10 \mathrm{keV}}$ abs (unabs)} & 
\colhead{$L_{X, 0.5-10 \mathrm{keV}}$ abs (unabs)} \\
 & ($10^{40}$ erg s$^{-1}$) & ($10^{40}$ erg s$^{-1}$) & ($10^{40}$ erg s$^{-1}$)
 }
\startdata
J0132-1027 & $1.2^{+0.4}_{-0.3} (1.3^{+0.4}_{-0.4})$ & $2.3^{+2.1}_{-1.2} (2.3^{+2.1}_{-1.2})$ & $3.3^{+2.5}_{-1.2} (3.5^{+2.6}_{-1.3})$ \\
J0839+4707 & $0.2^{+0.5}_{-0.1} (89.7^{+29.6}_{-29.0})$ & $165.0^{+44.7}_{-37.6} (245.0^{+61.2}_{-59.8})$ & $170.0^{+39.6}_{-40.1} (338.0^{+83.0}_{-84.3})$ \\
J1055+1520 & $9.7^{+3.2}_{-3.5} (11.6^{+3.3}_{-3.8})$ & $18.4^{+19.2}_{-8.8} (18.5^{+19.2}_{-8.8})$ & $30.1^{+22.5}_{-12.7} (32.3^{+21.1}_{-13.3})$ \\
J1117+6140 & $3.9^{+2.4}_{-1.9} (6.4^{+2.5}_{-2.9})$ & $7.7^{+14.6}_{-5.1} (7.8^{+14.6}_{-5.1})$ & $12.3^{+15.1}_{-7.1} (14.9^{+14.0}_{-8.2})$ \\
J1346+5228 & $0.8^{+0.4}_{-0.3} (1.0^{+0.4}_{-0.4})$ & $2.7^{+3.1}_{-1.4} (2.7^{+3.1}_{-1.5})$ & $3.7^{+3.5}_{-1.7} (3.8^{+3.6}_{-1.7})$ \\
J1654+1946 & $5.5^{+1.2}_{-1.2} (6.6^{+1.2}_{-1.4})$ & $11.9^{+2.1}_{-2.6} (12.0^{+2.1}_{-2.7})$ & $17.2^{+3.5}_{-3.2} (18.2^{+3.6}_{-3.5})$ \\
J2323+1405 & $2.9^{+1.2}_{-0.8} (7.0^{+1.3}_{-1.4})$ & $12.3^{+2.4}_{-2.6} (12.7^{+2.4}_{-2.6})$ & $15.4^{+3.3}_{-3.0} (19.8^{+3.9}_{-3.8})$ 
\enddata
\tablecomments{Column 2: absorbed (and unabsorbed) soft X-ray 0.5-2 keV luminosity. Column 3: absorbed (and unabsorbed) hard X-ray 2-10 keV luminosity. Column 4: absorbed (and unabsorbed) total X-ray 0.5-10 keV luminosity.}
\label{tbl-5}
\end{deluxetable*}

\subsection{HST/WFC3 F438W, F606W, and F160W Observations}
\label{hst}

The seven velocity-offset AGNs were also observed with {\it HST}/WFC3 (GO 13513, PI: Comerford), and the observations covered three bands: UVIS/F438W ($B$ band), UVIS/F606W ($V$ band), and IR/F160W ($H$ band).  The exposure times are summarized in Table~\ref{tbl-2}.  

Each band revealed different properties of the galaxies.  The F438W observations covered $H\delta$, $H\gamma$, and \oiiis for the $0.02 < z < 0.06$ galaxies and \oiidouble and $H\delta$ for the $0.09 < z < 0.12$ galaxies.  The F606W observations covered \hbn, \oiiidouble; \oidoublen; \niidouble; \han, and \siidouble for the $0.02 < z < 0.06$ galaxies; \hbn, \oiiidouble; and \oidouble for the $z=0.09$ galaxy; and $H\gamma$, \oiiisn, \hbn, \oiiidouble; and \oidouble for the $z=0.11$ galaxy.  The F160W observations primarily traced the stellar continuum, although they may also have included \feii 1.6436 $\mu$m emission for the $0.02 < z < 0.04$ galaxies and Pa$\beta$ emission for the $0.09 < z < 0.12$ galaxies.

To locate the stellar centroid of each galaxy, we fit a S\'ersic profile (plus a fixed, uniform sky component) to each galaxy's F160W image using GALFIT V3.0 \citep{PE10.1}.  We ran each fit on a square region of projected physical size 40 kpc on each side, with the angular size scale calculated from $z$ and assuming the cosmology stated in Section 1.   

The errors returned by GALFIT are purely statistical in that they are computed directly from the variance of the input images.  We note that in reality, the true radial profiles may deviate from the parametric model components used in GALFIT, particularly at large radii.  We previously examined this in \cite{CO15.1} by creating radial profiles of the S\'ersic fits to merger-remnant galaxies, where we found that, even with significant residuals at large radii, the S\'ersic component peaks are excellent tracers of the photometric peaks.

In our fitting procedure, we also attempted a two-S\'ersic component fit (over the same fitting region) to test for the presence of secondary nuclei and/or close interacting neighbors.  In these cases, we adopted the two-component model if the secondary component is detected at $>3\sigma$ significance above the background.  We found one system, SDSS J0839+4707, with a nearby neighbor galaxy.  GALFIT returned the positions of the sources and their integrated magnitudes, which we used to determine the spatial separation on the sky between the two galaxies and their merger mass ratio.  We approximated the merger mass ratio as the luminosity ratio of the two stellar bulges.

We also measured the centroid of emission for each galaxy, using Source Extractor \citep{BE96.1} on the F606W images.  According to the SDSS spectra, the \oiiiw emission line is the dominant line in the F606W image for each galaxy, within the central $3^{\prime\prime}$.  Therefore, the centroid of F606W emission within the central $3^{\prime\prime}$ is a proxy for the centroid of \oiiiw emission.  We ran Source Extractor with a detection threshold of $5\sigma$ above the background, and the errors on the positions are statistical.  

\begin{figure}
\begin{center}
\includegraphics[height=1.5in]{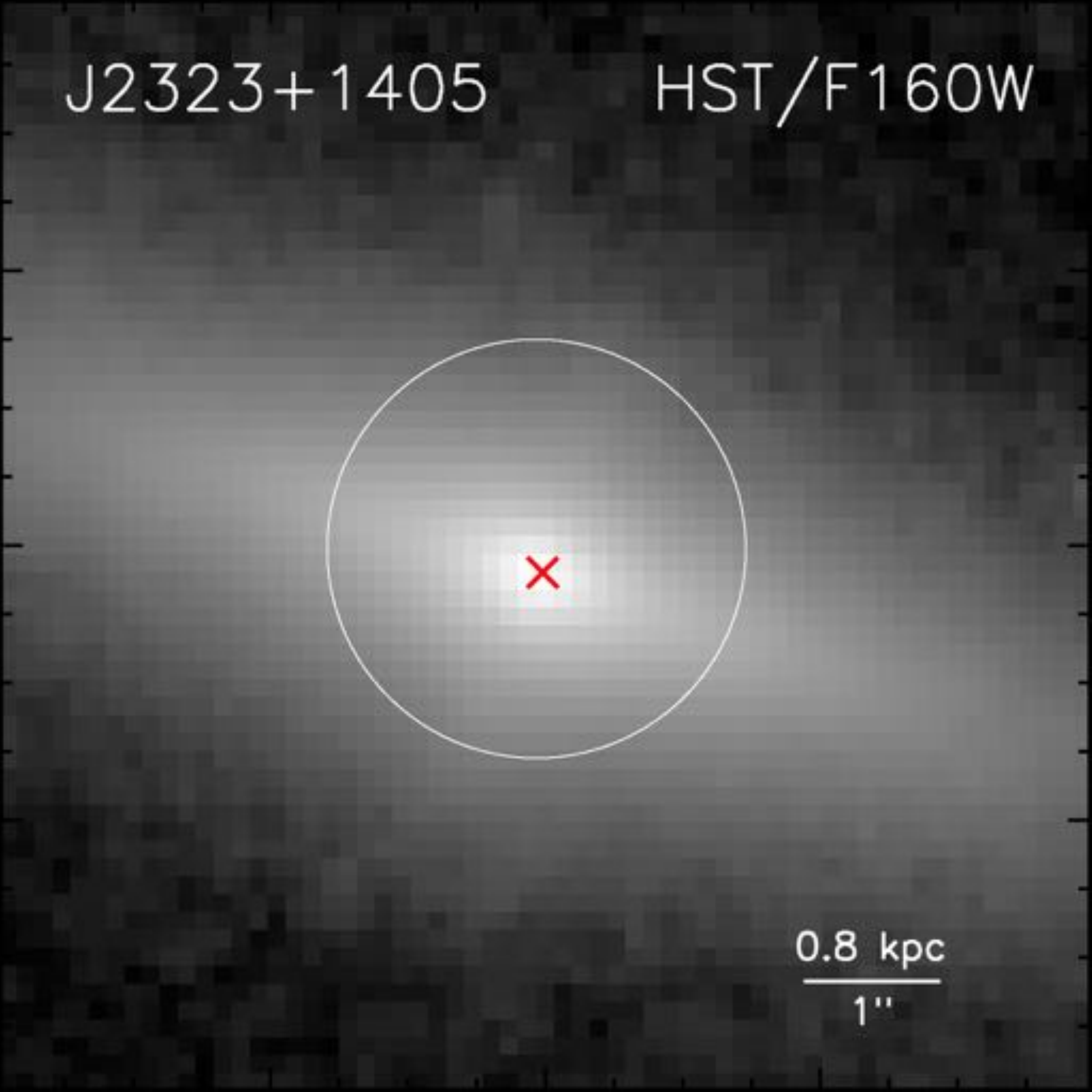}
\includegraphics[height=1.5in]{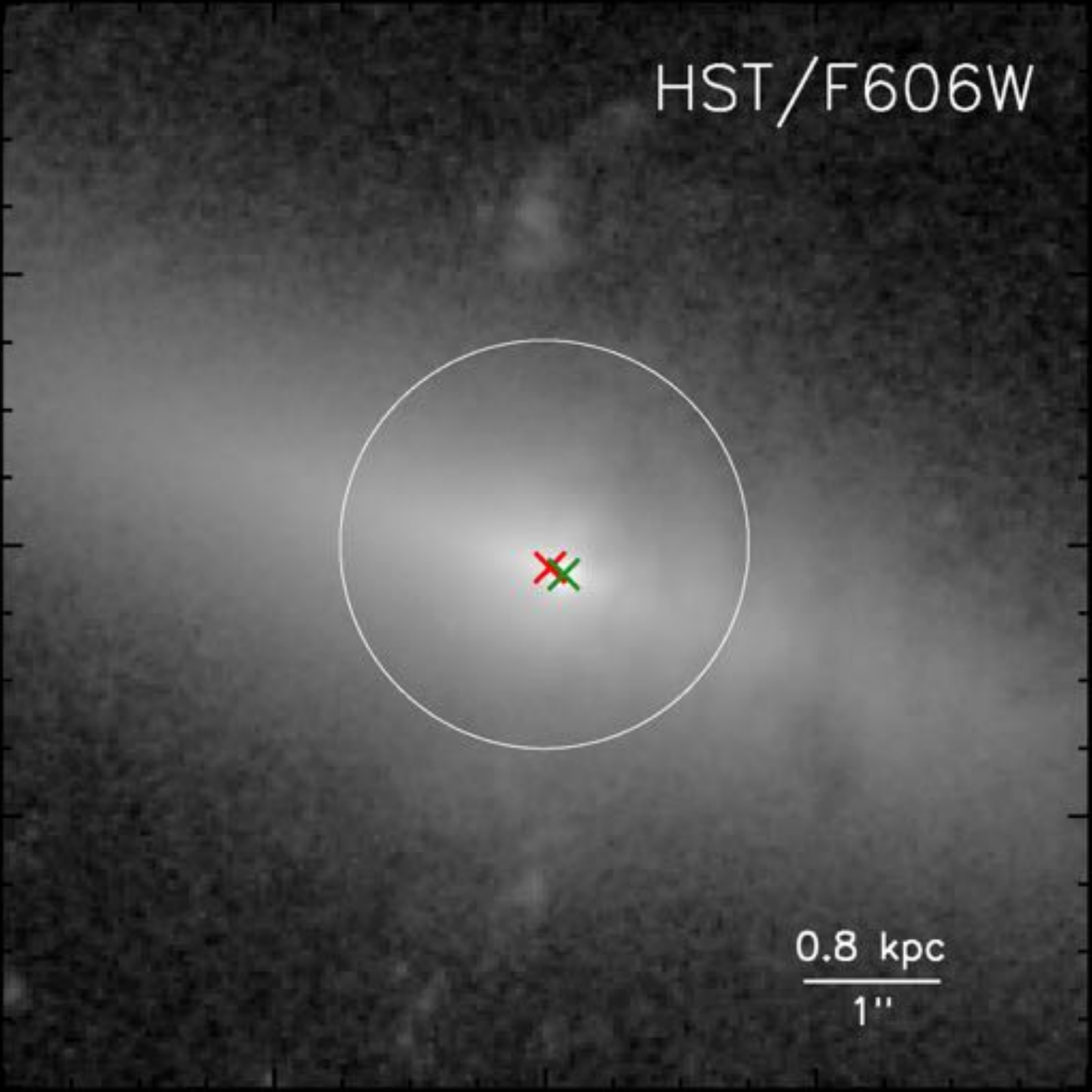}
\end{center}
\caption{{\it HST}/F160W (left) and {\it HST}/F606W (right) images of each galaxy.  The white circle illustrates the SDSS fiber, the red cross shows the galaxy's stellar centroid based on {\it HST}/F160W observations, and the green cross shows the centroid of the emission source detected in the {\it HST}/F606W observations.  All images are $5^{\prime\prime} \times 5^{\prime\prime}$, except for the J2323+1405 images, which are $8^{\prime\prime} \times 8^{\prime\prime}$.  In all panels, north is up and east is to the left.}
\label{fig:hstresults}
\end{figure}

The positions of the emission centroids, as well as their separations from the stellar centroids, are shown in Table~\ref{tbl-6} and Figure~\ref{fig:hstresults}.  We determined the spatial separation errors by combining the errors on the GALFIT positions in the F160W data, the Source Extractor positions in the F606W data, and the relative astrometric uncertainties in the F160W (10 mas) and F606W observations (4 mas; \citealt{DE16.1}).  The relative astrometric uncertainties dominate the errors, so that the spatial separation errors are all $0\farcs01$.  We found that all of the spatial separations between the emission centroids and the stellar centroids are greater than $3\sigma$ in significance.

\begin{deluxetable*}{lllll}
\tabletypesize{\scriptsize}
\tablewidth{0pt}
\tablecolumns{5}
\tablecaption{HST/F606W Positions of Each Source} 
\tablehead{
\colhead{SDSS Name} &
\colhead{RA$_{HST/F606W}$} &
\colhead{DEC$_{HST/F606W}$} &
\colhead{$\Delta \theta (^{\prime\prime})^a$} &
\colhead{$\Delta x$ (kpc)$^a$} 
}
\startdata
J0132$-$1027 & 01:32:58.922 & $-$10:27:07.01 & $0.078 \pm 0.011$ & $0.050 \pm 0.007$ \\
J0839+4707 & 08:39:02.937 & +47:07:56.02 & $0.193 \pm 0.011$ & $0.197 \pm 0.011$ \\
J1055+1520 & 10:55:53.638 & +15:20:27.96 & $0.124 \pm 0.012$ & $0.212 \pm 0.020$ \\
J1117+6140 & 11:17:29.218 & +61:40:15.31 & $0.179 \pm 0.011$ & $0.365 \pm 0.023$ \\
J1346+5228 & 13:46:40.802 & +52:28:36.23 & $0.252 \pm 0.011$ & $0.147 \pm 0.006$ \\
J1654+1946 & 16:54:30.734 & +19:46:15.48 & $0.152 \pm 0.011$ & $0.159 \pm 0.011$ \\
J2323+1405 & 23:23:28.004 & +14:05:30.03 & $0.111 \pm 0.011$ & $0.091 \pm 0.009$  
\enddata
\tablecomments{Columns 2 and 3: coordinates of the peak of the emission, measured from {\it HST}/WFC3/F606W observations.    Columns 4 and 5: angular and physical separations between the positions of the peak of the emission and the host galaxy's stellar nucleus.  All separations are $>3\sigma$ in significance.}
\tablenotetext{a}{The errors are dominated by the astrometric uncertainties, which are $0\farcs01$.}
\label{tbl-6}
\end{deluxetable*}

\begin{figure}
\begin{center}
\includegraphics[height=1.5in]{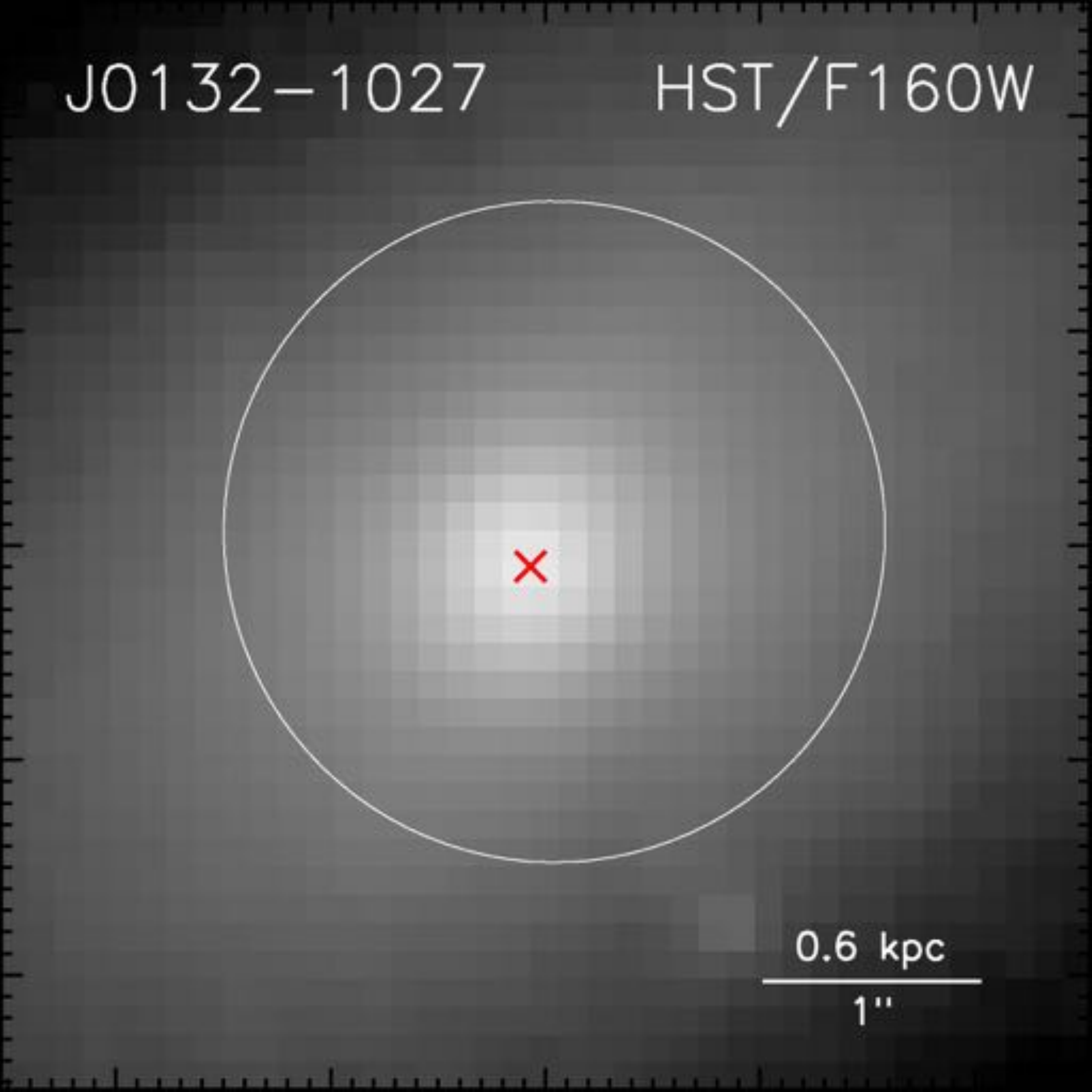}
\includegraphics[height=1.5in]{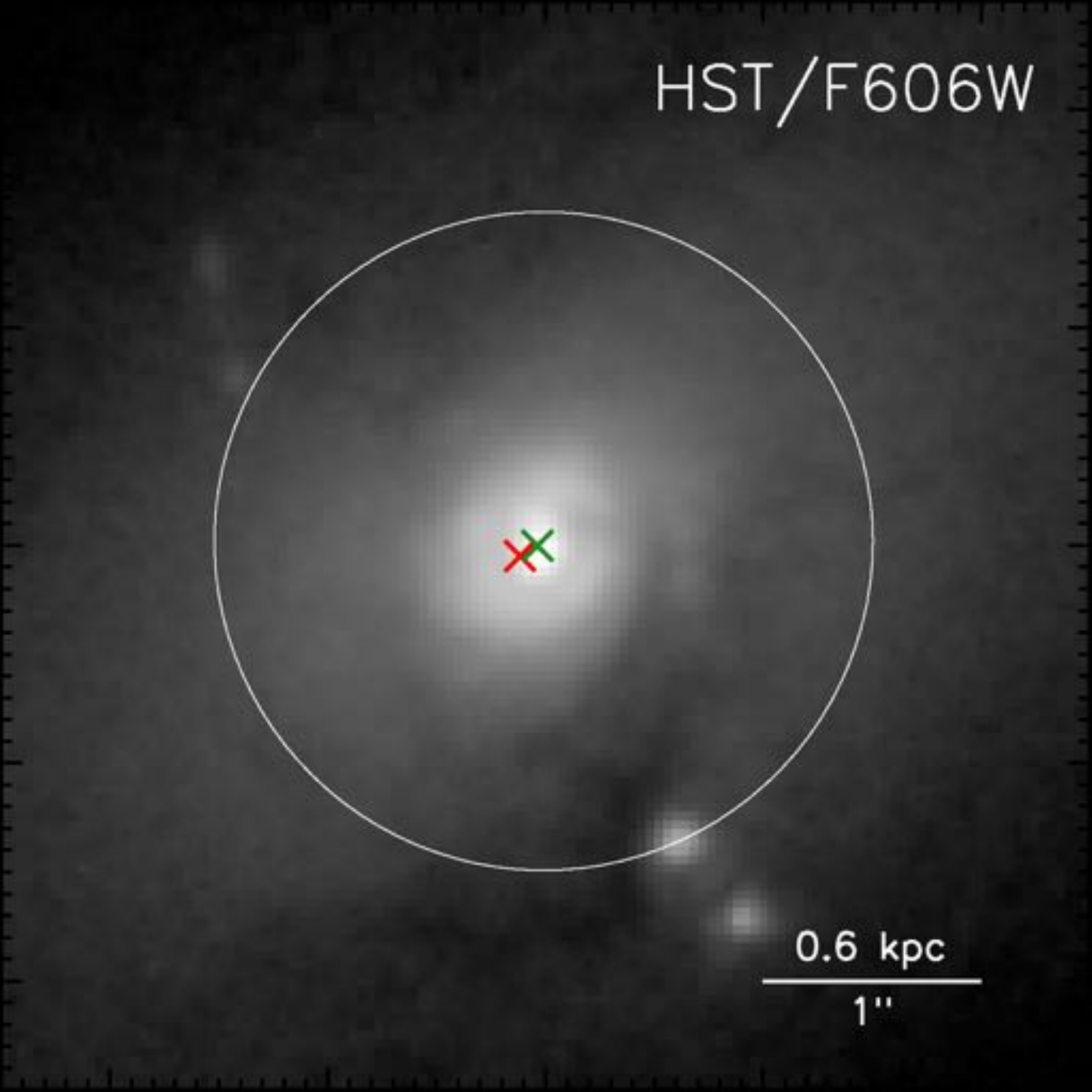}
\includegraphics[height=1.5in]{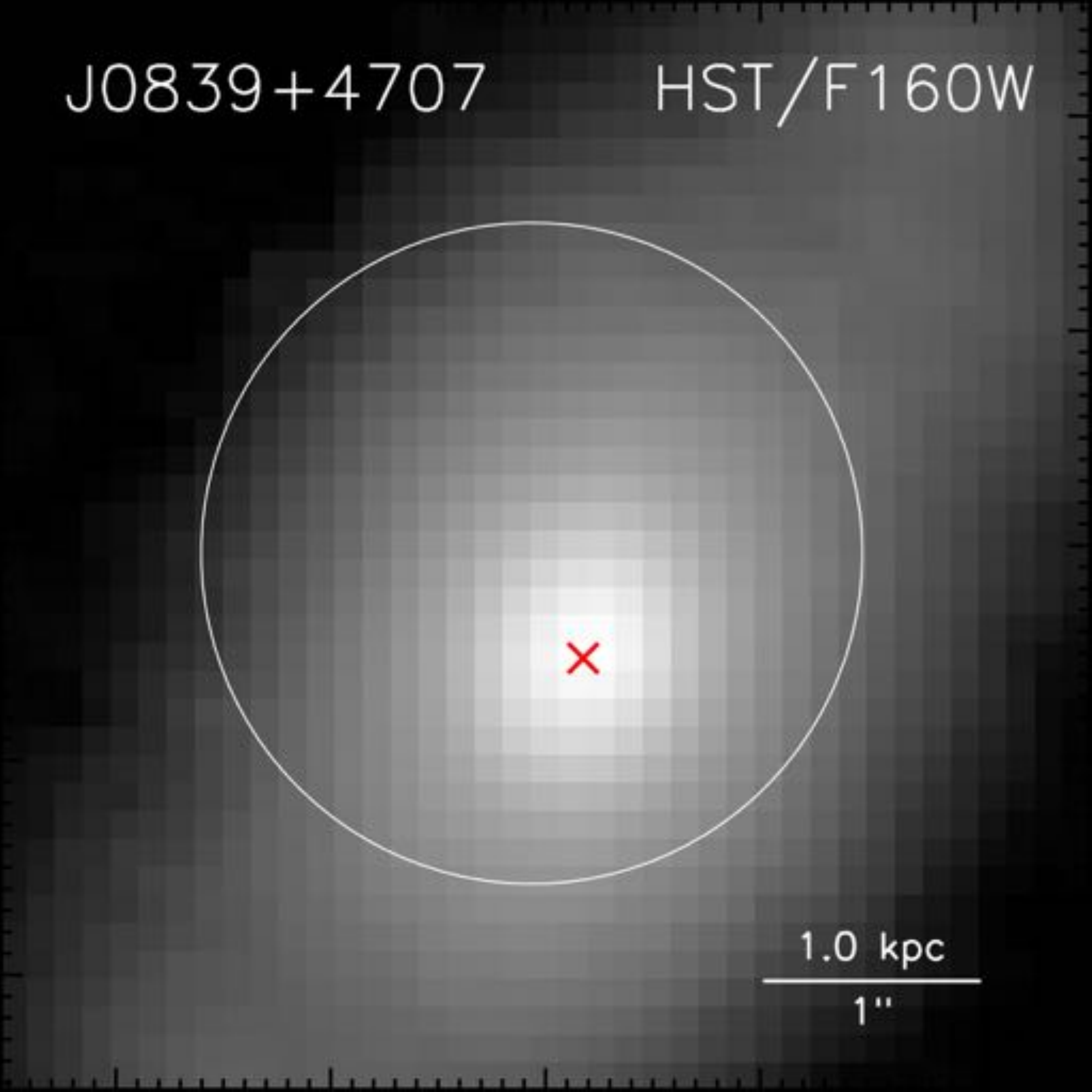}
\includegraphics[height=1.5in]{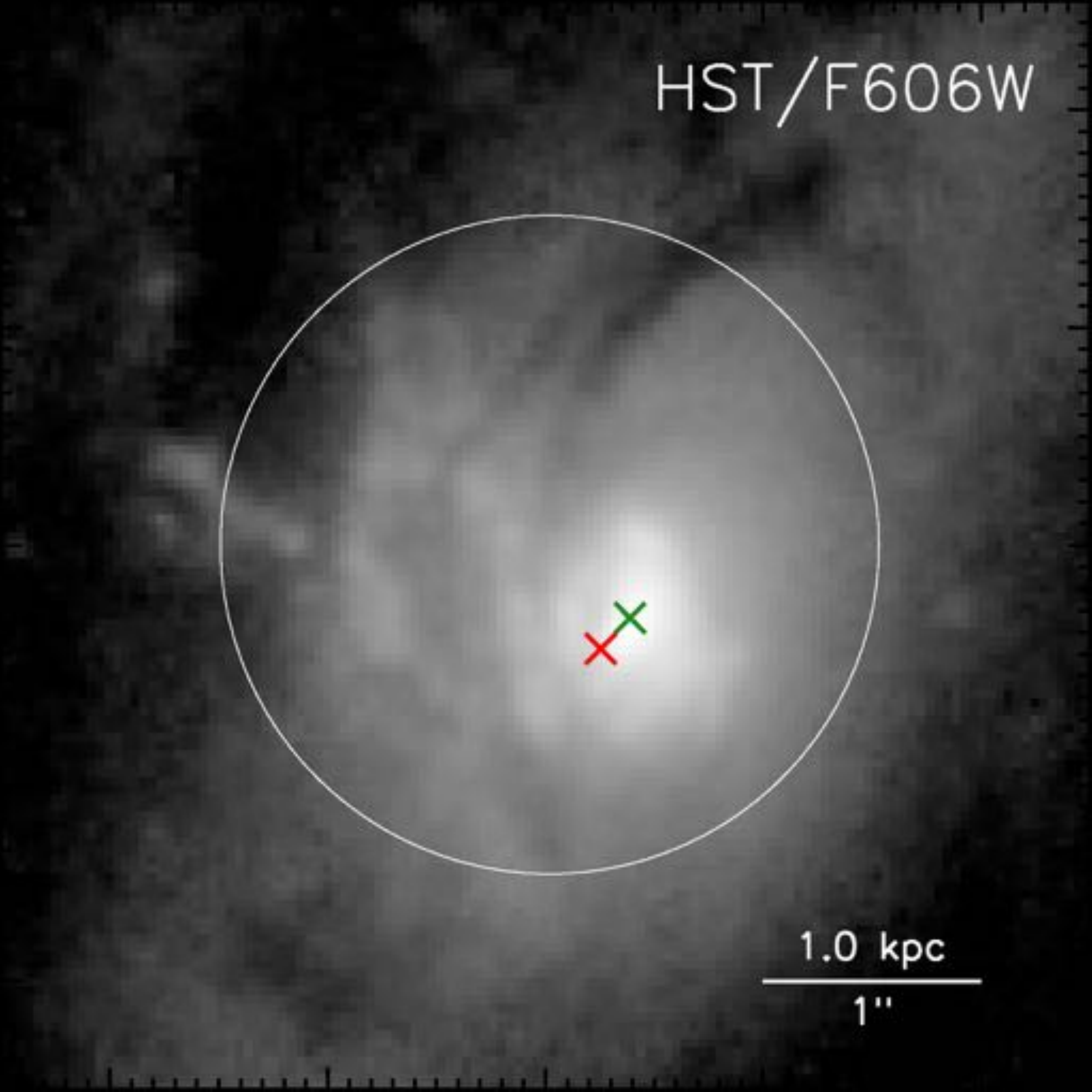}
\includegraphics[height=1.5in]{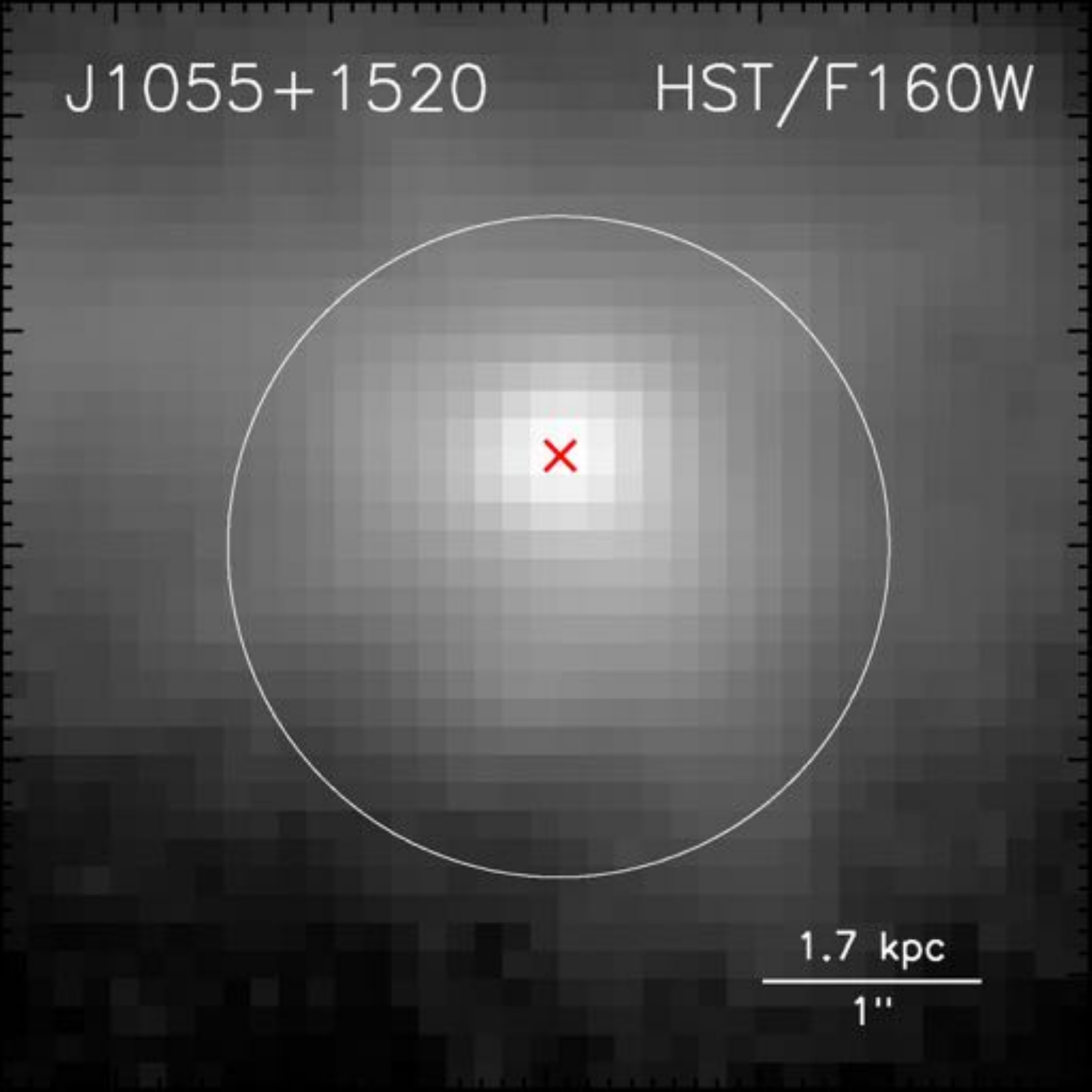}
\includegraphics[height=1.5in]{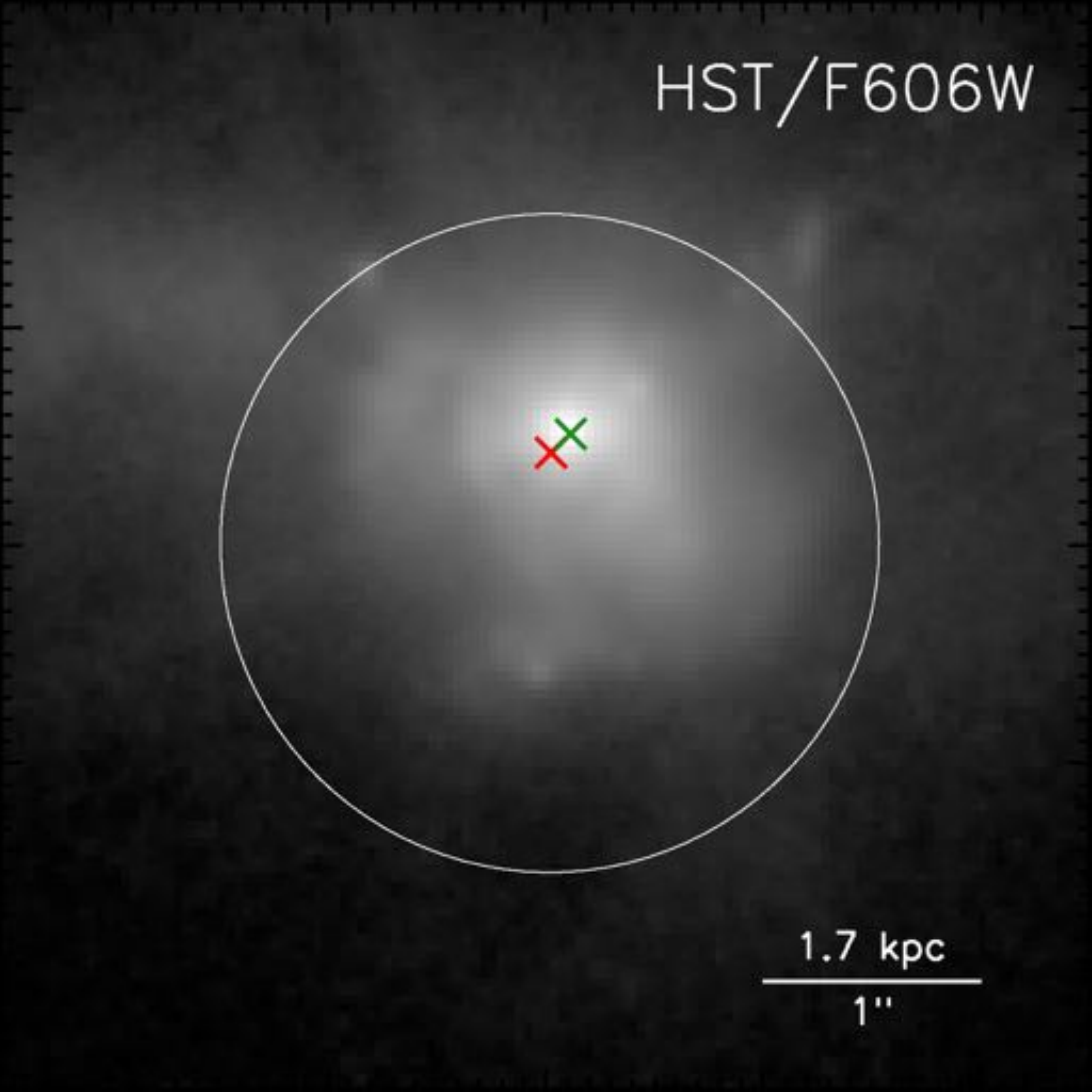}
\includegraphics[height=1.5in]{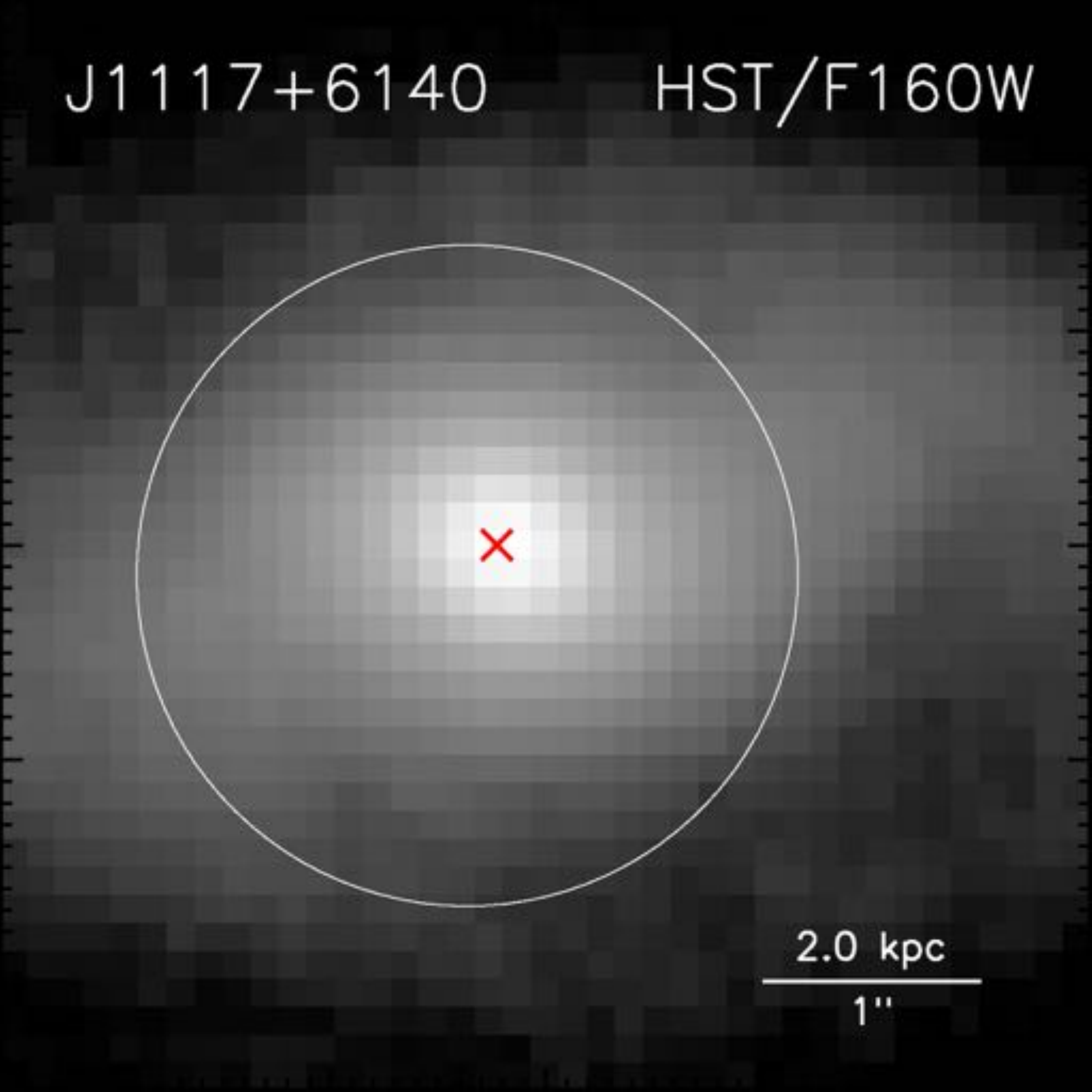}
\includegraphics[height=1.5in]{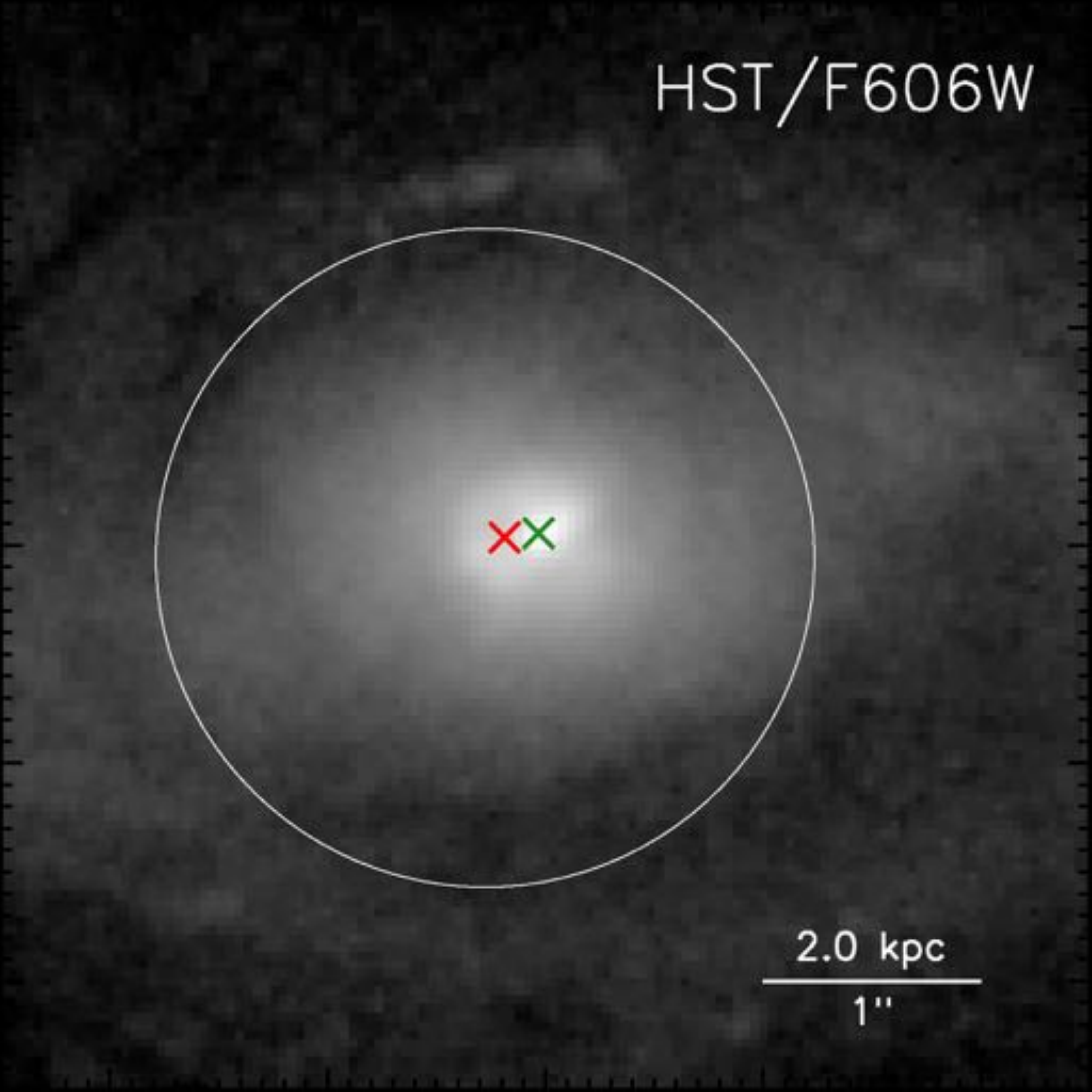}
\includegraphics[height=1.5in]{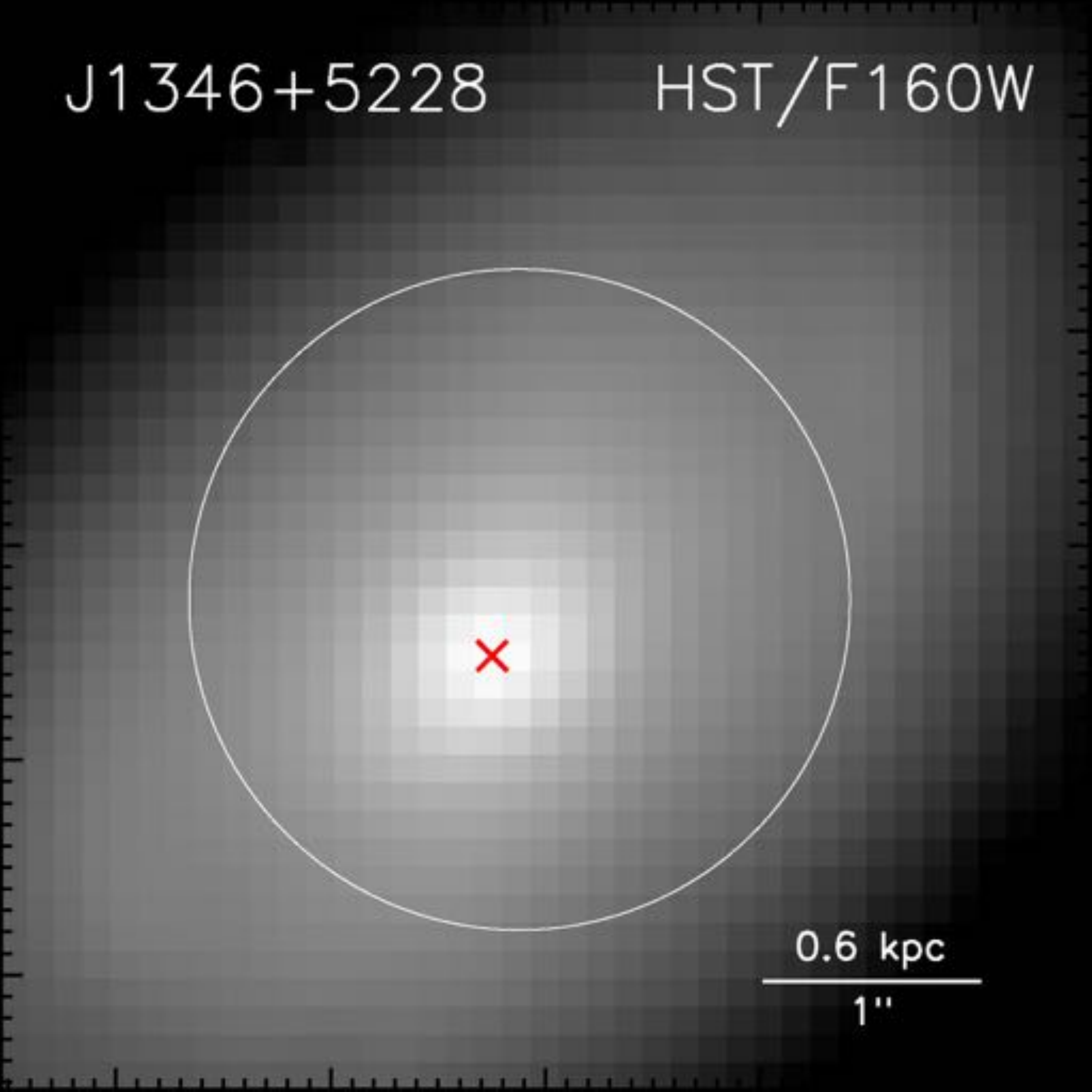}
\includegraphics[height=1.5in]{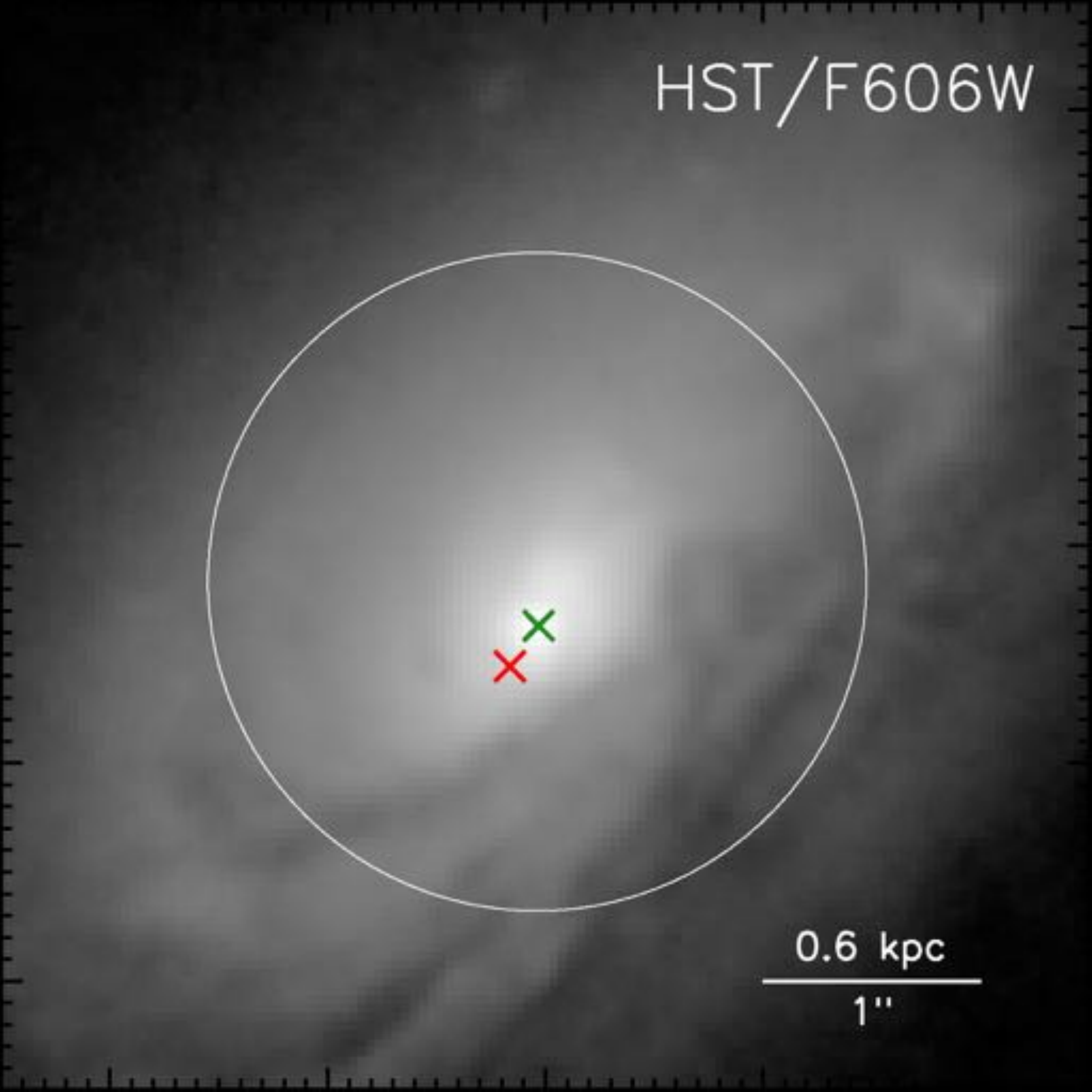}
\includegraphics[height=1.5in]{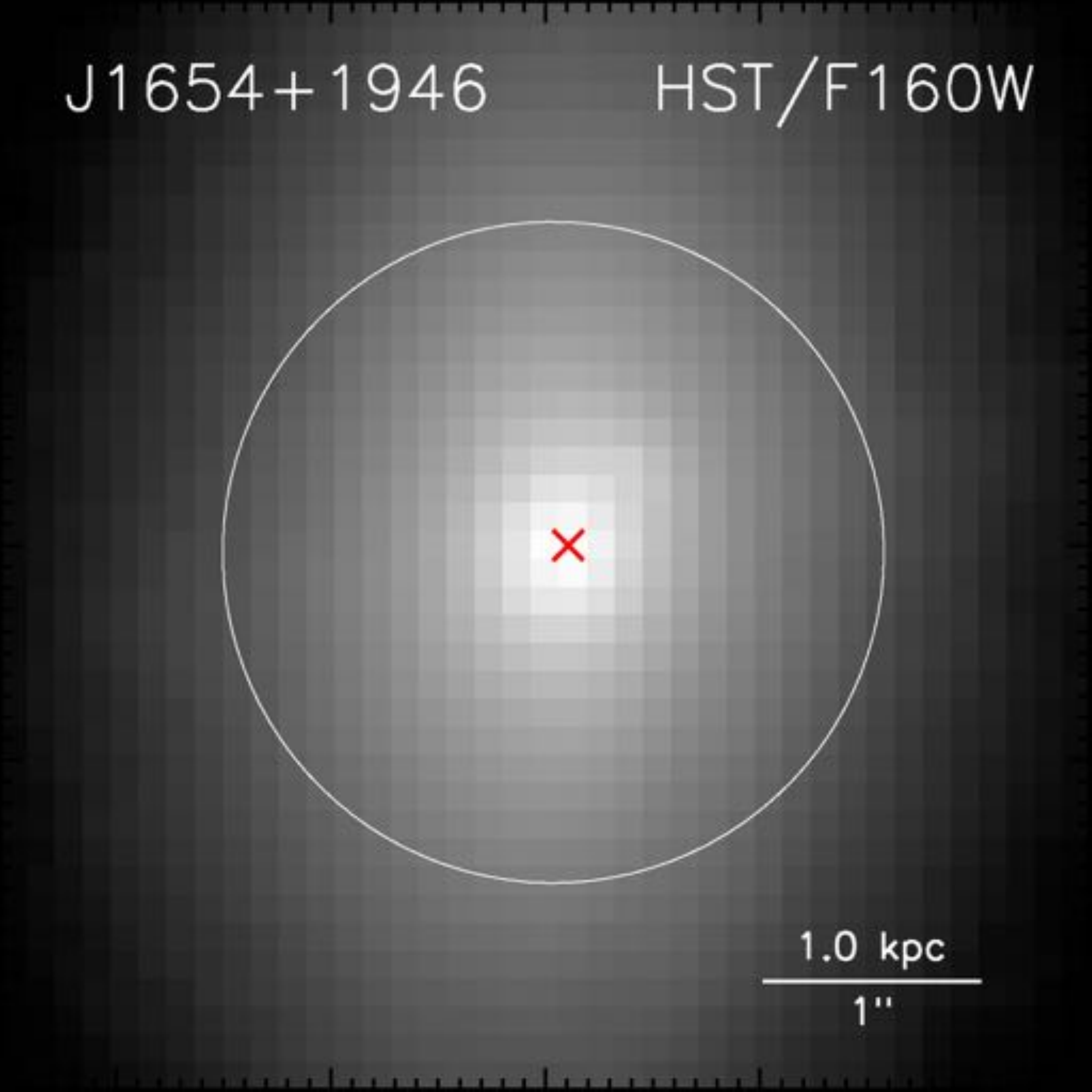}
\includegraphics[height=1.5in]{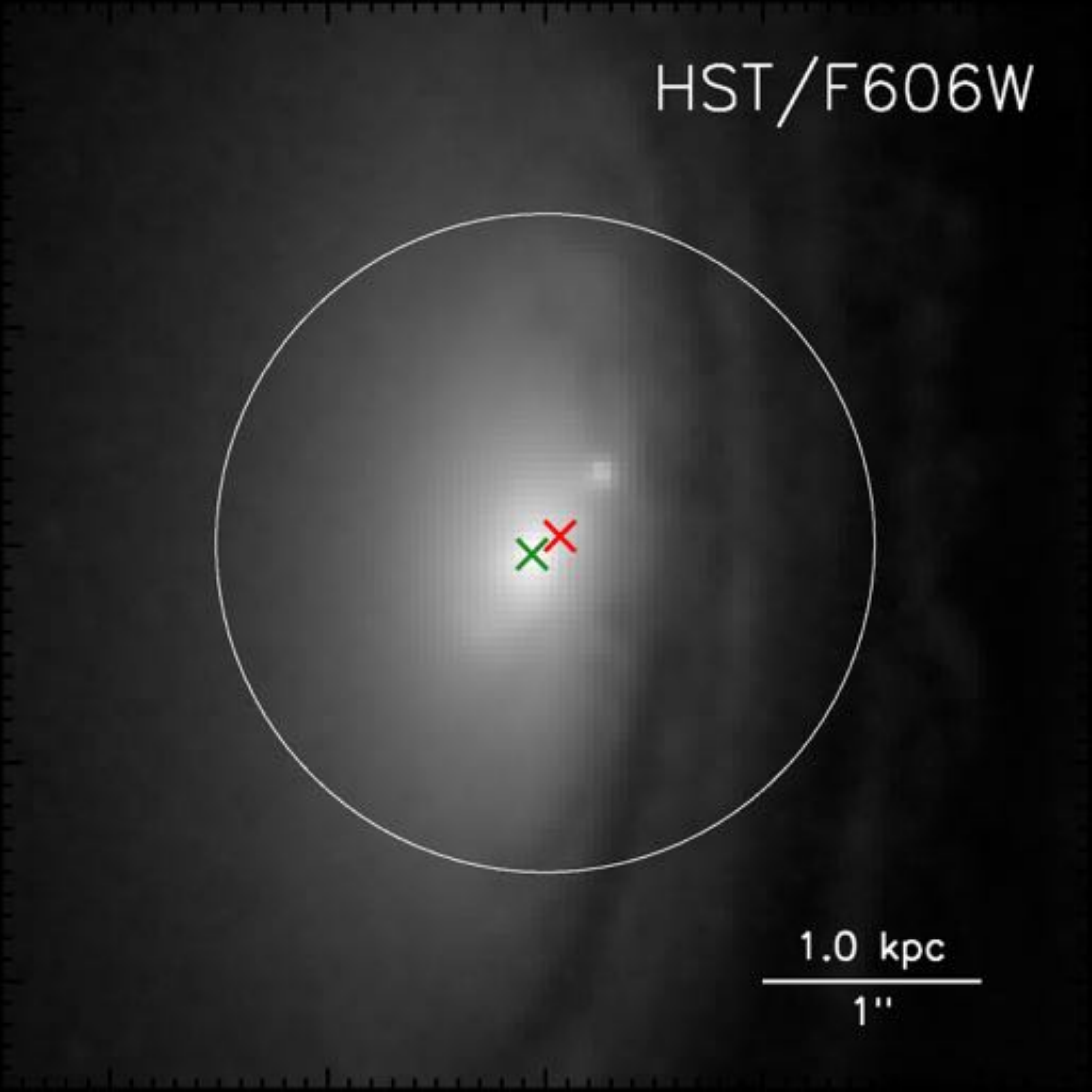}
\end{center}
\end{figure}

Finally, we measured the spatial separation between the X-ray AGN source and the center of the stellar bulge (Table~\ref{tbl-3}).  The error on each spatial separation incorporate the errors from the \texttt{beta2d} model fit to the {\it Chandra} data (Section~\ref{chandra}), the GALFIT fit to the {\it HST}/F160W data, and the astrometric uncertainty (Section~\ref{astrometry}).  The error budget is dominated by the uncertainty in aligning the {\it Chandra} and {\it HST} images, where the median astrometric uncertainty is $0\farcs5$.  Due in part to these large astrometric uncertainties, all of the spatial separations are less than $3\sigma$ in significance.

\subsection{Astrometry}
\label{astrometry}

To determine if any {\it Chandra} sources are significantly spatially offset from the stellar bulges seen in the {\it HST}/F160W data, we registered each pair of {\it HST}/F160W and $Chandra$ images and estimated their relative astrometric uncertainties.  Due to the small number of {\it Chandra}/ACIS sources and the relatively small {\it HST}/F160W field of view, we registered each image separately to SDSS ($u$, $g$, $r$, $i$, and $z$) and the 2MASS point source catalog \citep{CU03.1}.  Then, we combined the two transformations to register the {\it Chandra} and {\it HST} images.

We used \texttt{wavdetect} with a threshold of \texttt{sigthresh}=$10^{-8}$ to detect sources in {\it Chandra}, and Source Extractor with a threshold of $3\sigma$ to detect sources in SDSS, 2MASS, and {\it HST}.  Then, we matched sources in each pair of images using the \texttt{xyxymatch} task in IRAF.  Next, we used the \texttt{geomap} task in IRAF to calculate $X$ and $Y$ linear transformations for each matched pair ($X_{\mathrm{shift},j}$, $Y_{\mathrm{shift},j}$).  We took the final linear transformations in $X$ and $Y$ to be the error-weighted averages, $\overline{X_{\mathrm{shift}}}={\sum_{j=1}^{n} X_{\mathrm{shift},j}\times w_{j,X}}$ and $\overline{Y_{\mathrm{shift}}}={\sum_{j=1}^{n} Y_{\mathrm{shift},j}\times w_{j,Y}}$, where $n$ is the number of sources matched between two images and $w$ is the error weighting.

For each dimension, $X$ and $Y$, we combined in quadrature the errors on the {\it Chandra} and SDSS/2MASS source positions in each band.  We repeated this procedure to determine the uncertainty of the relative astrometry for the {\it HST} and SDSS/2MASS images. Then, we added the relative astrometric errors between {\it Chandra} and SDSS/2MASS and between {\it HST} and SDSS/2MASS in quadrature to determine the relative astrometric errors between the {\it Chandra} and {\it HST} images.  The final astrometric errors ($\overline{\Delta X}$, $\overline{\Delta Y}$) are then the error-weighted averages of these bands, shown in Table~\ref{tbl-7}.  These uncertainties range from $0\farcs2$ to $0\farcs8$, and they dominate the errors when we measure the spatial separations between sources in {\it Chandra} and {\it HST}.

\begin{deluxetable}{lllll} 
\tabletypesize{\scriptsize}
\tablewidth{0pt}
\tablecolumns{5}
\tablecaption{Astrometry Measurements} 
\tablehead{
\colhead{SDSS Name} &
\colhead{$n_{CS}$} & 
\colhead{$n_{HS}$} & 
\colhead{$\overline{\Delta X}$ ($\prime\prime$)} &
\colhead{$\overline{\Delta Y}$ ($\prime\prime$)} 
 }
\startdata  
J0132$-$1027 & 0,1,1,0,0,1 & 0,0,1,0,0,0 & 0.4656 & 0.2622 \\
J0839+4707 & 2,2,2,2,1,1 & 0,2,2,2,2,1 & 0.2612 & 0.2607 \\
J1055+1520 & 0,0,0,0,0,0 & 0,0,0,1,0,0 & 0.8382 & 0.8382 \\
J1117+6140 & 0,1,1,1,0,0 & 0,1,1,1,0,0 & 0.5299 & 0.7136 \\
J1346+5228 & 1,2,1,1,1,0 & 0,1,1,1,1,0 & 0.2837 & 0.3517 \\
J1654+1946 & 0,0,1,1,0,0 & 1,2,3,3,2,1 & 0.4721 & 0.4375 \\
J2323+1405 & 1,1,2,1,1,1 & 0,0,1,0,0,0 & 0.2035 & 0.2388
\enddata
\tablecomments{Column 2: number of sources matched between {\it Chandra} and SDSS $u$, $g$, $r$, $i$, $z$ and 2MASS images. Column 3: number of sources matched between {\it HST}/F160W and SDSS $u$, $g$, $r$, $i$, $z$ and 2MASS images. Columns 4 and 5: astrometric accuracy measurements based on matching these sources, in native $X$ and $Y$ coordinates of the {\it HST}/F160W image.}
\label{tbl-7}
\end{deluxetable}

\section{Results}

\subsection{The Galaxies Host Central AGNs, Where Shocks Produce Off-nuclear Peaks in Emission}
\label{shocks}

We use the {\it Chandra} observations to pinpoint the location of the AGN in each galaxy, and we find that each AGN's position is consistent with the host galaxy center to within $3\sigma$ (Table~\ref{tbl-3}).  Some of the AGNs may have small, but real, spatial offsets from the galaxy center, but the {\it HST}/F160W images do not show evidence of secondary stellar cores that would accompany these offset AGNs.  This leads us to conclude that each galaxy in our sample most likely hosts a central AGN, and not an offset AGN.

The emission line maps for each galaxy are probed by the {\it HST}/F606W observations, which are dominated by \oiiiwn.  We find that the emission line centroids are spatially offset from the host galaxy centers by 0.05 to 0.4 kpc, and that all of the spatial separations are greater than $3\sigma$ in significance (Table~\ref{tbl-6}).  For the three galaxies that were also observed with Keck/OSIRIS, in all three galaxies the spatial offsets of the emission in the OSIRIS data are consistent with those measured in the F606W data.

Such spatially-offset peaks in emission could be produced by photoionization of an off-nuclear cloud of gas.  Outflows and inflows can drive gas into off-nuclear dense regions, but this gas need not necessarily be excited by shocks  (e.g., \citealt{RO04.1,RO10.3}).  Spatially-offset peaks in emission can also be a signature of shocks.  Interacting gas clouds shock the gas, enhancing the ionized gas emission and producing an off-nuclear peak of emission within the narrow line region (e.g., \citealt{MA13.3}).  

To search for further evidence of shocks, we examine the optical line flux ratios $\oiiisn/ \oiiiwn$, $\oiiiwn / \hbn$, and $\oin /\oiiiwn$ measured from the SDSS spectrum of each galaxy.  Shocks driven into the surrounding gas clouds compress the gas, increasing its density and temperature.  The \oiiis emission line indicates a very high kinetic temperature, which is produced by shock wave excitation and is inconsistent with photoionized low-density clouds.  Consequently, the $\oiiis/ \oiiiwn$ line ratio is temperature sensitive and a good indicator of shock activity.  Shock heating can also be probed by the $\oiiiwn / \hbn$ line flux ratio (e.g., \citealt{SH79.1}).  The \oi emission line is another indicator of shocks (e.g., \citealt{DO76.1}), and $\oin /\oiiiwn$ is an ionization level-sensitive line flux ratio.

\begin{figure*}
\begin{center}
\includegraphics[width=3.5in]{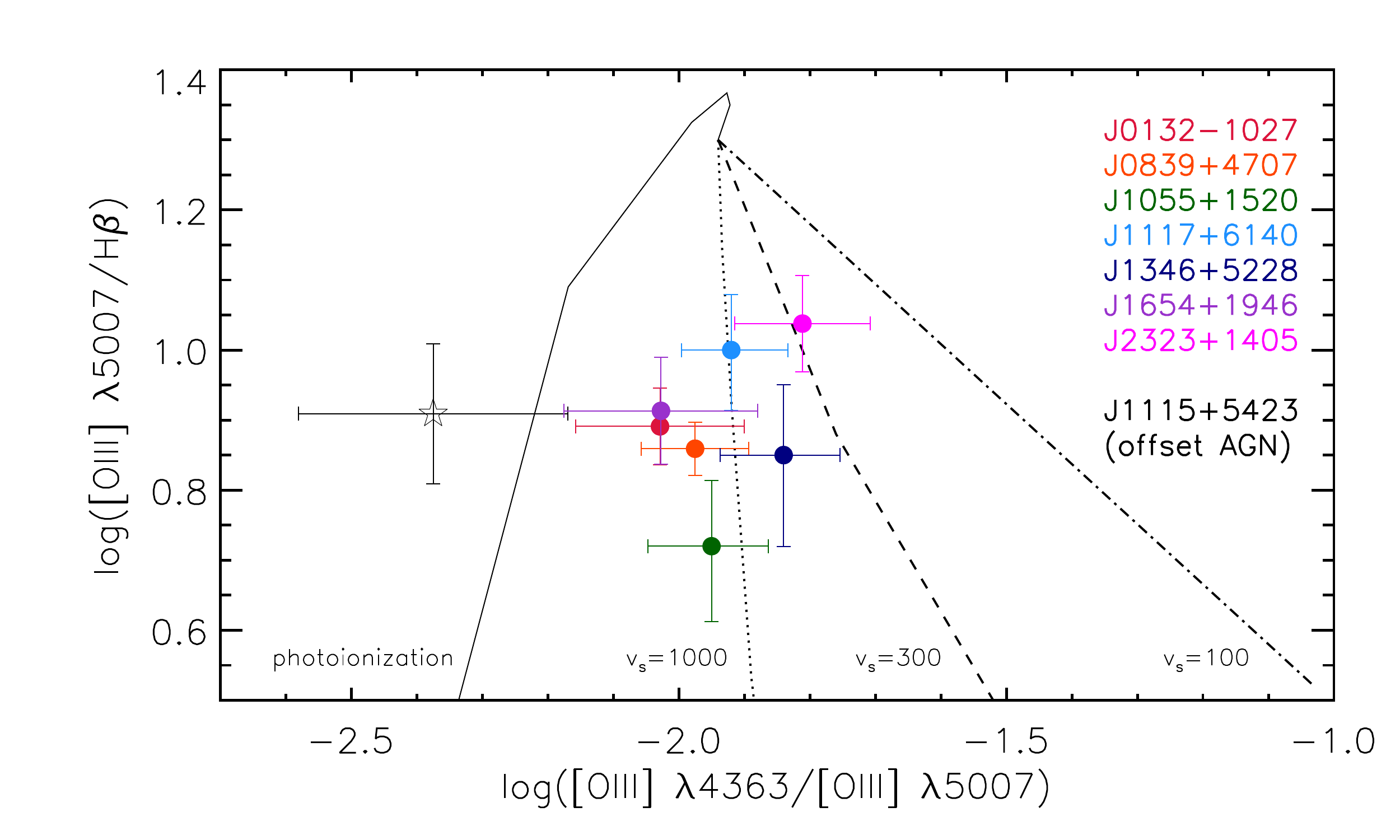}
\includegraphics[width=3.5in]{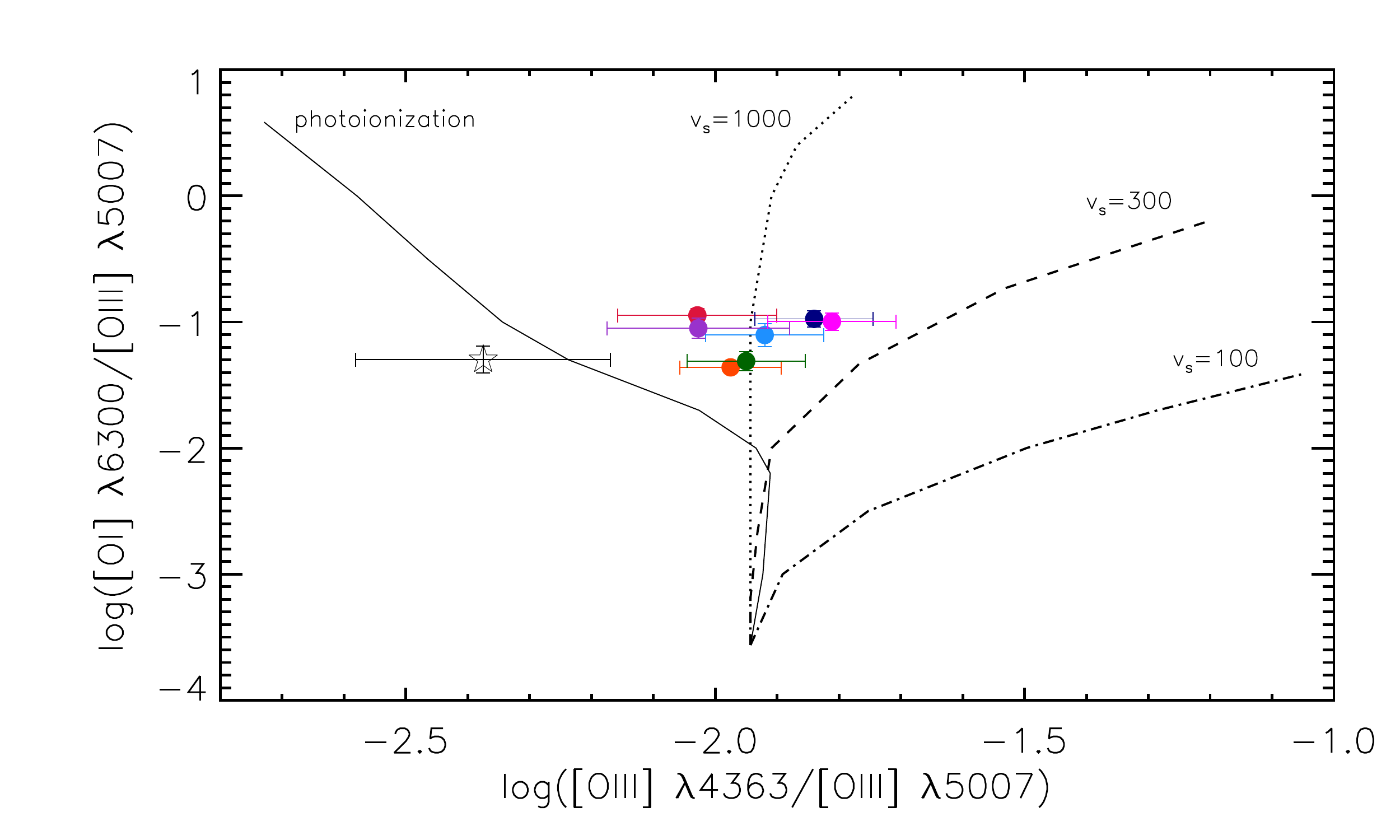}
\end{center}
\caption{Optical line flux ratios for the seven velocity-offset AGNs in this paper (filled colored circles), and a velocity-offset AGN that is an offset AGN (open black star).  Left: $\oiiiwn / \hbn$ vs. $\oiiisn/ \oiiiwn$, right: $\oin /\oiiiwn$ vs. $\oiiisn/ \oiiiwn$. The lines illustrate photoionization and shock models from \cite{MO02.1}: pure photoionization models (solid line, where the ionization parameter varies from $\log U=-4$ at the bottom to $\log U=-1$ at the top), and photoionization plus shock models with shock velocities $v_s=1000$ km s$^{-1}$ (dotted line),  $v_s=300$ km s$^{-1}$ (dashed line), and $v_s=100$ km s$^{-1}$ (dot-dashed line).  The offset AGN's emission lines are consistent with pure photoionization models, whereas the other velocity-offset AGNs (which do not have offset AGNs) have contributions from shocks.}
\label{fig:shocks}
\end{figure*}

We compare the $\oiiiwn / \hbn$ vs. $\oiiisn/ \oiiiwn$ line flux ratios, as well as the $\oin /\oiiiwn$ vs. $\oiiisn/ \oiiiwn$ line flux ratios, to models of pure AGN photoionization and combined AGN photoionization and shocks \citep{MO02.1}.   The pure photoionization models are computed with CLOUDY \citep{FE96.1} and use a spectral index $\alpha=-1$ of the ionizing continuum and an ionization parameter ranging from $-4 \leq \log U \leq -1$.  The hydrogen density is 100 cm$^{-3}$, which is typical for extended emission line regions \citep{MC90.1}, and the metallicity is solar.  The shock models are computed with MAPPINGSIII \citep{DO96.1}, and have a range of shock velocities $100 < v_s ($km s$^{-1}) < 1000$.  We find that none of the velocity-offset AGNs have line flux ratios consistent with pure photoionization, and that instead their spectra are explained by a combination of photoionization and shocks (Figure~\ref{fig:shocks}).  

To further explore the role of photoionization and shocks in these galaxies, we compare our data to the radiative shock models of \cite{AL08.1}. They assume solar abundance, a preshock density 1 cm$^{-3}$, magnetic parameters ranging from $10^{-4}$ to 10 $\mu$G cm$^{3/2}$, and shock velocities ranging from 200 to 1000 km s$^{-1}$, and they use MAPPINGSIII to model both the shock and its photoionized precursor.  For shocks with velocities $\gtrsim170$ km s$^{-1}$, the ionizing front is moving faster than the shock itself, and the ionizing front dissociates and spreads out to form a precursor \hii region in front of the shock.  Hence, a shocked region can have both shocked gas and photoionized gas.  We find that the line flux ratios of our seven velocity-offset AGNs are consistent with the shock plus precursor models of \cite{AL08.1}.

We conclude that all seven of the galaxies host both shocked gas and photoionized gas.  In the three galaxies observed with Keck/OSIRIS, the OSIRIS data show that the velocity-offset emission lines in the SDSS integrated spectra originate from the shocked off-nuclear emission peak in the gas \citep{MU16.1}.   The same is most likely true for the other four galaxies in our sample, and spatially resolved spectra would show it definitively.

\subsection{Sources of the Shocks in the Galaxies}

Here we explore the nature of the shocks in each of the seven galaxies individually. 

\subsubsection{Four AGN Outflows}
\label{outflows}

{\bf SDSS J0132$-$1027.}  This galaxy displays several colinear knots of emission (Figure~\ref{fig:hstresults}), which are often seen in radio jets driving collimated AGN outflows (e.g., \citealt{MI04.1, RO10.1,TO12.1}).  Indeed, SDSS J0132$-$1027 is detected in the FIRST radio survey \citep{BE95.1} with a 20 cm flux density of 1.6 mJy, and higher resolution radio observations would reveal whether it hosts a radio jet.  We also note that the southwestern most knot is also detected in the F160W observations.  While it is possible that this the stellar bulge of a minor merger, it seems an unlikely coincidence that the minor merger would be colinear with the other knots of emission.  Instead, the F160W observations may be tracing \feii 1.6436 $\mu$m emission, which is a common indicator of shocks (e.g., \citealt{AL97.1}) and could be produced as the jet drives into the interstellar medium. 

{\bf SDSS J1055+1520.}  The outflow in this galaxy has been modeled as a bicone using the Keck/OSIRIS observations of Pa$\alpha$, and the ratio of the outflow energy to the AGN bolometric luminosity was found to be $\dot{E}_{out}/L_{bol} = 0.06 \pm 0.015$ \citep{MU16.1}.  This exceeds the energy threshold for an outflow to drive a powerful two-stage feedback process that removes cold molecular gas from the inner parts of a galaxy and suppresses star formation, as found by theoretical studies ($\dot{E}_{out}/L_{bol} > 0.005$; \citealt{HO10.2}).  Further, the bicone is oriented with a position angle $138^\circ \pm 6^\circ$ east of north, which is consistent with the spatial orientation of the X-rays (Figure~\ref{fig:chandraresults}).  This hints that there may be spatially extended X-ray emission associated with the spatially extended ionized gas, and deeper X-ray observations would be required to confirm this.  SDSS J1055+1520 is not detected in FIRST, so its outflow is not radio jet driven.  The {\it HST} image also shows that the galaxy itself is asymmetric, which suggests that it may be a merger-remnant galaxy (Figure~\ref{fig:chandraresults}).

{\bf SDSS J1346+5228.} This galaxy's outflow was modeled as a bicone with the Keck/OSIRIS observations, and it is energetic enough ($\dot{E}_{out}/L_{bol}=0.01 \pm 0.002$; \citealt{MU16.1}) to suppress star formation in the galaxy (as was also the case for  
SDSS J1055+1520, above).  SDSS J1346+5228 is also detected in FIRST with a 20 cm flux density of 1.1 mJy, indicating that there may be a radio jet powering the outflow.

{\bf SDSS J2323+1405.}  This galaxy has symmetric emission line gas extending north and south of the galaxy center, out of the plane of the galaxy (Figure~\ref{fig:hstresults}).  This morphology is typical of AGN outflows (e.g., \citealt{MU96.1,SC03.3}), and we conclude that SDSS J2323+1405 most likely hosts an AGN outflow. 

\subsubsection{Two Inflows of Gas along a Bar}
\label{inflow}

\begin{figure*}
\begin{center}
\includegraphics[width=3in]{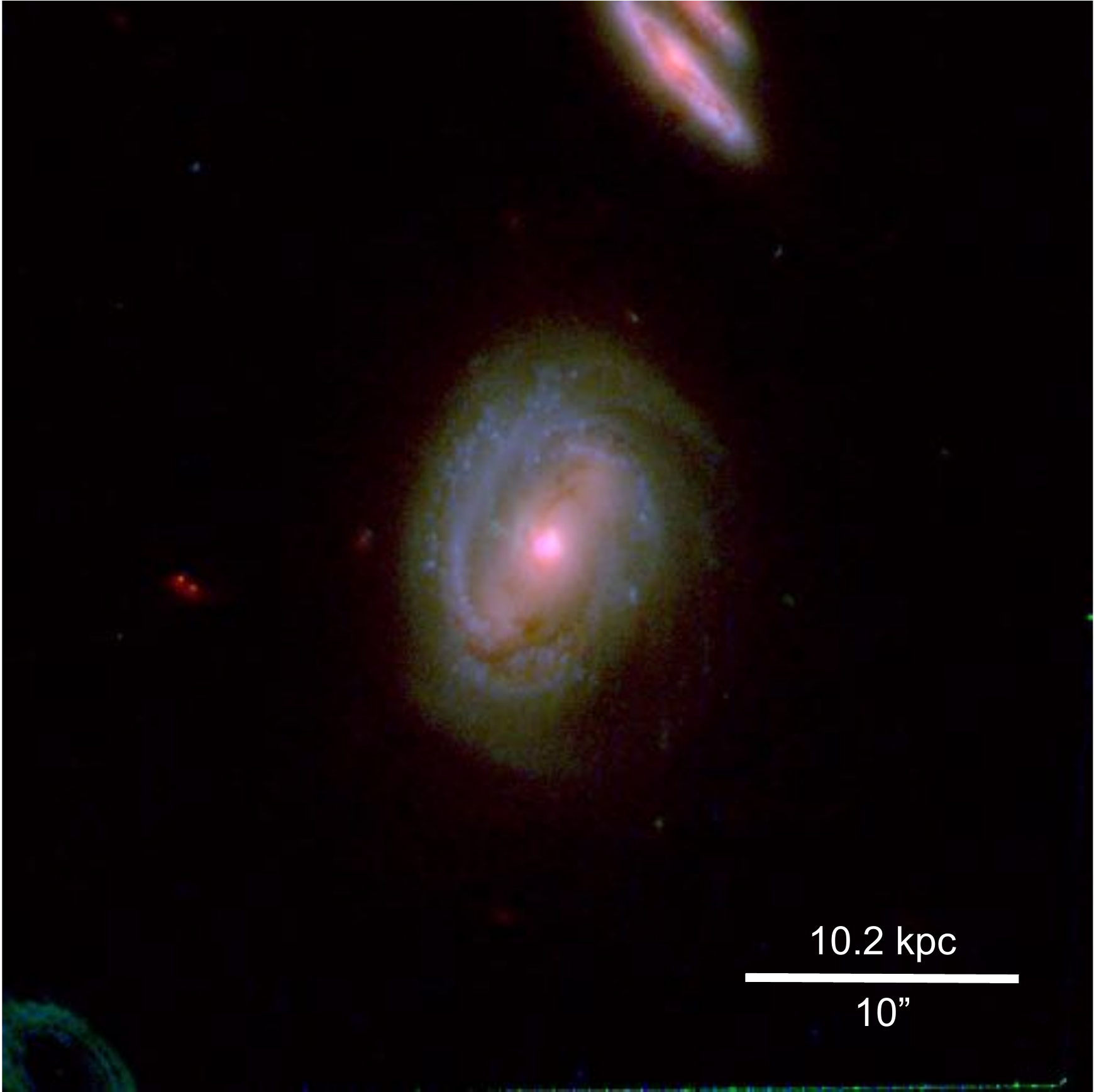}
\includegraphics[width=3in]{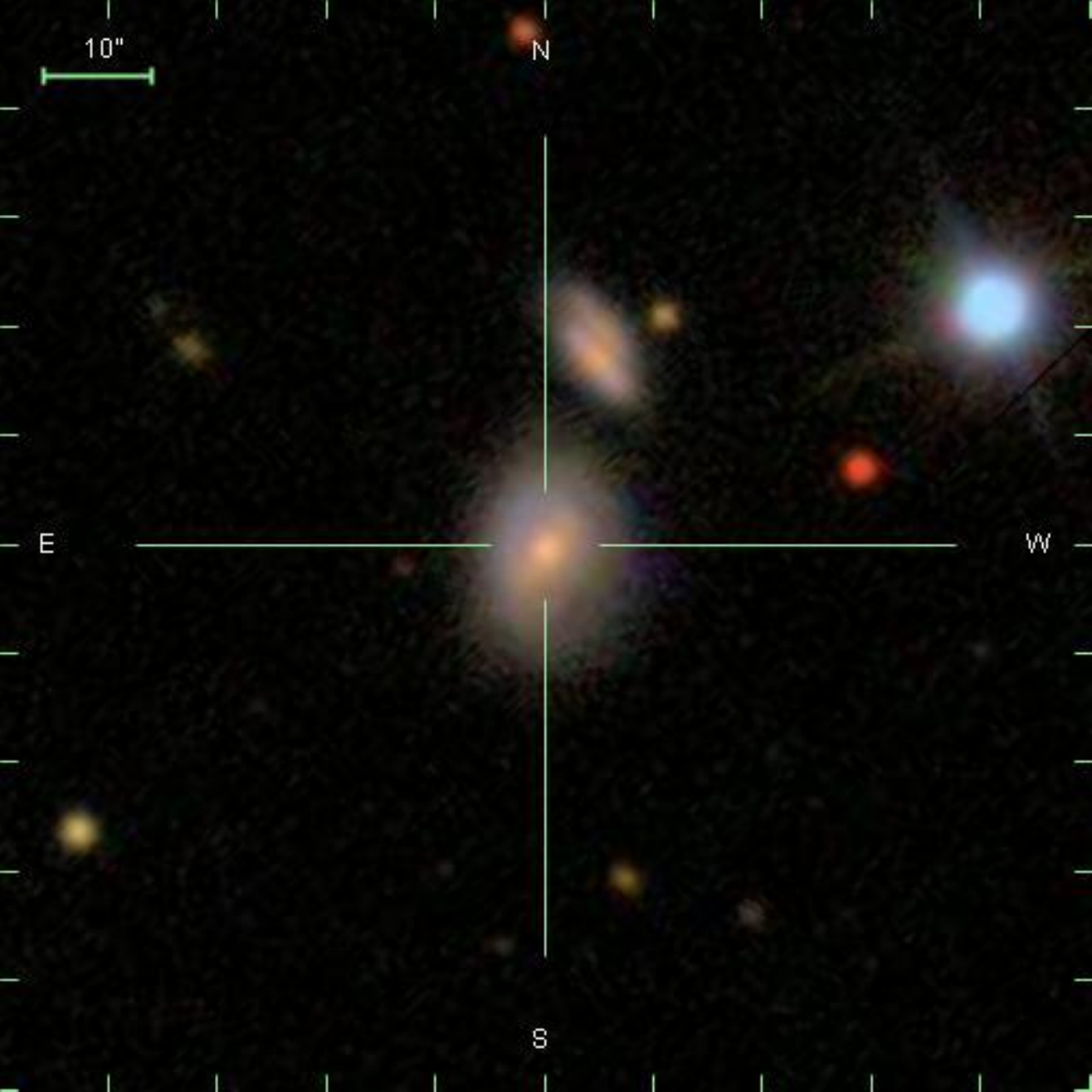}
\end{center}
\caption{Images of the galactic environment of SDSS J0839+4707.  Left: $40^{\prime\prime} \times 40^{\prime\prime}$ four-color image of {\it HST} F160W (red), F606W (green), F438W (blue), and {\it Chandra} restframe $0.5-10$ keV (purple, one-twelfth size pixels smoothed with a 16 pixel radius Gaussian kernel) observations.  The {\it HST} and {\it Chandra} images have been aligned using the astrometric shifts described in Section~\ref{astrometry}.  Right: $100^{\prime\prime} \times 100^{\prime\prime}$ SDSS $gri$ color composite image.  In both panels, north is up and east is to the left.}
\label{fig:0839companions}
\end{figure*}

{\bf SDSS J0839+4707.}  This galaxy has a stellar bar that is visible in Figure~\ref{fig:chandraresults}, and the peak of emission is spatially offset along the bar (Figure~\ref{fig:hstresults}).  SDSS J0839+4707 is also the only galaxy in our sample that has a close companion.  The companion galaxy, SDSS J083902.50+470813.9, is located 18.8 kpc ($18\farcs4$) to the northwest and has a redshift of $z=0.053454 \pm 0.000045$ (Figure~\ref{fig:0839companions}).  This corresponds to a velocity difference of $311.9 \pm 21.4$ km s$^{-1}$ redshifted away from the primary galaxy.  Emission line diagnostics of the companion's SDSS spectrum show that it is a star-forming galaxy \citep{BR04.1}.  Using the ratio of the stellar bulge luminosities as a proxy for the merger mass ratio, the merger ratio is 3.59:1 (SDSS J0839+4707 is the more massive galaxy).  There is no morphological evidence that SDSS J0839+4707 and its companion are interacting, though a future interaction may trigger new accretion onto the central AGN. 

{\bf SDSS J1117+6140.}  The OSIRIS observations of this galaxy reveal two kinematic components: a disturbed rotating disk on large scales and a counterrotating nuclear disk on the small scales of the central kpc \citep{MU16.1}.  The galaxy's stellar bar is apparent in Figure~\ref{fig:chandraresults}, and the peak of emission is spatially offset along the bar (Figure~\ref{fig:hstresults}). Based on the model of the counterrotating disk \citep{MU16.1}, the emission peak is located where the nuclear disk and the bar intersect.

\subsubsection{One Ambiguous System}

{\bf SDSS J1654+1946.}  The {\it HST} observations of this galaxy show no obvious signatures of an outflow, a bar, or a merger.  There is a knot of emission northwest of the galaxy center (Figure~\ref{fig:hstresults}), which could be a nuclear star cluster (e.g., \citealt{GE14.1}).  Since SDSS J1654+1946 is highly inclined (almost edge-on), we hypothesize that there could be a small nuclear bar that is too inclined to clearly see in the {\it HST} data.  Gas inflowing along this bar could be the cause of the off-nuclear peak in emission, though without evidence of this bar we classify this system as ambiguous.   

\subsection{Distinguishing between Velocity Offsets Produced by Shocks and by Offset AGNs}

We have determined that the velocity offsets in our seven targets are produced by shocks and not offset AGNs.  Now, for comparison, we consider a velocity-offset AGN that has been confirmed as an offset AGN: SDSS J111519.98+542316.65.  SDSS J1115+5423 is $z=0.07$ galaxy that is in the velocity-offset AGN catalog \citep{CO14.1} from which we selected the seven targets in this paper, and it is the only galaxy in that catalog that has been shown to be an offset AGN so far.  The emission lines in SDSS J1115+5423 are offset $-68.5 \pm 11.9$ km s$^{-1}$ from systemic.  By analyzing archival {\it Chandra} observations of this galaxy, \cite{BA16.1} found that it has a hard X-ray source with $L_{2-10 \mathrm{keV}}=4 \times 10^{43}$ erg s$^{-1}$ that is located $0.8 \pm 0.1$ kpc ($0\farcs64 \pm 0\farcs05$) from the host galaxy center.  This offset AGN is located within the $3^{\prime\prime}$ SDSS fiber and presumably is the source of the velocity-offset emission lines in the SDSS spectrum, which could be confirmed with  a spatially resolved spectrum of the system.  

Interestingly, SDSS J1115+5423's $\oiiiwn / \hbn$ vs. $\oiiisn/ \oiiiwn$ line flux ratios, as well as its $\oin /\oiiiwn$ vs. $\oiiisn/ \oiiiwn$ line flux ratios, are consistent with models of pure photoionization, in contrast to the seven velocity-offset AGNs studied here (Figure~\ref{fig:shocks}).  In the case of the offset AGN (SDSS J1115+5423), the emission lines are produced by photoionization from an AGN that is off-nuclear from the galaxy center but still within the SDSS fiber; this explains the velocity-offset emission lines observed in the SDSS spectrum.  On the other hand, in each of the seven velocity-offset AGNs studied here, the emission lines originate from a {\it central} (not offset) AGN.  Inflowing or outflowing gas is shocked, producing off-nuclear peaks in emission (still within the SDSS fiber) that result in the velocity-offset emission lines in the SDSS spectrum.  

Consequently, we suggest that it is possible to separate a sample of velocity-offset AGNs into offset AGNs and central AGNs (which have shocks resulting from inflows or outflows of gas) using the shocks vs. photoionization diagnostic line flux ratios $\oiiiwn / \hbn$ vs. $\oiiisn / \oiiiwn$, or $\oin /\oiiiwn$ vs. $\oiiisn / \oiiiwn$.  These line flux ratios are measurable with the SDSS spectrum alone; no follow-up observations are required.

\section{Conclusions}

We have presented {\it Chandra} and multiband {\it HST} observations of seven velocity-offset AGNs.  The seven AGNs are at $z<0.12$ and have SDSS spectra that show emission lines that are offset in line-of-sight velocity from systemic by 50 to 113 km s$^{-1}$.  To determine the nature of the velocity offset in each galaxy, we use the {\it Chandra} observations to determine the location of the AGN and the {\it HST} observations to identify the galaxy's stellar centroid and the location of the peak of the ionized gas emission.  Our main results are summarized as follows.

1.  All seven velocity-offset AGNs have central AGNs, yet each galaxy's peak in emission is spatially offset from the stellar centroid.  The spatial offsets range from 0.05 to 0.4 kpc, and they are all $>3\sigma$ in significance.  The spatially offset emission is produced by shocks, and the velocity offsets of the emission lines observed in the SDSS spectra originate from the spatially offset, shocked emission. 

2.  The shocks are produced by gas falling onto the AGN along a bar, or by AGN outflows propelling outward into the interstellar medium.  The seven velocity-offset AGNs are classified as follows: four outflows, two inflows of gas along a bar, and one ambiguous case (since this galaxy is nearly edge-on, it may have a bar that is difficult to see).

3.  All of the velocity-offset AGNs studied here fall in the regions of the $\oiiiwn / \hbn$ vs. $\oiiisn / \oiiiwn$ and $\oin /\oiiiwn$ vs. $\oiiisn / \oiiiwn$ diagrams that are consistent with a combination of photoionization and shock contributors.  However, a comparison velocity-offset AGN (where the velocity offset is caused by an offset AGN in a galaxy merger) is consistent with models of pure photoionization and no shocks.  We suggest that these emission lines, measured from the SDSS spectrum alone, may efficiently separate the velocity-offset AGNs produced by offset AGNs (photoionization only) from those produced by central AGNs with shocked gas in inflows or outflows (photoionization plus shocks). 

Additional follow-up observations, including spatially resolved spectroscopy, X-ray observations, and radio observations, of a large sample of velocity-offset AGNs could test the hypothesis that the $\oiiiwn / \hbn$, $\oiiisn / \oiiiwn$, and $\oin /\oiiiwn$ line flux ratios distinguish the offset AGNs from the central AGNs with shocks.  The offset AGNs could then be used for studies of AGN fueling during galaxy mergers (e.g., \citealt{BA17.1}), while the central AGNs with outflows may be particularly effective drivers of feedback.  Since the outflows selected from velocity-offset AGNs are outflows {\it with shocks}, these outflows have already been pre-selected to be interacting with their host galaxies.  In fact, we found that the two outflows in our sample that were modeled as bicones are energetic enough to drive cold molecular gas out of the galaxy's inner regions and regulate star formation.  Thus, AGN outflows with velocity offsets may be a rich source of examples of feedback.

\acknowledgements We thank the anonymous referee for comments that have improved the clarity of this paper.  Support for this work was provided by NASA through Chandra Award Number GO4-15113X issued by the Chandra X-ray Observatory Center, which is operated by the Smithsonian Astrophysical Observatory for and on behalf of NASA under contract NAS8-03060.  Support for HST program number GO-13513 was provided by NASA through a grant from the Space Telescope Science Institute, which is operated by the Association of Universities for Research in Astronomy, Inc., under NASA contract NAS5-26555.

The scientific results reported in this article are based in part on observations made by the Chandra X-ray Observatory, and this research has made use of software provided by the Chandra X-ray Center in the application packages CIAO, ChIPS, and Sherpa.  The results reported here are also based on observations made with the NASA/ESA Hubble Space Telescope, obtained at the Space Telescope Science Institute, which is operated by the Association of Universities for Research in Astronomy, Inc., under NASA contract NAS 5-26555. These observations are associated with program number GO-13513.

{\it Facilities:} \facility{{\it CXO}}, \facility{{\it HST}}

\bibliographystyle{apj}

\end{document}